\documentclass[10pt,twocolumn,a4paper]{article}
\usepackage[margin=18mm,top=17mm,bottom=19mm,columnsep=7mm]{geometry}
\usepackage{fontspec}
\usepackage{amsmath,amssymb}
\usepackage{graphicx,booktabs,array}
\usepackage{microtype}
\usepackage{caption}
\usepackage{dblfloatfix}
\usepackage{flushend}

\usepackage{titlesec}
\titleformat{\section}{\large\bfseries\raggedright}{\thesection}{0.5em}{}
\titleformat{\subsection}{\normalsize\bfseries\raggedright}{\thesubsection}{0.5em}{}
\titleformat{\subsubsection}{\normalsize\bfseries\raggedright}{\thesubsubsection}{0.5em}{}
\titlespacing*{\section}{0pt}{2ex plus .5ex minus .3ex}{.8ex}
\titlespacing*{\subsection}{0pt}{1.6ex plus .4ex minus .3ex}{.7ex}
\titlespacing*{\subsubsection}{0pt}{1.4ex plus .4ex minus .3ex}{.6ex}
\usepackage[numbers,sort&compress]{natbib}
\usepackage{xurl}
\usepackage[hidelinks,unicode]{hyperref}
\hypersetup{
  pdftitle={Cloudra: Geometric Reasoning and Scientific Validation for Autonomous Reaction Mechanism Exploration},
  pdfauthor={Chengyi Yang, Qianxin Zhou, Yang Li}
}
\numberwithin{equation}{section}
\title{Cloudra: Geometric Reasoning and Scientific Validation for Autonomous Reaction Mechanism Exploration}
\author{Chengyi Yang \and Qianxin Zhou \and Yang Li}
\date{\small 2 November 2026}
\makeatletter
\renewcommand{\maketitle}{%
  \twocolumn[{%
    {\centering
      {\LARGE\bfseries\@title\par}\vspace{0.6em}
      {\normalsize Chengyi Yang\,\textsuperscript{1,*}, Qianxin Zhou\,\textsuperscript{2,\textdagger}, and Yang Li\,\textsuperscript{1,\textdagger}\par}
\vspace{0.4em}
{\small \textsuperscript{1}\,College of Chemistry and Pharmaceutical Engineering, Changsha University of Science and Technology, Changsha 410114, China\par
\textsuperscript{2}\,State Key Laboratory of Photocatalysis on Energy and Environment, College of Chemistry, Fuzhou University, Fuzhou 350002, China\par}
\vspace{0.3em}
{\footnotesize \textsuperscript{*}\,Corresponding author: Chengyi Yang. E-mail: \href{mailto:202509240114@csust.edu.cn}{\nolinkurl{202509240114@csust.edu.cn}}.\par}
{\footnotesize \textsuperscript{\textdagger}\,Qianxin Zhou and Yang Li contributed equally as second authors.\par}
\vspace{0.6em}
      {\small\@date\par}\vspace{1.2em}}
    \begin{minipage}{\textwidth}
    \small
    \noindent\textbf{Abstract}\par\smallskip
    Autonomous reaction-mechanism exploration requires proposing reaction steps, preparing three-dimensional structures, and determining which reactant and product a computed saddle point connects. Large language model (LLM) agents can misjudge spatial relationships and confuse a computed structure with the species or step they intended to study.

We present Cloudra, a multi-agent framework that constrains mechanistic output through a Delivery DAG of chemical objects. Species nodes carry geometries and bonding topology; elementary-step edges carry transition states and reaction connections between species. Cheminformatics algorithms establish molecular identities and atom correspondences. Deterministic routines measure and transform structures to prepare candidate geometries. Frequency and IRC calculations determine the stationary points found and the species they connect. Agents use these objects to test proposed steps, compare pathways, and revise the mechanism.

CloudraBench evaluates the framework's end-to-end mechanism-calculation capability across 50 tasks covering elementary reactions, multistep mechanisms, stereochemical competition, and computational method limitations. Across 200 runs under four workflow configurations, the primary Guided workflow achieves a mean composite score of 0.805, compared with 0.584 for the same-model Standard baseline. On Hard-tier tasks, the Guided mean increases from 0.689 under Standard to 0.958. Under Guided, the self-consistency component reaches its maximum score in 24 of the 40 non-Impossible tasks, compared with seven under Standard. Behavioral analyses examine how the agents use bond and dihedral measurements, optimized structures, and IRC endpoints to revise local mechanistic assignments.

    \end{minipage}\par\vspace{1.2em}}]%
}
\makeatother

\begin{document}
\maketitle
\setlength{\abovedisplayskip}{6pt plus 1pt minus 1pt}
\setlength{\belowdisplayskip}{6pt plus 1pt minus 1pt}
\setlength{\abovedisplayshortskip}{6pt plus 1pt minus 1pt}
\setlength{\belowdisplayshortskip}{6pt plus 1pt minus 1pt}

\section{Introduction}

\subsection{Physical requirements for autonomous reaction mechanism research}

Reaction-mechanism validation verifies that molecular geometries, reaction pathways, and quantum-chemical results correspond to the declared species and elementary steps. Frequency analysis identifies minima and first-order saddles, and intrinsic reaction coordinate (IRC) calculations identify the species connected by each saddle. The computed endpoints determine the saddle's mechanistic assignment.\citep{ref09,ref10}

Automated reaction-network exploration organizes calculations through chemical objects. Within SCINE, Chemoton 2.0 groups structures into compounds and elementary steps into reactions, and validates optimized saddles through IRC calculations, molecular-graph comparisons, and product optimization.\citep{chemoton2022} AutoRXN combines autonomous reaction exploration with density-functional calculations, coupled-cluster energy refinement, and automated multireference diagnostics.\citep{autorxn2023}

Computational chemistry agents automate several parts of this work. El Agente Q \citep{ref01}, Aitomia \citep{ref02}, and Gráfico \citep{ref06} orchestrate quantum-chemical calculations and typed workflows. Estructural \citep{ref04} and ARCHE \citep{ref05} provide tools for molecular fragment editing, atom mapping, and transition-state geometry preparation; Quntur \citep{ref03} addresses reaction kinetics and transition-state research. Extending these capabilities to autonomous mechanism exploration requires reliable geometric operations and physical interpretation of the species and pathways obtained.

Aligning stereochemical orientations, assembling reactant fragments, and preparing pathway endpoints require three-dimensional geometric reasoning. Language models can make errors when they infer spatial relationships or directly manipulate Cartesian coordinates. Quntur \citep{ref03}, for example, reports the need for human assistance with endpoint alignment in complex pathway and transition-state preparation. In Cloudra, an agent can ask the program to rotate the atoms on one side of a bond, measure the resulting dihedral angle, and compare the new geometry with another transition-state guess.

\subsection{Cloudra: autonomous mechanism exploration through chemical objects}

Tool-oriented workflow graphs use calculation calls as nodes and execution dependencies as edges. Cloudra is a multi-agent system for small-molecule mechanism exploration that represents the delivered reaction mechanism through species and elementary-step objects in a Delivery DAG. Agents express candidate species and elementary steps in this graph, choose geometric operations and subsequent calculations from the associated structures, energies, and pathway results, and deliver the resulting mechanism through the same objects.

\begin{figure}[tbp]
\centering
\includegraphics[width=\linewidth]{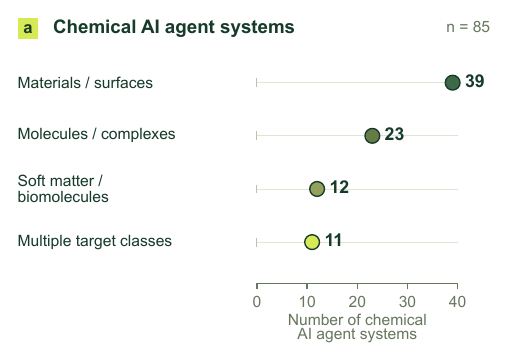}
\caption{Physical targets in a structured literature sample of 85 chemical AI agent systems: 39 materials/surface, 23 finite-molecule/complex, 12 soft-matter/biomolecular, and 11 cross-class systems.}
\label{fig:01}
\end{figure}

\begin{figure}[tbp]
\centering
\includegraphics[width=\linewidth]{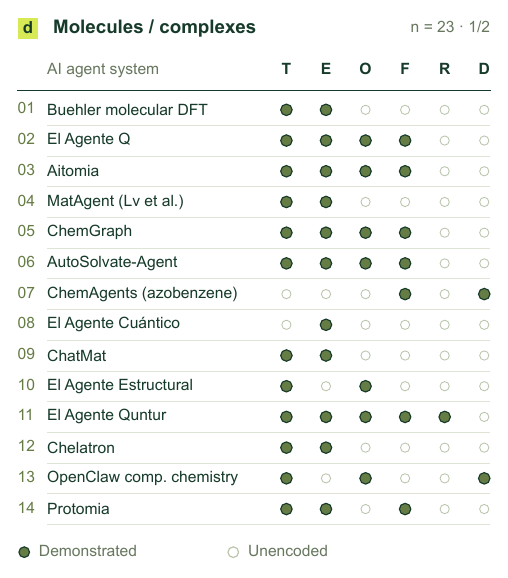}
\caption{Documented tasks in the finite-molecule/complex class (panel 1 of 2; 23 chemical AI agent systems in total). T: structure construction/editing; E: static energy/properties; O: geometry optimization; F: vibrations/thermochemistry; R: explicit pathways/TS; D: molecular/quantum dynamics. Filled cells denote positive evidence; blanks leave capability coverage unspecified.}
\label{fig:02}
\end{figure}

\begin{figure}[tbp]
\centering
\includegraphics[width=\linewidth]{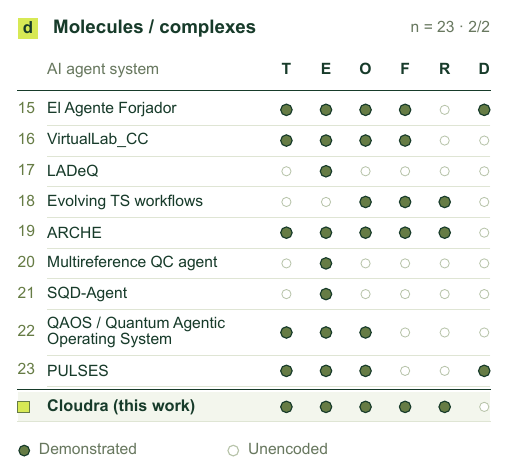}
\caption{Chemical AI agent systems in the finite-molecule/complex class, continued (panel 2 of 2), using the codes in Figure~\ref{fig:02}. Cloudra (this work) is benchmarked at the bottom.}
\label{fig:03}
\end{figure}

Figure~\ref{fig:01} summarizes a structured literature sample of 85 chemical AI agent systems: 39 concern materials or surfaces, 23 concern finite molecules or complexes, 12 concern soft matter or biomolecules, and 11 cover multiple categories of objects. Figures~\ref{fig:02} and \ref{fig:03} show the documented tasks in the finite-molecule/complex category and compare Cloudra's capabilities with those of the existing systems. The accompanying literature materials contain the complete matrix and the sources for each entry.

Cloudra uses deterministic routines to calculate and modify molecular coordinates. MCS matching selects molecular fragments; atom mapping matches reactant and product atoms and identifies atom pairs whose bonds change. The program measures bond lengths, bond angles, dihedral angles, and stereochemistry, and rotates or translates selected atoms to prepare candidate structures.

The Delivery DAG represents the mechanism as a graph of chemical objects. Each species node carries molecular geometries, bonding topology, stereochemistry, and associated calculation results. Each elementary-step edge carries the transition state and the reaction path connecting adjacent species. Topology and stereochemistry comparisons identify species, while atom mapping tracks corresponding atoms across reactant and product structures. Frequency calculations characterize the stationary points, and IRC endpoint calculations determine their reaction connections. The program checks that the frequency calculation and both IRC directions refer to the same first-order saddle, and that optimizations from the IRC terminal frames reach the declared reactant and product species. The Supervisor reviews these objects against the task and identifies calculations still needed and steps requiring further investigation.

\subsection{Contributions}

Cloudra's central design connects LLM research decisions and mechanism delivery through chemical objects. The Delivery DAG specifies the species and steps under investigation and the calculation results required to support them. Molecular topology and stereochemistry identify species; atom mappings track atoms across structures. Geometry operations construct candidates, and frequency and IRC calculations assign the resulting stationary points and reaction connections. The agents use these objects to extend and revise the mechanism as the study progresses.

The geometric procedures include MCS-based fragment selection, graph traversal to select the atoms that move, and transition-state fragment assembly. They allow the agents to rotate selected groups, adjust fragment positions, measure the resulting structures, and prepare new guesses after an unsuccessful transition-state search.

We evaluate the framework's end-to-end mechanism-calculation capability using CloudraBench, which measures graph coverage, physical self-consistency, and task-defined target accuracy from native computational outputs. Its 50 tasks cover elementary and multistep reactions, stereochemistry, and the applicability limits of the allowed methods. The experiments comprise 48 independent runs with three repeats on four tasks and a completed study of 200 runs under four workflow configurations across all 50 tasks. Standard uses the same model backbone as the primary comparison; Record and Brief-errors examine evidence organization and diagnostic feedback.

\section{System Design and Methods}

\subsection{System Overview and Division of Responsibilities}
Cloudra investigates reaction mechanisms by constructing candidate molecular configurations, locating and characterizing stationary points, tracing elementary reaction paths, and comparing competing channels under the specified reaction conditions and computational requirements. A workflow graph organizes tool calls and computational tasks by their execution dependencies. Cloudra's Delivery DAG represents the mechanism being delivered: its nodes are species occurrences, and its edges are elementary reaction steps. Agents must express candidate species and their reaction connections in this chemical graph. They choose subsequent geometric operations, calculations, and exploration branches from their hypotheses and returned results.

Canonical topology and stereochemical comparisons identify the species represented by a structure, while atom mapping identifies corresponding atoms across reactant and product configurations. Geometric operators rotate, translate, and assemble molecular fragments. Frequency calculations determine stationary-point curvature, and IRC calculations followed by endpoint optimization and frequency analysis determine which species a saddle point connects.

The Coordinator proposes mechanistic hypotheses and candidate elementary steps, and selects which branches to investigate. Computational agents prepare three-dimensional structures with the geometric operators and submit quantum-chemical calculations for their assigned steps. If a geometry diverges, a saddle-point optimization fails to converge, or a path bifurcates, the computational agent examines the structures and calculation diagnostics to decide whether to continue the step. If the step appears blocked, it reports the diagnosis to the Coordinator, which revises the hypothesis or explores another branch. The Supervisor reviews the species, steps, and calculation results in the DAG against the task requirements. Its review notes are stored separately from the facts parsed from calculation outputs.

\begin{figure*}[!tbp]
\centering
\includegraphics[width=\textwidth]{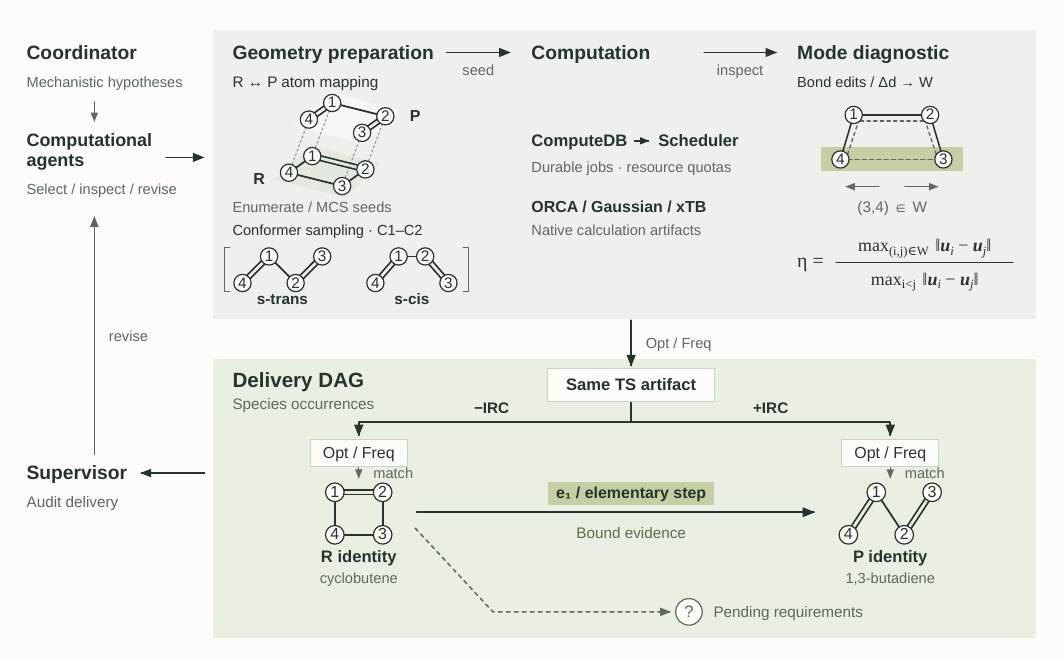}
\caption{Cloudra's mechanism exploration workflow. Atom mapping and conformer sampling prepare reactant and product structures; ComputeDB and the scheduler manage the calculations. Mode diagnostics measure motion on monitored atom pairs. Bidirectional IRCs start from the same transition state, and their outer frames undergo endpoint optimization and frequency analysis. The resulting structures are compared with the species declared in the Delivery DAG. The Supervisor reviews the graph and identifies calculations to complete and assignments to resolve.}
\label{fig:04}
\end{figure*}

\subsubsection{Asynchronous Computation Scheduling and Resource Management}
The calculation database (\texttt{ComputeDB}) stores each job's status. Separate capacity pools limit concurrent calculations on local workers, remote SSH clusters, and Slurm queues. Before dispatch, the scheduler computes a SHA-256 digest from the Cartesian coordinates, spin, charge, and level of theory. The scheduler reuses a matching active job or cached result. Operating-system exclusive file locks (\texttt{flock}) track running processes and concurrency; quantum-chemistry processes holding active locks remain running during monitor exceptions. When a calculation ends, a scheduler notification triggers asynchronous output reading by the computational agent.

\subsection{Molecular Representations and Geometric Operators}
A molecular input consists of an attributed graph $H=(A,B)$ and Cartesian coordinates $\mathbf{X}\in\mathbb{R}^{n\times 3}$. Here $A$ is the atom set, $B$ is the bond set, and each row of $\mathbf{X}$ gives one atom position. The graph stores atom and bond attributes. Geometric operators change the coordinates and return measurements of the resulting structure. Replacing atoms or functional groups also changes $H$.

\subsubsection{Atom Correspondence and Reaction Centers}
Atom attributes retain element, isotope, formal charge, hydrogen count, radical-electron count, and declared stereochemistry; bond attributes retain order and stereochemical labels. Existing coordinates are retained. ETKDGv3 \citep{ref12,etkdg2020,ref13} embeds the remaining components, which are then assembled if necessary. Complete, unique atom-map labels on both endpoints define the correspondence directly and must agree in element and isotope. Otherwise, let $A_{\mathrm{P}}$ and $A_{\mathrm{R}}$ be the product and reactant heavy-atom sets. Candidate bijections $\pi:A_{\mathrm{P}}\to A_{\mathrm{R}}$ satisfy $z_i^{\mathrm{P}}=z_{\pi(i)}^{\mathrm{R}}$ and $\iota_i^{\mathrm{P}}=\iota_{\pi(i)}^{\mathrm{R}}$, where $z$ is atomic number and $\iota$ is isotope label. Each candidate is scored by

\begin{equation}\label{eq:methods-01}
\begin{aligned}
J(\pi)&=2(N_{\mathrm{f}}+N_{\mathrm{b}})+N_{\mathrm{o}}\\
&\quad{}+4\Delta_H+8\Delta_q+8\Delta_r+2N_{\mathrm{s}}.
\end{aligned}
\end{equation}

Let $B_{\mathrm{P}}^\pi$ denote product bonds expressed in reactant atom indices. Bond formation and cleavage are the set differences $B_{\mathrm{P}}^\pi\setminus B_{\mathrm{R}}$ and $B_{\mathrm{R}}\setminus B_{\mathrm{P}}^\pi$, with cardinalities $N_{\mathrm{f}}$ and $N_{\mathrm{b}}$. The count $N_{\mathrm{o}}$ includes pairs in $B_{\mathrm{P}}^\pi\cap B_{\mathrm{R}}$ whose bond orders differ. For $a\in\{H,q,r\}$, the mismatch is $\Delta_a=\sum_{i\in A_{\mathrm{P}}}|a_i^{\mathrm{P}}-a_{\pi(i)}^{\mathrm{R}}|$, where $H$, $q$, and $r$ are the atom hydrogen count, formal charge, and radical-electron count. $N_{\mathrm{s}}$ counts stereochemical mismatches. Exactly one declared tetrahedral label contributes one mismatch, increasing $J$ by 2. When both corresponding atoms carry tetrahedral labels, the comparison requires four corresponding neighbors, including a single hydrogen where applicable; it corrects for neighbor-order parity before comparing the labels. For two-sided labels, only comparable environments enter mismatch scoring.

For element–isotope classes of sizes $n_g$, the search space contains $N_\Pi=\prod_g n_g!$ bijections. All are enumerated when $N_\Pi\le 10^5$. Above this limit, MCS seeds use element matching, unrestricted bond-order matching, ring-to-ring matching, and complete rings \citep{ref13}. The MCS timeout is 10 s, and each endpoint contributes at most 64 substructure matches. Unmatched atoms are paired within element–isotope classes after sorting by hydrogen count, degree, and input index. Candidates with the same formed, broken, and order-changed bonds and the same score are merged, keeping the lexicographically smallest mapping for each group. Representatives are sorted by score and then atom-map sequence; the first 16 form the heavy-atom candidate set $C$. For nonempty $C$, define optimal score $J_{\min}=\min_{\pi\in C}J(\pi)$ and tolerance set $\mathcal{N}=\{\pi\in C:J(\pi)\le J_{\min}+2\}$. An unlabeled mapping is reported as unique when $|\mathcal{N}|=1$, ambiguous when $|\mathcal{N}|>1$, and inconclusive when $C=\emptyset$.

Each retained heavy-atom mapping is expanded to all atoms using matches compatible with both endpoint graphs. Compatible explicit hydrogen labels are preserved first; remaining hydrogens are paired by isotope on corresponding heavy atoms. For remaining isotope groups of sizes $h_g$, the remaining $N_H=\prod_g h_g!$ permutations are enumerated when $N_H<257$. An expansion with $N_H\ge 257$ requires compatible explicit hydrogen labels that resolve the permutation space; unresolved expansions are reported as inconclusive. Complete maps are deduplicated, ranked by score and atom-map sequence, and evaluated with the same +2 ambiguity criterion. The selected all-atom bijection reorders the product into reactant atom order. Formed and broken bonds identify the atom pairs monitored during path interpolation; bond-order changes are recorded with the transformed graph.

\subsubsection{Structural Measurements and Rigid Body Operations}
The system calculates bond lengths, bond angles, dihedral angles, ring conformations, and close contacts from the coordinates. For coordinates $\mathbf{x}_i$, the interatomic distance is $d_{ij}=\|\mathbf{x}_j-\mathbf{x}_i\|$; the bond angle at vertex $j$ is obtained from the normalized dot product of the two bond vectors. Distances are compared with reference bond lengths and van der Waals contact scales to flag geometric anomalies. For coordinate-only inputs, connectivity is determined at entry. Local coordinate transformations retain this bond graph and repeat the measurements before and after moving atoms.

Default fragment selection on either side of an acyclic bond uses graph connectivity: the specified bond $a-b$ is temporarily removed, and the connected component containing $b$, denoted $\mathcal{F}_b$, is selected. If the endpoints remain connected after bond removal, movement within a ring framework requires an explicit atom group. Let $\mathcal{F}$ denote the explicitly specified moving atom group or the component $\mathcal{F}_b$ obtained by bond removal. To rotate the selected fragment about an axis, Rodrigues' formula is applied to every $i\in\mathcal{F}$:

\begin{equation}\label{eq:methods-02}
\begin{aligned}
\mathbf{x}_i'&=\mathbf{a}+\mathbf{R}(\hat{\mathbf{u}},\theta)(\mathbf{x}_i-\mathbf{a}),\\
\mathbf{R}&=\cos\theta\,\mathbf{I}+(1-\cos\theta)\hat{\mathbf{u}}\hat{\mathbf{u}}^{\mathsf{T}}\\
&\quad+\sin\theta\,[\hat{\mathbf{u}}]_\times.
\end{aligned}
\end{equation}

Here, $\mathbf{a}$ is a point on the axis, $\hat{\mathbf{u}}$ its unit direction, $\theta$ the rotation angle, $\mathbf{I}$ the identity matrix, and $[\hat{\mathbf{u}}]_\times \mathbf{v}=\hat{\mathbf{u}}\times\mathbf{v}$. The operation fixes the coordinates of atoms outside the selected set. Stretching bond $a-b$ translates $\mathcal{F}_b$ as a whole:

\begin{equation}\label{eq:methods-03}
\mathbf{x}_i'=\mathbf{x}_i+(L-d_{ab})\frac{\mathbf{x}_b-\mathbf{x}_a}{d_{ab}},\qquad i\in\mathcal{F}_b.
\end{equation}

Here $L$ is the target bond length; coordinates outside $\mathcal{F}_b$ remain fixed. Chirality inversion rotates two separable branches around their center to exchange branch directions. Group substitution removes the specified-side fragment, docks the new group along the original attachment-bond direction, and updates the graph. To compare structures before and after optimization, the system differences distance-perceived connectivity sets under the same element sequence and classifies gained or lost contacts as intra- or interfragment changes.

\subsubsection{Branch Crowding and Reactant Preorganization}
After checking hybridization, ring structure, and substituent graph inequivalence, the system calculates crowding around each candidate axis from the atoms in its substituent branches:

\begin{equation}\label{eq:methods-04}
B_s=\sum_{k\in T_s} c_k r_k \rho^{\ell_k},\qquad\rho=0.5,\quad 0\le \ell_k\le 5.
\end{equation}

Here $r_k$ represents the van der Waals radius of atom $k$. $T_s$ is generated by a last-in-first-out depth-first traversal starting from the substituent root at depth zero, with the adjacent axis-end atom marked visited. On popping an unvisited atom, the algorithm records its traversal depth $\ell_k$ and pushes its unvisited neighbors if $\ell_k<5$. The weighting coefficient is $c_k=1.5$ for a root of graph degree at least 3, and $c_k=1$ otherwise.

The descriptor uses standard van der Waals radii in Å: H 1.20, B 1.92, C 1.70, N 1.55, O 1.52, F 1.47, P 1.80, S 1.80, Cl 1.75, Br 1.85, and I 1.98 (1.70 fallback for unlisted elements). The four branch values are rounded to three decimal places and summed to obtain the axis score $B_{\mathrm{axis}}$; $B_{\mathrm{axis}}\ge 9.0$ flags the axis as congested for subsequent conformational inspection.

During multicomponent preorganization, the primary fragment is fixed at the origin. Remaining fragments undergo orientation sampling on the unit sphere. With $K$ samples, direction $s$ is

\begin{equation}\label{eq:methods-05}
\begin{aligned}
\hat{\mathbf{u}}_s&=(a_s\cos s\gamma,\,y_s,\,a_s\sin s\gamma),\\
a_s&=\sqrt{1-y_s^2},\\
y_s&=1-\frac{2s}{K-1},\\
\gamma&=\pi(3-\sqrt{5}).
\end{aligned}
\end{equation}

Here $\gamma=\pi(3-\sqrt{5})\approx 2.3999\text{ rad}$ ($137.5^\circ$) represents the Golden Angle, generating a Fibonacci spherical spiral grid for uniform distribution of directions. The axial twist about $\hat{\mathbf{u}}_s$ is $2\pi s/K$ (default $K=32$). Translation along the interfragment vector resolves steric overlap. For a set $W$ of guided atom pairs with target distance windows $[l_{ij},u_{ij}]$, placement $\mathbf{X}$ is scored by

\begin{equation}\label{eq:methods-06}
S(\mathbf{X})=\max_{(i,j)\in W}\{0,\,l_{ij}-d_{ij}(\mathbf{X}),\,d_{ij}(\mathbf{X})-u_{ij}\}.
\end{equation}

Configurations with low $S(\mathbf{X})$ undergo distance-restrained MMFF94 force-field optimization\citep{mmff941996} to bring the specified atom pairs into their target distance windows while relaxing nonbonded contacts. Candidates are retained only if the force-field optimization converges, the distances fall within their windows, stereochemistry is preserved, and fragment contacts satisfy $d_{i^*j^*}\le 1.7(r_{i^*}+r_{j^*})$. The retained configurations are ranked by unconstrained single-point energy and clustered with a heavy-atom Kabsch RMSD threshold of 0.5 Å to provide starting structures.

\subsubsection{Endpoint Alignment and IDPP Interpolation}
Let $\mathbf{P}$ and $\mathbf{R}$ denote the centered, atom-matched product and reactant coordinates. Kabsch alignment\citep{ref07} minimizes $\|\mathbf{P}\mathbf{Q}-\mathbf{R}\|_{\mathrm{F}}^2$ over proper rotations (\mbox{$\mathbf{Q}^{\mathsf{T}}\mathbf{Q}=\mathbf{I}$}, \mbox{$\det\mathbf{Q}=1$}).

\begin{samepage}
Given the singular-value decomposition $\mathbf{P}^{\mathsf{T}}\mathbf{R}=\mathbf{U}\boldsymbol{\Sigma}\mathbf{V}^{\mathsf{T}}$, the optimal rotation is

\begin{equation}\label{eq:methods-07}
\mathbf{Q}=\mathbf{U}\operatorname{diag}\!\bigl(1,1,\det(\mathbf{U}\mathbf{V}^{\mathsf{T}})\bigr)\mathbf{V}^{\mathsf{T}}.
\end{equation}
\end{samepage}

The resulting matrix $\mathbf{Q}$ is the proper rigid-body rotation matrix. The determinant correction factor $\det(\mathbf{U}\mathbf{V}^{\mathsf{T}})$ enforces $\det\mathbf{Q}=+1$, ruling out improper reflections and preserving chiral configuration.

For $M$ images including endpoints, define $\lambda_k=k/(M-1)$ and initialize $\mathbf{X}_k=(1-\lambda_k)\mathbf{R}+\lambda_k\mathbf{P}\mathbf{Q}$. The image-dependent pair potential (IDPP) method \citep{ref08} interpolates each image's target pair distances between the endpoints and optimizes its coordinates with the potential

\begin{equation}\label{eq:methods-08}
\begin{aligned}
d_{ij,k}^*&=(1-\lambda_k)d_{ij}^{\mathrm{R}}+\lambda_k d_{ij}^{\mathrm{P}},\\
V_k&=\sum_{i<j}\frac{(d_{ij,k}-d_{ij,k}^*)^2}{d_{ij,k}^4}.
\end{aligned}
\end{equation}

The inverse-fourth-power weighting $d_{ij,k}^{-4}$ imparts stiffness to short covalent bonds, prioritizing the preservation of basic covalent distances while allowing long-range nonbonded contacts to relax. Standard mode couples intermediate images with improved-tangent nudged elastic band (NEB) springs and optimizes with MDMin, evaluating convergence by the maximum force residual across movable atoms.\citep{neb2000,ase2017} Lightweight mode evaluates target distances with an objective denominator $\max(d_{ij,k}^*,0.1\,\text{Å})^4$ and relaxes intermediate images independently using Adam optimization.\citep{adam2015}

Active monitoring pairs $W$ incorporate bond-state changes, pairs bonded at both endpoints with $|\Delta d|\ge 0.15$ Å, and the five atom pairs exhibiting the largest endpoint distance variations. Within the monitoring set $W$, the primary reaction coordinate and its midpoint distance are
\[
\begin{aligned}
p^*&\in\operatorname{argmax}_{p\in W}|d_p^{\mathrm{P}}-d_p^{\mathrm{R}}|,\\
m&=(d_{p^*}^{\mathrm{R}}+d_{p^*}^{\mathrm{P}})/2.
\end{aligned}
\]
The initial transition-state guess is extracted from the first image $k$ satisfying $(d_{p^*,k}-m)(d_{p^*}^{\mathrm{R}}-m)\le 0$, or the image minimizing $|d_{p^*,k}-m|$. The selected configuration serves as the initial seed for saddle-point optimization.

\subsection{Chemical Species and Reaction Steps in the Delivery DAG}

\subsubsection{Species Occurrences and Elementary Steps}
The Delivery DAG is $G=(V,E)$. Each vertex $v\in V$ represents a species at one position in the reaction sequence, and each directed edge $e=(v_i,v_j)\in E$ represents an elementary step. The mapping $\kappa:V\to\mathcal{S}$ gives the canonical chemical identity of each species occurrence. An initial and a regenerated catalyst can therefore occupy separate nodes while sharing the same chemical identity.

Each species is described by molecular geometry, bonding topology, and stereochemistry. Its incoming and outgoing step edges describe its reaction connections to adjacent species. Canonical molecular graphs and stereochemical labels are compared when assigning structures to species; atom mapping identifies corresponding atoms in reactant and product structures. Species nodes record stationary-point geometries, frequencies, and absolute energies; step edges record transition-state geometries, imaginary-mode displacement vectors, and reaction paths. Frequency and IRC calculations characterize these stationary points and paths, with the physical criteria defined below.

For each species or step, $K$ lists the required calculation stages. Standard elementary steps require the sequence $K=\{\text{ts\_opt}, \text{frequency}, \text{short\_irc}\}$, with thermochemical corrections added when required by the task rubric. Calculation outputs are archived as immutable records identified by content digests. Each parsed result is recorded as $f=(c,o,r,d)$: $c$ names the check, $o\in\{+,-,?\}$ records support, contradiction, or uncertainty, $r$ identifies the source calculation output, and $d$ contains the measured values.

For a species or step $t$, let $F_t$ contain its parsed results, $R$ its required property checks, and $O$ its optional checks. The physical validation criterion is

\begin{equation}\label{eq:methods-09}
\begin{aligned}
S_t&=\mathbf{1}[R\ne\emptyset]\\
&\quad{}\wedge\bigwedge_{c\in R}\mathbf{1}[\exists f\in F_t:f.c=c,\,f.o=+]\\
&\quad{}\wedge\mathbf{1}[\nexists f\in F_t:f.c\in R\cup O,\,f.o=-].
\end{aligned}
\end{equation}

When $R\ne\emptyset$, each required check must have a supporting result, and every current result for a check in $R\cup O$ must indicate support or uncertainty. When $R=\emptyset$, only the calculation-stage requirements are checked.

\subsubsection{Curvature and Reaction Mode Diagnostics}
Stationary-point curvature is evaluated from the mass-weighted Hessian:

\begin{equation}\label{eq:methods-10}
\begin{aligned}
\mathbf{H}_{\mathrm{mw}}&=\mathbf{M}^{-1/2}\nabla_{\mathbf{x}}^2 U\,\mathbf{M}^{-1/2},\\
\mathbf{H}_{\mathrm{mw}}\mathbf{v}_k&=\lambda_k\mathbf{v}_k.
\end{aligned}
\end{equation}

In the vibrational subspace orthogonal to translational and rotational degrees of freedom, negative eigenvalues $\lambda_k<0$ correspond to imaginary harmonic frequencies $\nu_k<0$. The number of imaginary frequencies is $n_{\mathrm{imag}}=\sum_k \mathbf{1}[\nu_k<0]$. Ground-state reaction intermediates require $n_{\mathrm{imag}}=0$, whereas transition states require strictly $n_{\mathrm{imag}}=1$.

The geometry used for the frequency calculation is compared with the optimized stationary-point geometry. For coordinate matrices $\mathbf{X}$ and $\mathbf{Y}$ centered at the origin, the geometry match requires
\[
\begin{aligned}
\delta_\pi(\mathbf{X},\mathbf{Y})
&=\max_i\|\mathbf{Y}_{\pi(i)}\mathbf{Q}_\pi-\mathbf{X}_i\|\\
&\le 3\times 10^{-6}\text{ Å},
\end{aligned}
\]
evaluated across element-conserving permutation mappings $\pi$ and optimal Kabsch rotations $\mathbf{Q}_\pi$.

For a first-order saddle point with imaginary normal-mode displacement vector $\mathbf{u}$, the mode-alignment diagnostic compares the largest relative-displacement magnitude among monitored reaction-center pairs $W$ with the largest among all atom pairs:

\begin{equation}\label{eq:methods-11}
\eta=\frac{\max_{(i,j)\in W}\|\mathbf{u}_i-\mathbf{u}_j\|}{\max_{i<j}\|\mathbf{u}_i-\mathbf{u}_j\|}.
\end{equation}

The diagnostic $\eta\in[0,1]$ measures reaction-center participation relative to the largest pair displacement, using a comparison ratio of 0.5. It flags spectator motion such as methyl wagging or solvent libration for inspection. The alignment flag is advisory.

\subsubsection{IRC Direction and Endpoint Consistency}
Intrinsic reaction coordinate (IRC) trajectories evolve from the saddle point along the steepest-descent path in mass-weighted coordinates $\mathbf{q}=\mathbf{M}^{1/2}\mathbf{x}$ \citep{ref09,ref10}:

\begin{equation}\label{eq:methods-12}
\begin{aligned}
\frac{\mathrm{d}\mathbf{q}_{\pm}}{\mathrm{d}s}&=-\frac{\nabla_{\mathbf{q}}U}{\|\nabla_{\mathbf{q}}U\|},\\
\mathbf{q}_{\mathrm{init},\pm}&\approx\mathbf{q}_{\mathrm{TS}}\pm\epsilon\mathbf{v}_-.
\end{aligned}
\end{equation}

Cloudra checks that both IRC directions start from the verified saddle point and extracts the terminal frame from each direction. For paths between endpoint minima, each terminal frame is used to start an unconstrained geometry optimization. The resulting stationary point must have zero imaginary frequencies ($n_{\mathrm{imag}}=0$), and its canonical isomeric SMILES must equal that of the declared reactant or product, including topology and stereochemistry.

For elementary steps yielding dissociated molecular components, each dissociated product must independently converge to a verified minimum. Over the monitored pair set $W$, let $\mathbf{q}_{\mathrm{TS}}$, $\mathbf{q}_{\mathrm{E}}$, and $\mathbf{q}_{\mathrm{R}}$ denote distance vectors at the transition state, terminal IRC frame, and reactant reference. With covalent cutoffs $c_{ij}=1.25(r_i+r_j)$ and tolerance $\tau=0.10$ Å, define $W_{\mathrm{c}}=\{(i,j)\in W:d_{ij}^{\mathrm{R}}\le c_{ij}\}$. The interval $[a_p,b_p]$ has endpoints
\[
\begin{aligned}
a_p&=\min(d_p^{\mathrm{TS}},d_p^{\mathrm{R}})-\tau,\\
b_p&=\max(d_p^{\mathrm{TS}},d_p^{\mathrm{R}})+\tau.
\end{aligned}
\]
The Boolean predicate $\operatorname{is\_between}$ is satisfied when both conditions hold:
\[
\begin{array}{@{}l@{}}
\|\mathbf{q}_{\mathrm{E}}-\mathbf{q}_{\mathrm{R}}\|_2
<\|\mathbf{q}_{\mathrm{TS}}-\mathbf{q}_{\mathrm{R}}\|_2,\\
a_p\le d_p^{\mathrm{E}}\le b_p
\quad\text{for all }p\in W_{\mathrm{c}}.
\end{array}
\]
Dissociated-product paths must satisfy both the component-minimum checks and $\operatorname{is\_between}$.

\subsubsection{Evidence Aggregation and Energies}
For elementary step $e$, let $K_e$ list its required calculation stages. Set $P_{e,k}=1$ when the calculation results assigned to stage $k$ satisfy its physical criteria, and $P_{e,k}=0$ otherwise. All required stages must pass:

\begin{equation}\label{eq:methods-13}
A_e=\bigwedge_{k\in K_e}P_{e,k}.
\end{equation}

When explicit property checks $R$ are declared, the step must also satisfy $A_e\wedge S_e$. Each stage is recorded as supported, pending, skipped, or rejected. A documented exemption or method limitation is thus recorded separately from a completed calculation.

The electronic activation barrier is computed directly from verified stationary-point energies:

\begin{equation}\label{eq:methods-14}
\Delta E_e^\ddagger=E_{\mathrm{TS},e}-E_{\mathrm{ref},e}.
\end{equation}

Energies share a consistent quantum chemical level of theory. For thermochemical profiles, zero-point energy, enthalpy, and Gibbs free-energy corrections are evaluated from harmonic partition functions under declared temperature and pressure specifications.

\subsection{Calculation Diagnostics and Retries}
The system distinguishes execution errors, numerical nonconvergence, and physical evidence problems. Execution errors include invalid input syntax, environment faults, and parsing failures. Numerical checks record self-consistent field (SCF) and geometry optimization convergence. Physical evidence problems include saddles with $n_{\mathrm{imag}}\ne 1$, IRC endpoints awaiting verification, or molecular graphs corresponding to a different species. Imaginary-mode participation and geometric interval measurements indicate which atomic motions or distances to inspect.

Let $T$ contain the species and elementary steps requested by the task. Extending $P_{e,k}$ to each $t\in T$, the calculation-stage requirements to be resolved are

\begin{equation}\label{eq:methods-15}
U=\{(t,k):t\in T,\, k\in K_t,\, P_{t,k}=0\}.
\end{equation}

For each $(t,k)\in U$, diagnostic feedback identifies the affected species or step and the problem: SCF nonconvergence, normal-mode misalignment, a missing IRC-to-endpoint calculation link, or a molecular graph mismatch. After a failed calculation, the system retains its output, updates $U$, and reports the geometry and convergence information to the agent.

Calculations that depend on a failed stage are rescheduled; earlier attempts and raw outputs are retained. The system reorders atom indices, assigns each calculation result to its species or step, and updates the graph status. The model revises the chemical constraints, dihedral angles, or electronic-structure method for the next attempt.

\subsection{Implementation Defaults}
The implementation described here corresponds to Cloudra revision 70daa5a.\citep{ref14} Numerical arrays and linear algebra use NumPy\citep{numpy2020}; quantum-chemical quantities are extracted with cclib alongside engine-specific native-log readers\citep{cclib2008}; and benchmark graph representation and matching use NetworkX\citep{networkx2008}. Preorganization uses 32 rigid orientations, up to 8 guided placements, minimum interfragment separation 1.0 Å, fragment-contact multiplier 1.7, and heavy-atom RMSD clustering threshold 0.5 Å. MMFF94 relaxation uses up to 2000 iterations per phase with a distance restraint force constant of 100.0. Standard IDPP employs 10 images optimized via MDMin (100 steps, force tolerance $f_{\max}=0.1$, NEB spring constant $k=0.1$); lightweight IDPP uses Adam optimization (4000 steps, step size 0.02, gradient tolerance $10^{-7}$). Path convergence criteria evaluate the maximum force residual on intermediate path images.\par

The mapping coefficients and the $J_{\min}+2$ ambiguity window are fixed ranking heuristics. Charge and radical mismatches carry larger costs than hydrogen-count and bond-edit mismatches. Branch crowding uses the fixed decay $\rho=0.5$, traversal through depth five, and congestion flag $B_{\mathrm{axis}}\geq9.0$.

Cloudra also provides a conformer-search tool with RDKit and external CREST backends.\citep{crest2020} The optional CREST backend samples conformers or noncovalent complexes, followed by quantum-chemical optimization and physical validation.

\subsection{CloudraBench construction and experimental design}

\subsubsection{Task coverage and tier rationale}

\begin{figure*}[!tbp]
\begin{minipage}{\textwidth}
\centering
\includegraphics[width=\textwidth]{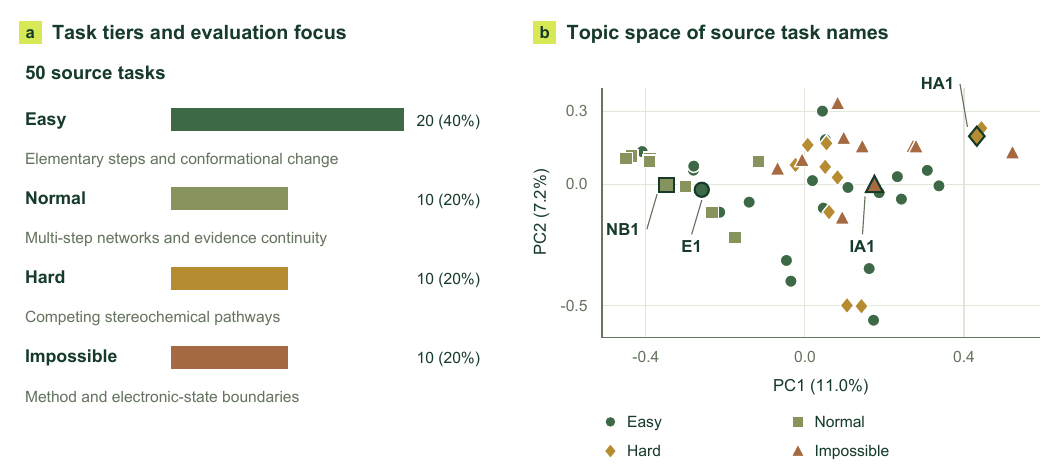}
\caption{CloudraBench task composition and research themes. (a) Counts and evaluation focus of the four task tiers. (b) Two-dimensional PCA, computed with scikit-learn\citep{sklearn2011}, of BAAI/bge-m3 embeddings\citep{bgem32024} of all 50 original task names. E1, NB1, HA1, and IA1 are the four tasks used in the repeated-run experiment. Colors and marker shapes encode the predefined task tiers.}
\label{fig:05}
\end{minipage}
\end{figure*}

CloudraBench evaluates end-to-end mechanism calculations on 50 tasks: 20 Easy tasks examine elementary transformations, conformational changes, and basic execution reliability; 10 Normal tasks examine intermediate identification and multistep continuity; 10 Hard tasks examine competing pathways and stereochemical identity; and 10 Impossible tasks examine the limits of the allowed methods and justified negative conclusions (Figure~\ref{fig:05}). Related reactions can appear in different tiers when they require different pathways, configurations, or electronic states. Tasks were selected using literature sources and reference calculations, considering both their ability to distinguish workflow performance and their computational cost. An accompanying catalogue lists each task's name, input conditions, and allowed methods.

\subsubsection{Reference answers and evaluation protocol}

Each task fixes the chemical conditions, charge, spin, deliverables, and allowed methods. Easy, Normal, and Hard tasks permit xTB\citep{gfn2xtb2019}, PM7\citep{pm72013}, and B3LYP-D3(BJ)/def2-SVP\citep{becke1993,lyp1988,stephens1994,d32010,d3bj2011,def22005}; Impossible tasks are restricted to xTB and PM7. Computational evidence is read from native ORCA, Gaussian, or MOPAC logs.\citep{orca2025,gaussian16,mopac2026,ref11} Reference answers specify species, elementary steps, stereochemical identities, and applicable numerical or negative targets. Numerical barriers are electronic energy differences at a consistent level of theory; stable-minimum references and actual IRC connectivity are verified separately. For each accessible reaction step requiring verification, we use TS frequency analysis, bidirectional IRCs, and endpoint optimization and frequency analysis. We also check that each endpoint optimization starts from the actual terminal IRC frame. Species-property and electronic-state questions are checked against their specified targets.

\subsubsection{Scoring and study configurations}

Graph structure, self-consistency, and accuracy are evaluated separately and combined with weights of 0.1, 0.3, and 0.6. Graded physical support and runtime delivery completeness are also reported. A correctly identified method limitation is scored as a supported negative conclusion. Each task lists reference-graph availability, available reference calculations, task-specific exceptions, and unresolved validation checks.

\subsubsection{Repeated-task study, primary comparison, and workflow ablations}

The repeated-task study (P1b) compares Guided, Standard, Record, and Brief-errors on E1, NB1, HA1, and IA1, with three independent runs per combination (48 task-level runs). The repeated runs used DeepSeek-v4.1-flash (1017).\citep{deepseekv42026} E1, NB1, HA1, and IA1 concern elimination preorganization, multistep decarboxylation, stereochemical ring opening, and electronic-structure method limits, respectively; their archived trajectories supply the subsequent local diagnostic cases. Its elapsed-time unit is a four-task batch at concurrency two, with three batches per configuration. The full-benchmark study (P2) retains one run for each of the 50 tasks and four configurations (200 task-level runs). All P2 configurations used DeepSeek-v4.1-flash (1017), called with the archived API alias \texttt{deepseek-v4-flash} through the official DeepSeek endpoint and streaming OpenAI-compatible chat completions, with requested minimal reasoning effort. The primary comparison is Guided against Standard using this same backbone; Standard is labelled Backbone in the figures. The P2 execution versions and request settings are listed in Supplementary Information Section S1.

Standard hides explicit DAG tools, removes enforcement of physical checks, and supplies neutral computational feedback. Guided uses the research graph, scientific checks, and detailed diagnostics. Record provides tools and records with optional physical validation; Brief-errors retains guidance while shortening diagnostics. All configurations receive the same task statements and method restrictions. The archive stores each run's code version, budget and stopping reason, together with the source files for its calculations.

Before scoring, human curators reconstruct each run's mechanism graph from its conversations, tool records, reports, and calculation files, following the agent's stated mechanistic intent and conclusions. They restore file references and record the source of each recovered field. Reconstruction uses only that run's existing calculations and conclusions. The evaluator searches its sandbox for native calculation evidence and scores the graph against Gold targets under common rules. The final dataset contains all 200 runs and reports Graph, Self-consistency, and Accuracy alongside each composite score. Paired differences and 95\% percentile intervals use 100,000 task-cluster bootstrap samples, retaining the four configurations of each task together (seed 20261031). Computational effort is measured on the 15 selected tasks with records for all four configurations, using quantum-chemical job counts and cumulative job-hours.

\section{Results and Discussion}


\begin{figure*}[!tbp]
\centering
\includegraphics[width=\textwidth]{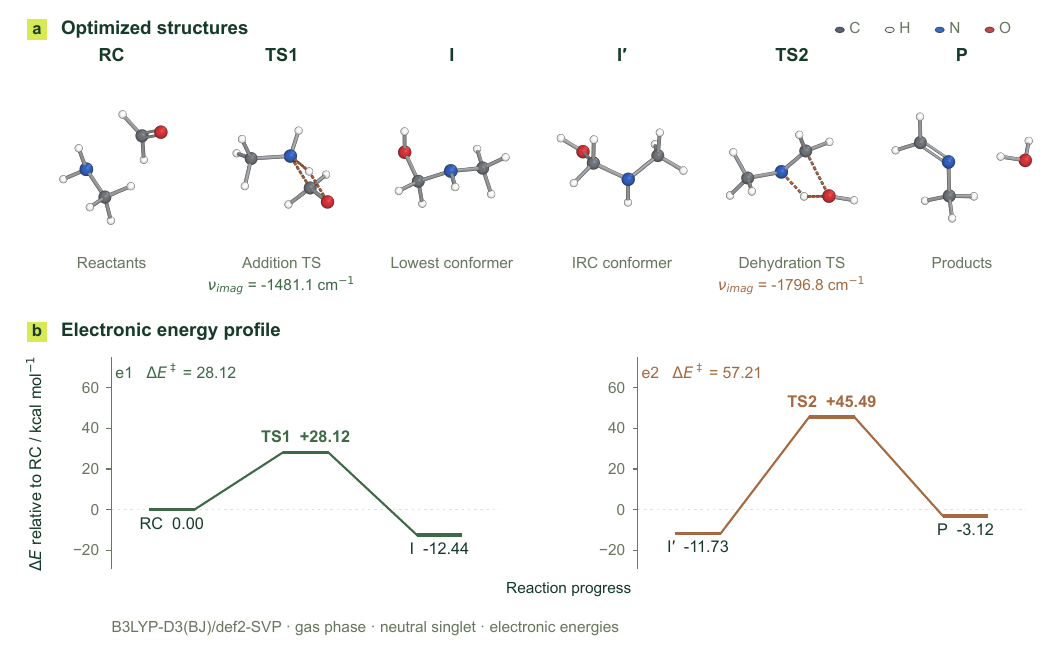}
\caption{Two requested steps in the archived Guided/NI1 study: methylamine addition to formaldehyde and carbinolamine dehydration. Structures and electronic energies come from this run's B3LYP-D3(BJ)/def2-SVP gas-phase calculations. Each saddle has one imaginary frequency, and its short IRC branches lead through endpoint optimization/frequency analysis to minima with zero imaginary frequencies. Both energy panels use RC as the common reference.}
\label{fig:ni1-chemical-objects}
\end{figure*}

\begin{figure*}[!tbp]
\begin{minipage}{\textwidth}
\centering
\includegraphics[width=\textwidth]{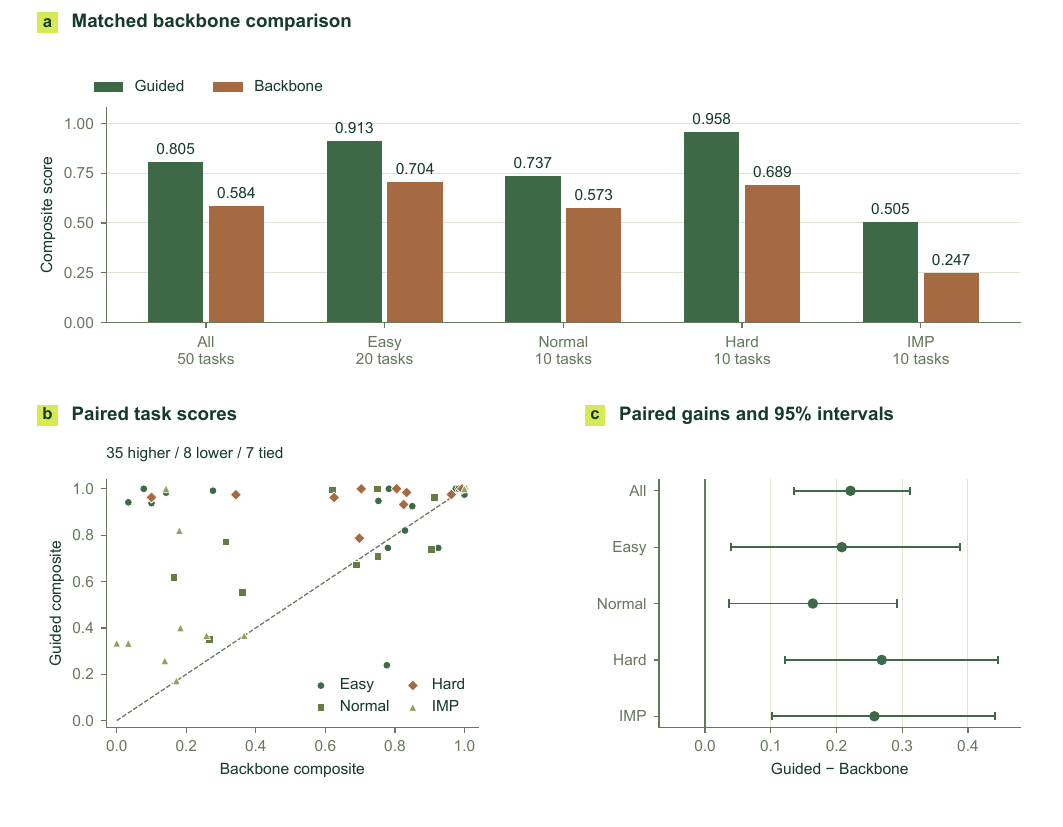}
\caption{P2 benchmark outcomes under the archive-adjudicated track. (a) Guided and the same-backbone Standard baseline across all 50 tasks and each tier. (b) All 50 paired task outcomes. (c) Mean paired score differences with 95\% percentile intervals from 100,000 task-level bootstrap resamples. Each task has one run per configuration; Easy contains 20 tasks and the other tiers contain 10 each.}
\label{fig:bench-overview}
\end{minipage}
\end{figure*}

\begin{figure*}[!tbp]
\begin{minipage}{\textwidth}
\centering
\includegraphics[width=\textwidth]{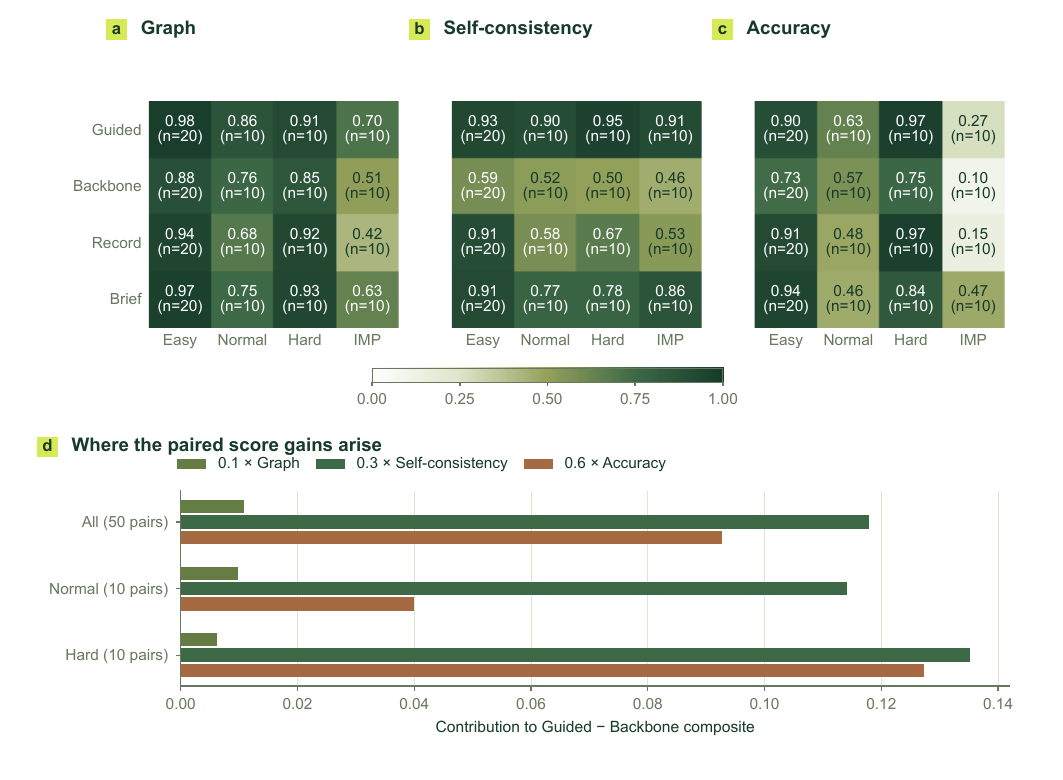}
\caption{CloudraBench score layers across all 200 organized runs. (a--c) Means of Graph, Self-consistency, and Accuracy in each arm and tier, using 50 tasks per arm overall, 20 Easy tasks, and ten tasks in each other tier. (d) Guided--Backbone mean differences decomposed with the fixed weights 0.1/0.3/0.6, using all 50 pairs, ten Normal pairs, and ten Hard pairs. The component values come from each run's final evaluation; their weighted sum gives the published composite score to the reported precision.}
\label{fig:bench-components}
\end{minipage}
\end{figure*}

\begin{figure*}[!tbp]
\begin{minipage}{\textwidth}
\centering
\includegraphics[width=\textwidth]{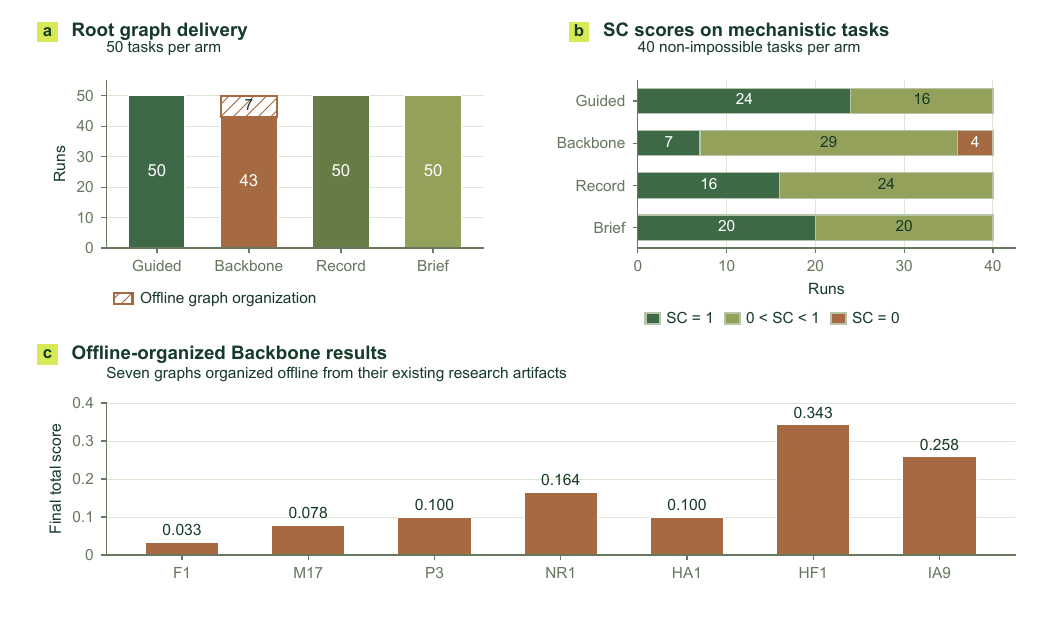}
\caption{Runtime graph delivery and final scientific support in P2. (a) Original root mechanism graphs across 50 tasks per arm: Guided 50, Backbone 43, Record 50, and Brief 50. The hatched segment marks seven additional Backbone graphs assembled offline from the runs' own research materials; the organized set contains 50 graphs per arm. (b) Final continuous self-consistency (SC) scores for all 40 non-Impossible tasks per arm, separating SC = 1, partial positive credit, and SC = 0. Full-score counts are 24, 7, 16, and 20; Backbone has four zero-score tasks. (c) Organized composite scores for the seven Backbone tasks whose root graphs were reconstructed offline, ranging from 0.0333 to 0.3425 (mean 0.1537).}
\label{fig:bench-delivery}
\end{minipage}
\end{figure*}

\begin{figure*}[!tbp]
\begin{minipage}{\textwidth}
\centering
\includegraphics[width=\textwidth]{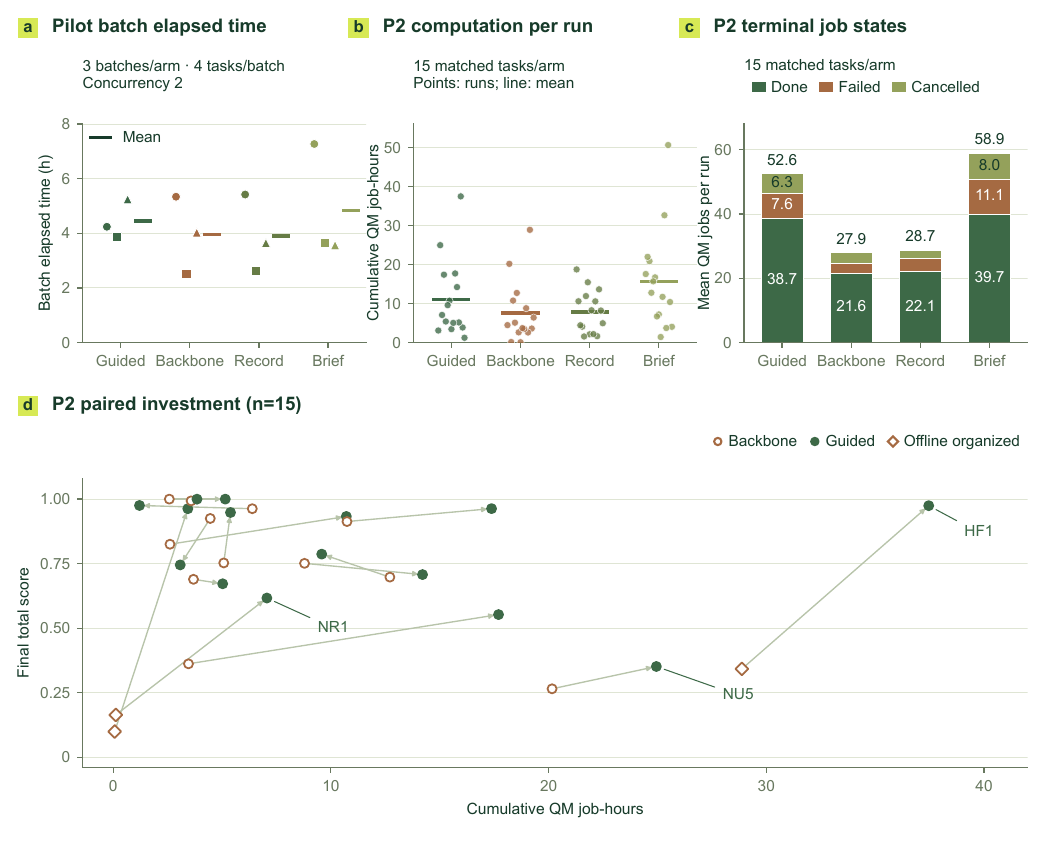}
\caption{Research elapsed time and computational investment. (a) Twelve P1b four-task batches at concurrency two, three per arm; lines mark means. (b) Cumulative QM process durations for the 15 P2 tasks shared by all arms. (c) Mean completed, failed, and cancelled jobs in that cohort. (d) Same-task Backbone-to-Guided changes in investment and final score; diamonds mark offline-organized Backbone graphs. Job-hours sum process walltime; panel a measures batch latency. Backbone/M1 effort is reconstructed from native CPU time divided by four.}
\label{fig:bench-speed}
\end{minipage}
\end{figure*}

\begin{figure*}[p]
\begin{minipage}{\textwidth}
\centering
\includegraphics[width=\textwidth]{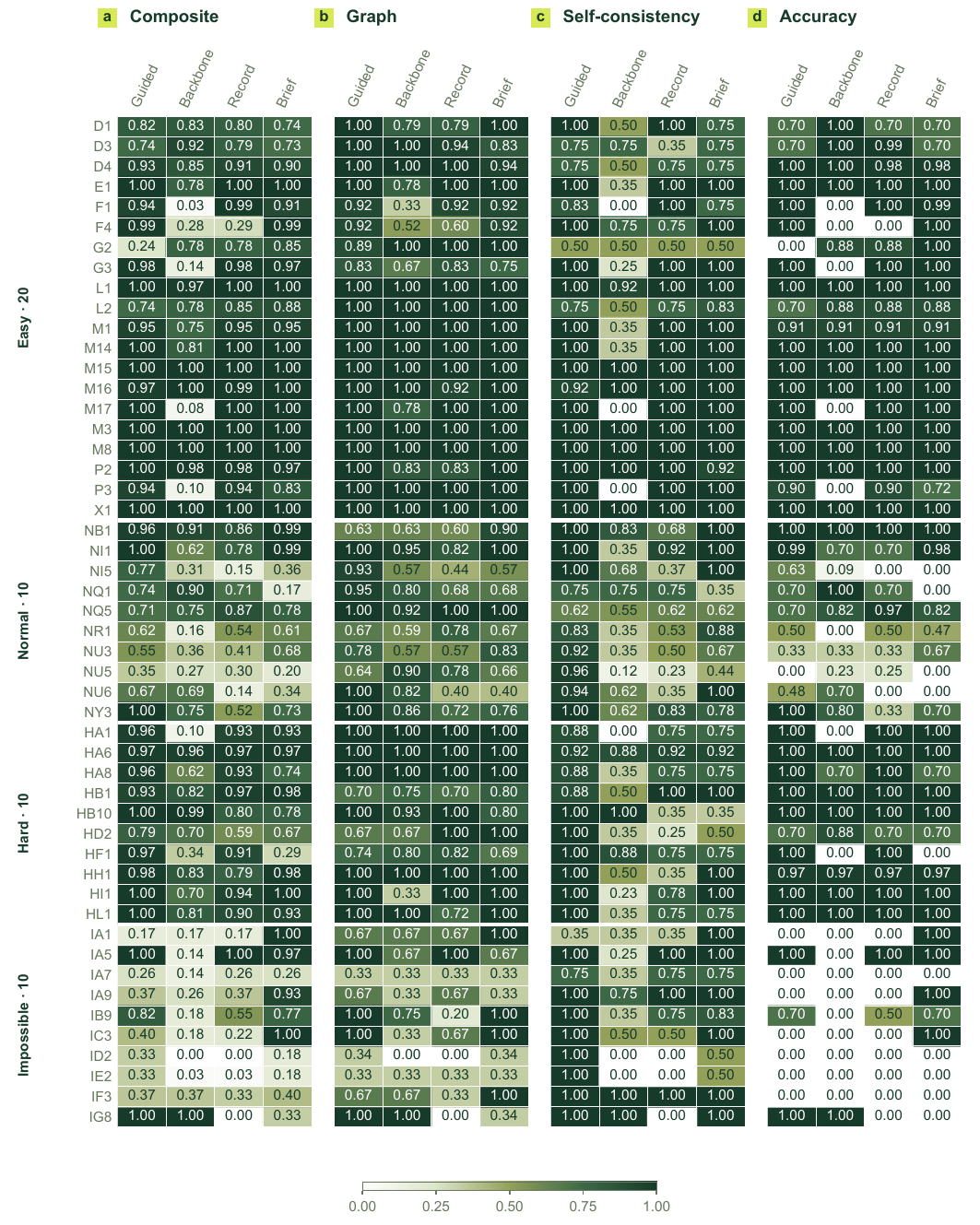}
\caption{Complete CloudraBench results for all 50 tasks and four configurations. The aligned blocks show composite, Graph, Self-consistency, and Accuracy on the same 0--1 scale; all 800 cells are present. Values are rounded to two decimals; the source table retains full precision.}
\label{fig:bench-tasks}
\vspace{6pt}
\end{minipage}
\end{figure*}

\begin{figure*}[!tbp]
\begin{minipage}{\textwidth}
\centering
\includegraphics[width=\textwidth]{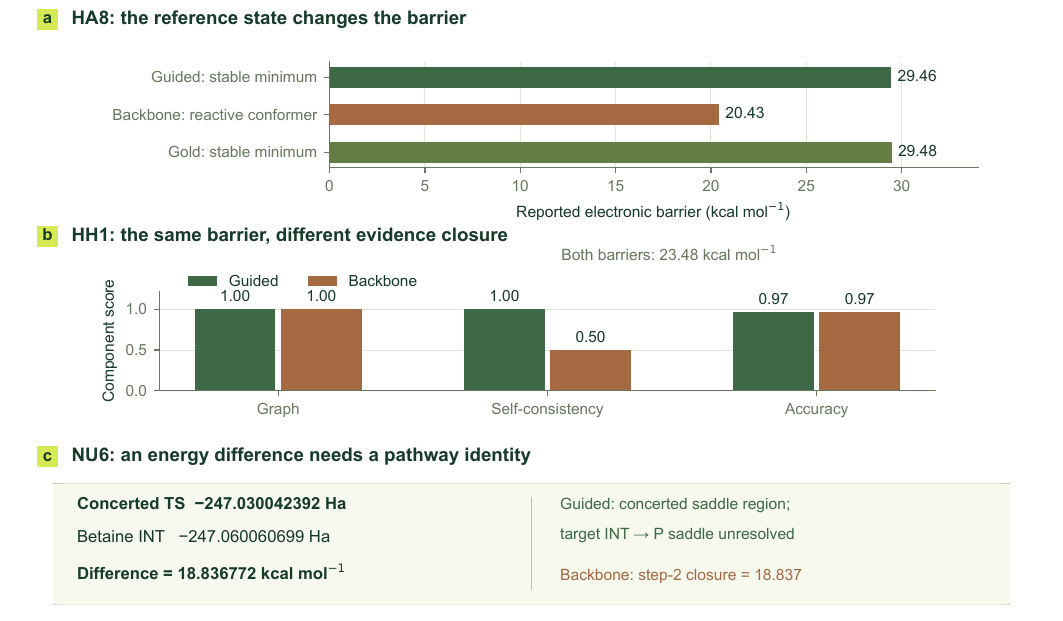}
\caption{Energy references and pathway assignments in matched Guided--Backbone tasks. (a) HA8 barriers use a stable-minimum or reactive-conformer reference and are compared with campaign Gold. (b) Both HH1 runs report 23.48 kcal/mol with similar accuracy but different self-consistency scores. (c) NU6's recorded energy difference is 18.836772 kcal/mol: Guided reports the cost of reaching the concerted saddle region and leaves the target ring-closure saddle unresolved; Backbone reports 18.837 as the second-step ring-closure barrier.}
\label{fig:bench-cases}
\end{minipage}
\end{figure*}

\begin{figure*}[!tbp]
\begin{minipage}{\textwidth}
\centering
\includegraphics[width=\textwidth]{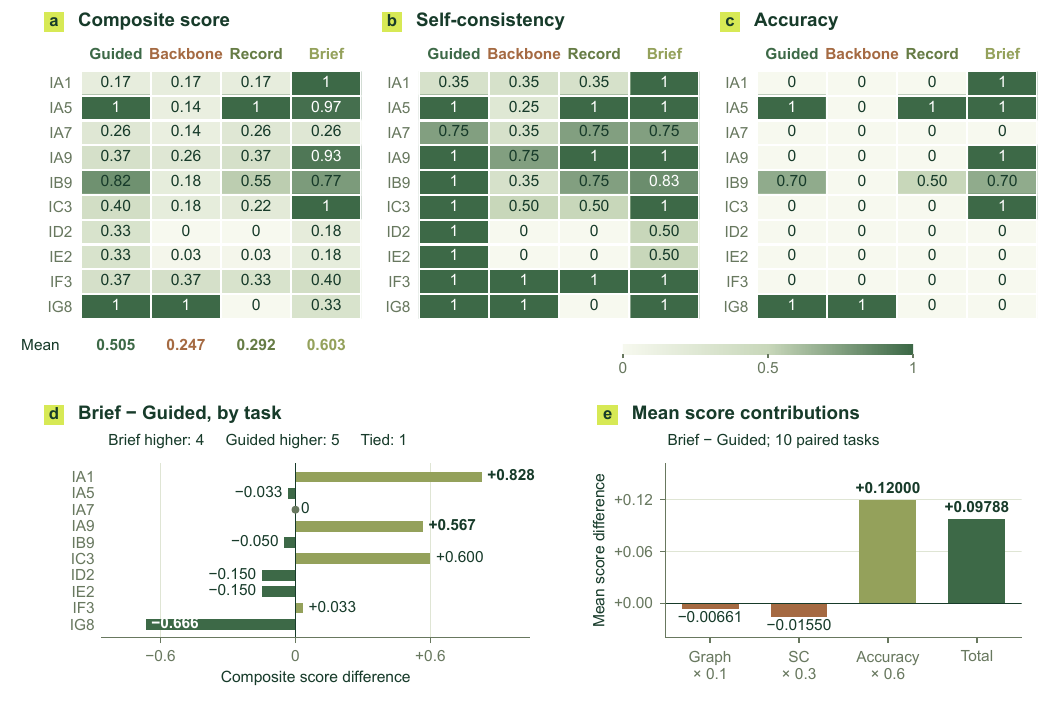}
\caption{Impossible-tier outcomes after offline organization and scoring. (a--c) Composite score, self-consistency (SC), and accuracy for ten tasks and four arms, using a shared 0--1 color scale. Backbone denotes Standard. Brief averages 0.603 and Guided 0.505. (d) Guided scores higher on 5 tasks, Brief on 4, and 1 tie. (e) The +0.09788 mean difference comprises accuracy +0.12000, Graph $-0.00661$, and SC $-0.01550$ after weighting. The paired task-bootstrap 95\% interval is $-0.160$ to $+0.361$.}
\label{fig:bench-imp}
\end{minipage}
\end{figure*}

\begin{figure*}[!tbp]
\begin{minipage}{\textwidth}

\centering
\includegraphics[width=\textwidth]{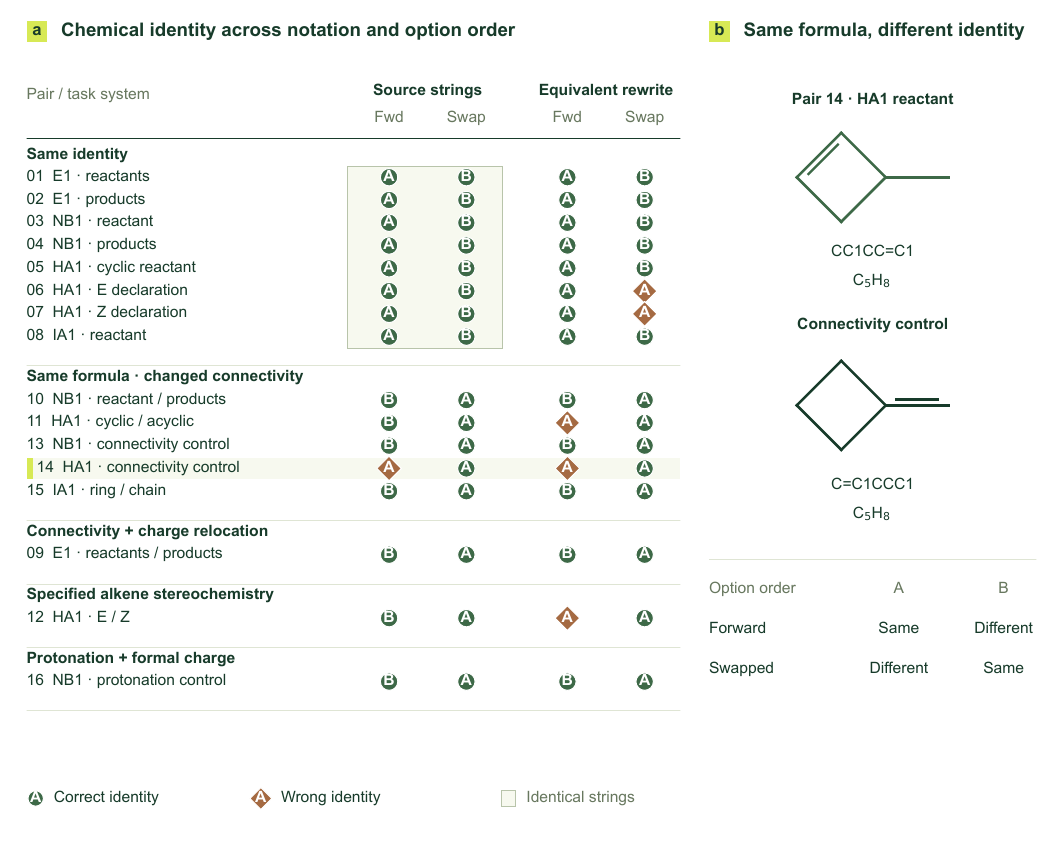}
\caption{Molecular-identity judgments under changes in molecular representation and option order. Each of the 16 molecular pairs has four input conditions. Circles and diamonds indicate correct and incorrect judgments, respectively; the letters retain the model's actual answers. Shaded regions denote identical-string conditions. The right panel shows the pair-14 connectivities \texttt{CC1CC=C1} and \texttt{C=C1CCC1}, both C\textsubscript{5}H\textsubscript{8}, and the meanings of the answer options.}
\label{fig:06}
\end{minipage}
\end{figure*}

\begin{figure*}[!tbp]
\begin{minipage}{\textwidth}
\begin{minipage}[t]{.65\textwidth}
\vspace{0pt}
\centering
\includegraphics[width=\linewidth]{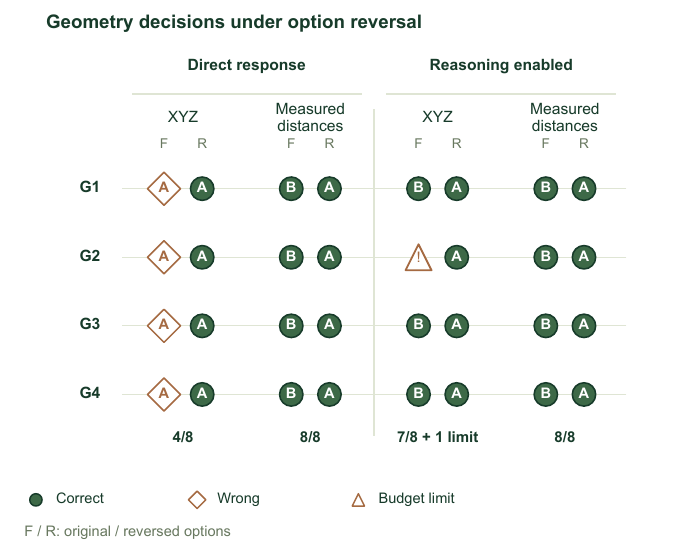}
\caption{Judgments of which of two specified atom-pair distances is shorter, using coordinates or measured distances. Four source structures are compared under direct response and reasoning-enabled conditions, each with original (F) and reversed (R) option order. Circles indicate correct judgments, diamonds incorrect judgments, and the triangle a response that reached the output budget. Letters retain the actual answers.}
\label{fig:07}
\end{minipage}\hfill
\begin{minipage}[t]{.31\textwidth}
\vspace{0pt}
\centering
\includegraphics[width=\linewidth]{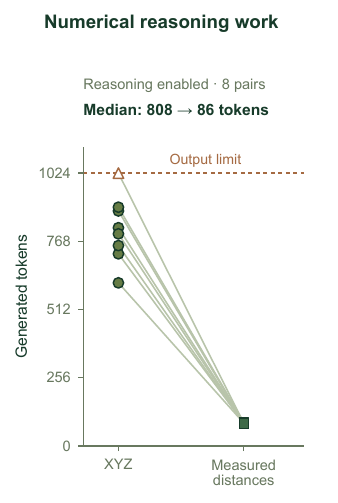}
\caption{Generated-token counts with reasoning enabled. Each line connects Cartesian-coordinate input and measured-distance input for the same source structure and option order; eight paired comparisons are shown. The dashed line marks the output limit.}
\label{fig:08}
\end{minipage}
\end{minipage}
\end{figure*}

\begin{figure*}[!tbp]
\begin{minipage}{\textwidth}
\centering
\includegraphics[width=.92\textwidth]{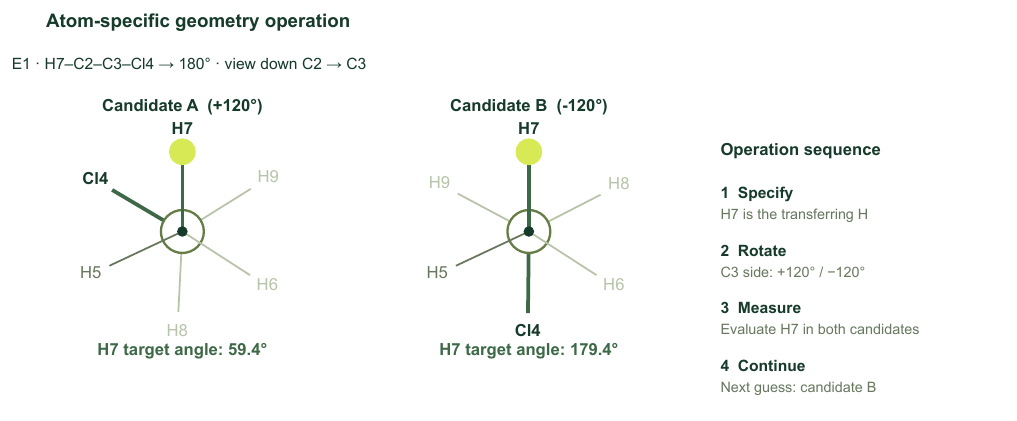}
\caption{Target-specific geometry operations in E1. Newman projections retain the directions of the original candidate coordinates viewed down C2\(\to\)C3. H7 is the specified transferring atom. Candidate A and candidate B give target angles of 59.4° and 179.4°, respectively. The operation sequence is shown alongside the projections.}
\label{fig:09}
\vspace{5pt}
\begin{minipage}[t]{.48\textwidth}
\vspace{0pt}
\centering
\includegraphics[width=\linewidth]{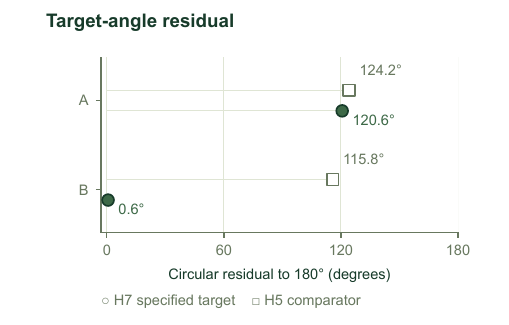}
\caption{Circular angular residuals from the specified 180° target for candidates A and B. H7 is the specified target atom; H5 is the comparator.}
\label{fig:10}
\end{minipage}\hfill
\begin{minipage}[t]{.48\textwidth}
\vspace{0pt}
\centering
\includegraphics[width=\linewidth]{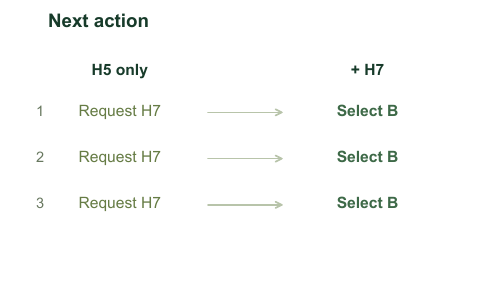}
\caption{Next-action choices under three local option permutations. When only H5 measurements are supplied, the model requests H7 measurements; with H7 supplied, it selects candidate B.}
\label{fig:11}
\end{minipage}
\end{minipage}
\end{figure*}

\begin{figure*}[!tbp]
\begin{minipage}{\textwidth}
\centering
\includegraphics[width=.92\textwidth]{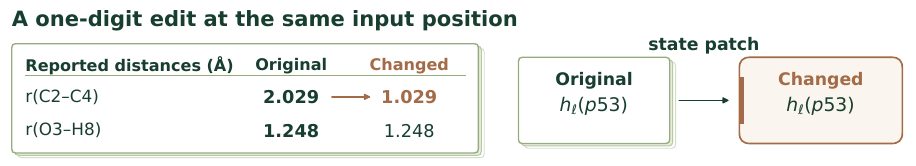}
\caption{Intervention on a single reported distance. A one-digit edit changes the C2–C4 distance from 2.029 to 1.029 Å while O3–H8 remains 1.248 Å. The original-input residual state at token position 53 is inserted into the changed-input run at a specified layer. Token positions are zero-based.}
\label{fig:12}
\vspace{4pt}
\centering
\includegraphics[width=.92\textwidth]{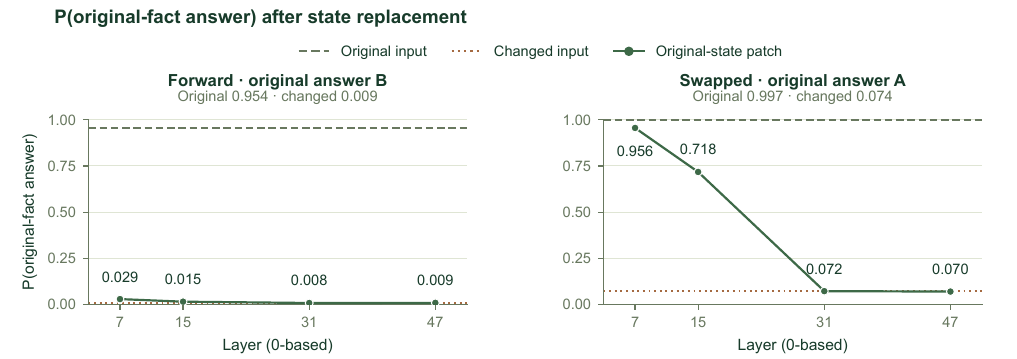}
\caption{Original-fact answer probability after state replacement. Curves show full-vocabulary probability mass for the original-fact answer under forward (B) and swapped (A) option order, including bare and leading-space token spellings. Horizontal lines show original-input and changed-input baselines. Layer indices are zero-based.}
\label{fig:13}
\vspace{4pt}
\centering
\includegraphics[width=.92\textwidth]{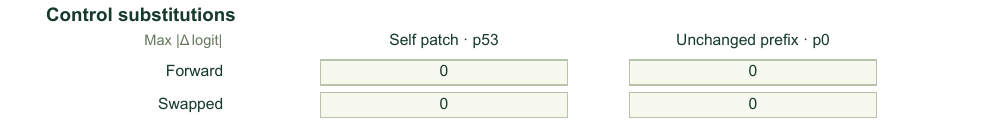}
\caption{Control substitutions for the state-intervention experiment. Same-state self replacement at p53 and replacement of the unchanged prefix at p0 preserve all vocabulary logits in both option orders. The table reports the maximum absolute logit difference. Token positions are zero-based.}
\label{fig:14}
\end{minipage}
\end{figure*}

\begin{figure*}[!tbp]
\begin{minipage}{\textwidth}

\begin{minipage}[t]{.48\textwidth}
\vspace{0pt}
\centering
\includegraphics[width=\linewidth]{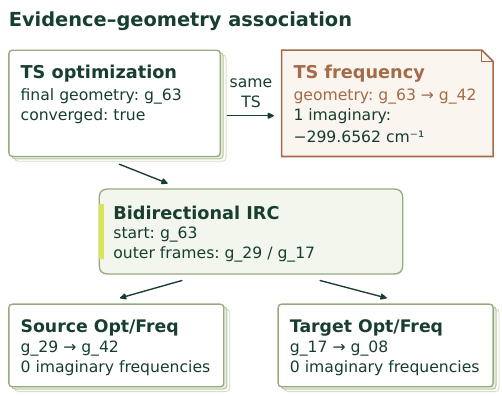}
\caption{Changing the geometry assigned to the E1 frequency calculation. The frequency geometry reference changes from g\_63 to g\_42 while local termination, convergence, energy, and frequency values remain fixed. Arrows connect the TS, IRC, and endpoint optimization/frequency calculations.}
\label{fig:15}
\end{minipage}\hfill
\begin{minipage}[t]{.48\textwidth}
\vspace{0pt}
\centering
\includegraphics[width=\linewidth]{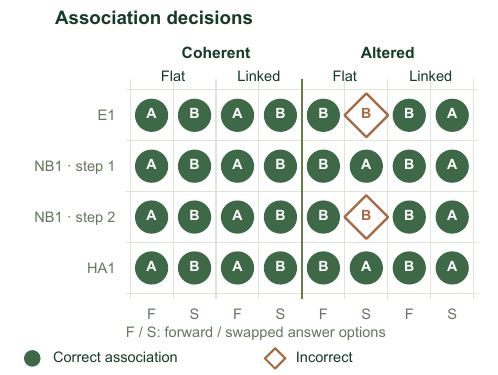}
\caption{Judgments of whether calculation references support the specified step. The 32 inputs cross four source chains, coherent/altered links, flat/linked presentation, and forward/swapped answer order. Saved A/B responses and correctness markers are retained.}
\label{fig:16}
\end{minipage}
\end{minipage}
\end{figure*}

\begin{figure*}[!tbp]
\begin{minipage}{\textwidth}
\centering
\includegraphics[width=.96\textwidth]{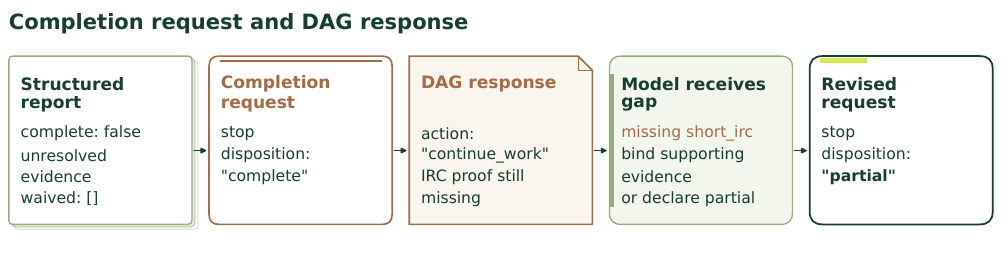}
\caption{Completion requests and DAG feedback in the original IA1 trajectory. A report with unresolved evidence is followed by a completion request, a response identifying missing short\_irc proof, feedback to the model, and a revised partial-completion request.}
\label{fig:17}
\end{minipage}
\end{figure*}

\subsection{A two-step mechanism expressed through chemical objects}
\label{sec:ni1-chemical-objects}

The Guided NI1 run computes methylamine addition to formaldehyde and carbinolamine dehydration to an imine--water complex (Figure~\ref{fig:ni1-chemical-objects}). Its Delivery DAG connects the optimized reactant complex, carbinolamine, and product complex through two first-order saddles. Calculations used B3LYP-D3(BJ)/def2-SVP in the gas phase, with charge zero and singlet multiplicity.

The addition saddle has one imaginary frequency at $-1481.06$ cm$^{-1}$ and an electronic barrier of 28.12 kcal/mol relative to the reactant complex. Its two short IRC branches lead, after endpoint optimization and frequency analysis, to the reactant complex and carbinolamine, both with zero imaginary frequencies. The dehydration saddle has one imaginary frequency at $-1796.84$ cm$^{-1}$ and an electronic barrier of 57.21 kcal/mol relative to its IRC-side carbinolamine minimum. Its two branches similarly yield carbinolamine and the imine--water complex. Each IRC followed 24 points in each direction before reaching the point limit; the endpoint calculations establish the resulting minima.

Interconversion between the carbinolamine conformers remains uncomputed. Supplementary Information Section~S4 provides the conformer references, absolute energies, native calculation records, and figure inputs.

\subsection{End-to-end benchmark evaluation}

CloudraBench measures end-to-end mechanism-calculation capability: constructing candidate mechanisms, carrying out the required quantum-chemical calculations, and reaching task-consistent positive or negative conclusions. We report composite scores together with their graph, self-consistency, and accuracy components. Offline postprocessing reconstructs mechanism graphs from each run's own research materials and compares the computed species, reaction channels, electronic states, and barrier references with the task definition. Before calculating Accuracy, we match each reported reaction channel and electronic state to the target specified in the task. All four workflows use the same postprocessing rules. The resulting 200-run table contains all three components for every run, with an overall composite mean of 0.7176. Figures~\ref{fig:bench-overview}--\ref{fig:bench-tasks} report runtime graph production, TS--IRC--endpoint evidence, and agreement with the task's answer.

\subsubsection{Cloudra improves end-to-end mechanism-calculation scores with the same backbone}

Guided obtained a mean composite score of 0.8050, compared with 0.5836 for Standard. Across the 50 task pairs, the mean improvement was 0.2214, with a 95\% task-bootstrap interval of [0.1349, 0.3124]. Guided scored higher on 35 tasks, lower on eight, and equally on seven; the median paired difference was 0.1142 (Figure~\ref{fig:bench-overview}).

Guided's mean over its ten lowest-scoring tasks was 0.3374, compared with 0.0930 for Standard.

The most consistent difference occurred in the Hard stereochemistry tasks. Guided averaged 0.9577 and Standard 0.6888, a paired increase of 0.2689 [0.1213, 0.4459], with Guided scoring higher on all ten tasks. These tasks require assigning each barrier and computed pathway to the corresponding stereoisomer or competing channel.

On Normal tasks, Guided averaged 0.7370 versus 0.5731, an increase of 0.1640 [0.0365, 0.2922], with seven wins and three losses. Easy-task means were 0.9125 and 0.7044, respectively; the paired median difference was 0.0209, showing a different pattern from the broadly positive Hard comparison. Impossible-task means were 0.5051 and 0.2473. Their interpretation depends on whether the run addressed the requested method-limited negative target, so these tasks are examined separately below.

\subsubsection{Self-consistency contributes most to the composite-score gain}

The three components score different parts of the calculation (Figure~\ref{fig:bench-components}). Graph measures the delivered coverage and identities of the requested species and steps. Self-consistency measures the support linking saddle points, frequencies, IRC starts, and endpoint optimization/frequency calculations. Accuracy measures agreement with the task's numerical or negative target.

Across all 50 Guided--Standard task pairs, the weighted graph contribution to the mean improvement was +0.0108, self-consistency contributed +0.1179, and accuracy +0.0927, accounting together for the composite gain of 0.2214. For the 20 Normal and Hard tasks, the mean gain was 0.2164, with contributions of +0.0080 from graph, +0.1247 from self-consistency, and +0.0837 from accuracy. Within the ten Normal tasks, the corresponding contributions were +0.0098, +0.1141, and +0.0400; within the ten Hard tasks they were +0.0063, +0.1353, and +0.1274.

On Hard tasks, the weighted accuracy improvement is comparable in size to the self-consistency improvement.

HH1 provides a direct example. Guided and Standard both reported a binaphthyl racemization barrier of 23.48 kcal/mol, close to the 23.01 kcal/mol reference. Both had graph scores of 1 and accuracy scores of approximately 0.973. Their self-consistency scores were 1.0 and 0.5, producing composite scores of 0.9838 and 0.8338, respectively. The archived Standard study contained candidate channels and IRC inputs; the required endpoint connections remained unverified. Guided attached the path evidence to the corresponding saddle and endpoints. HI1 shows a related pattern: accuracy was approximately 1 in both runs, while self-consistency increased from 0.2333 to 1.0 and the composite score from 0.7032 to 0.9999.

\subsubsection{Mechanism-graph delivery and self-consistency scores}

Runtime root-graph production is recorded separately from the physical evidence found in each run (Figure~\ref{fig:bench-delivery}). During the original runs, Standard delivered root graphs for 43 of 50 tasks, while Guided, Record, and Brief-errors each delivered 50. The seven Standard tasks requiring offline root-graph reconstruction were F1, HA1, HF1, IA9, M17, NR1, and P3. Offline organization assembled their graphs from the corresponding runs' existing materials, producing a complete 50-task set for every configuration.

The seven Standard studies averaged 0.1537 after organization, compared with 0.8285 for Guided on the same tasks, and contributed 42.7\% of the total Guided--Standard difference. Their individual scores ranged from 0.0333 to 0.3425.

For the 40 non-Impossible tasks in each configuration, Guided had self-consistency exactly equal to 1 on 24 tasks (60.0\%) and partial positive credit on 16. Standard had seven full-score (17.5\%), 29 partial-score, and four zero-score tasks. Record had 16 full-score (40.0\%) and 24 partial-score tasks; Brief-errors had 20 full-score (50.0\%) and 20 partial-score tasks. Figure~\ref{fig:bench-delivery} presents these complete evidence-score distributions alongside the separate runtime graph-delivery counts.

These percentages count full self-consistency scores for delivered species and steps. A uniform audit covering every species and step requested by each task remains pending (Supplementary Information Section~S4).

\subsubsection{Guided outperforms Record with the same available tools}

Record averaged 0.7038 and Brief-errors 0.7781 across all 50 tasks. Guided exceeded Record by 0.1012 [0.0396, 0.1695], with 28 wins, seven losses, and 15 ties. On all 20 Normal and Hard tasks together, the mean Guided--Record difference was 0.1468: graph contributed +0.0083, self-consistency +0.0919, and accuracy +0.0466. The largest contribution came from self-consistency, which scores the frequency, IRC, and endpoint calculations supporting the reported steps.

\subsubsection{Research batch times and quantum-chemical job-hours}

Figure~\ref{fig:bench-speed} presents the timing information from the repeated-task study alongside the computational effort available for P2. For the repeated study, the measurement unit is a complete four-task batch at a concurrency of two, with three observed batches per configuration. Mean batch elapsed times were 4.451, 3.958, 3.894, and 4.827 hours for Guided, Standard, Record, and Brief-errors, respectively. Guided batches took 12.5\% longer than Standard and 7.8\% less time than Brief-errors on average. The figure shows all 12 observed batches, retaining their variation and original outcomes. These measurements include the batch's research lifecycle, waiting for calculations, and continuation after results return.

P2 effort is summarized over the 15 deeply inspected tasks shared by all four configurations. Guided used 11.08 job-hours per task, Standard 7.55, Record 7.88, and Brief-errors 15.59. The corresponding final composite means were 0.8124, 0.6495, 0.7360, and 0.7273. Guided used 46.7\% more cumulative job time than Standard while increasing the mean score by 0.1629; relative to Brief-errors, it used 28.9\% less job time and scored 0.0851 higher. Guided was both faster in cumulative job time and higher-scoring than Brief-errors on seven of the 15 paired tasks.

Across the common tasks, Guided submitted 789 recorded quantum-chemical jobs, Standard 419, Record 431, and Brief-errors 883. Recorded failed-job counts were 114, 45, 62, and 167, respectively, with cancellations recorded separately. The four difficult studies NU3, NU5, NU6, and NR1 accounted for 163.5 job-hours across their 16 runs, 24.3\% of the 673.82 job-hours in the available 63-run effort archive. Some required steps in these four studies remained unresolved after additional attempts. In the archived Standard missing-graph cases, calculations could also outlive the research session, leaving subsequent results outside an active study.

Guided exceeded Brief-errors by 0.0268 overall, with an interval of [$-0.0486$, 0.0996], 26 wins, eleven losses, and 13 ties. The comparison varied by research demand: Guided's mean advantage was 0.1514 on Normal and 0.1295 on Hard, whereas Brief-errors had the higher Impossible mean.

\subsubsection{Uncapped electronic-barrier errors in a common numerical subset}

To express numerical deviations in chemical units, we recovered electronic barriers directly from archived native TS and reactant-reference energies. A common subset contains 26 reference edges from 23 positive tasks with recoverable values in all four configurations and components matching the final publication records. Edge assignments follow the task's specified reaction and energy-reference definitions. Table~\ref{tab:uncapped-barrier-errors} reports errors against the frozen task reference values, using each edge once per configuration and applying the energy conversion in Section~2.

\begin{table}[tbp]
\centering\small
\caption{Uncapped electronic-barrier errors for the same 26 reference edges in all four configurations, in kcal/mol.}
\label{tab:uncapped-barrier-errors}
\begin{tabular}{lrrr}
\toprule
Workflow & MAE & RMSE & Median $|\Delta|$ \\
\midrule
Guided & 1.17 & 2.86 & 0.057 \\
Standard & 1.54 & 4.64 & 0.057 \\
Record & 1.58 & 5.71 & 0.073 \\
Brief-errors & 1.13 & 2.31 & 0.099 \\
\bottomrule
\end{tabular}
\end{table}

Guided has a lower MAE and RMSE than Standard in this subset, while Brief-errors has the lowest MAE and RMSE. SI Section~S4 gives every included value, the coverage of the 53 positive numerical reference edges, and a smaller subset with additional Gold-input method and medium corroboration. The numerical ordering varies with that coverage.

\subsubsection{Per-task comparisons of barriers, pathways, and calculation evidence}

Figure~\ref{fig:bench-tasks} places all 50 tasks and four configurations alongside their complete score components. NB1 and NI1 also illustrate the scores obtained on multistep mechanisms. NB1 scored 0.9628 under Guided, 0.9128 under Standard, and 0.9900 under Brief-errors; Guided and Brief-errors both had self-consistency 1. NI1 scored 0.9958 under Guided and 0.6196 under Standard, with self-consistency 1 and 0.35. In contrast, NQ5 favored Record over Guided, and NU3 favored Brief-errors.

The archived records explain several barrier differences (Figure~\ref{fig:bench-cases}). In HA8, Guided reported 29.46 kcal/mol and Standard 20.43 kcal/mol against a reference of 29.48 kcal/mol. The archive traces this 9.03 kcal/mol difference to the energy reference: the lowest all-s-trans reactant minimum in Guided and a reactive cisoid conformer in Standard. The IRC identifies the local structures connected by the saddle; the reported activation energy uses the stable-minimum reference specified by the task.

NU6 exposes a related issue in path identity. Guided's recorded concerted saddle had an energy of $-247.030042392$ Ha and the betaine minimum $-247.060060699$ Ha, giving a difference of 18.836772 kcal/mol. Guided described this as an energy cost involving the concerted saddle region while retaining the target betaine-to-product ring-closure saddle as unresolved. Standard delivered 18.837 kcal/mol as the second-step ring-closure barrier. The two reports assign this nearly identical energy difference to different processes. Guided's attempts from interpolated betaine structures and constrained scan peaks repeatedly relaxed toward the direct reactant-to-product saddle region. The saddle's path identity therefore required another assessment from its optimized geometry and pathway.

In G2, Guided scored 0.2389 and Standard 0.7769. Guided's native logs supported an approximately 0.845-kcal/mol double-H-transfer barrier. After three rejections of numerical barrier declarations on a companion edge, the agent removed the numerical fields. Supervisor feedback prompted restoration; a fourth rejection and a subsequent saddle-evidence check preceded their final removal, together with the saddle energy/log bindings. The final graph retained the values in text and received accuracy 0. Brief-errors retained the corresponding numerical result and scored 0.8500.

The accuracy values of 0.7 in D1 and D3 followed the specified cap on numerical-deviation penalties. D1's barrier difference from Standard traced to the reactant-conformer reference; D3's mainly reflected different saddle energies. The Supplementary Information provides the workflow-level components and native calculations for these case studies.

\subsubsection{Impossible tasks distinguish local support from the requested negative conclusion}

The Impossible tier tests whether research identifies the requested target as physically inaccessible under the permitted xTB/PM7 methods. Brief-errors averaged 0.6029, Guided 0.5051, Standard 0.2473, and Record 0.2925 (Figure~\ref{fig:bench-imp}). Brief-errors exceeded Guided on four tasks and scored lower on five, with one tie. The Brief-errors-minus-Guided mean was +0.09788, with a 95\% task-bootstrap interval of [$-0.160$, $+0.361$]. Its weighted contributions were $-0.00661$ from graph, $-0.01550$ from self-consistency, and $+0.12000$ from accuracy. Task-level Brief-errors-minus-Guided differences were IA1 $+0.8283$, IA9 $+0.5666$, IC3 $+0.6000$, IF3 $+0.0333$, IA7 $0$, IA5 $-0.0333$, IB9 $-0.0500$, ID2 $-0.1500$, IE2 $-0.1500$, and IG8 $-0.6661$.

In IA1, concerning cyclobutadiene's square-geometry automerization target, Brief-errors had graph, self-consistency, and accuracy scores of 1, while Guided's accuracy was 0. In IA9, concerning dioxetane thermolysis, Brief-errors and Guided both had self-consistency 1, with accuracy 1 and 0, respectively. IG8 showed the opposite accuracy ordering for the chromium-dimer target. After target correspondence, Guided, Record, and Brief-errors received full negative-target Accuracy credit on IA5 (thermal disrotatory ring opening), and Brief-errors received full credit on IC3 (open-shell singlet phenylnitrene); Guided/IC3 retained Accuracy 0 with Graph and self-consistency both 1. Across the six tasks with the same accuracy score in Brief-errors and Guided, Brief-errors averaged 0.0583 lower. The higher Brief-errors mean therefore arose from the requested conclusions on specific tasks, while Guided retained stronger local support on several others.

Across all 40 Impossible runs, eight had accuracy 1 and ten combined self-consistency 1 with accuracy 0. IF3 illustrates the latter pattern: Brief-errors and Guided both had Accuracy 0 with self-consistency 1. Guided and Brief-errors both received partial accuracy of 0.7 in IB9, whose requested network contains positive and negative components.

\subsection{Molecular identity, distance comparison, and fragment rotation}

We next examine whether the model distinguishes molecular identities, compares distances between specified atoms, detects mismatched calculation references, and identifies missing IRC calculations. Archived research trajectories supply the molecular structures and calculation records. Controlled local studies with Qwen3.8:27b vary the supplied information, answer order, and selected internal states to test their effects on each judgment.

\subsubsection{Equivalent SMILES and answer order affect molecular-identity judgments}
\label{sec:identity-judgments}

A geometric modification first requires identifying the molecule, atoms, and reaction center to which it applies. A molecule can be represented by different SMILES traversals, while a molecular formula can describe different connectivities. To examine how the model handles these distinctions, we constructed 16 molecular-identity pairs covering identical identities, connectivity changes within the same molecular formula, changes in connectivity and charge location, double-bond stereochemistry, and protonation changes. Each pair was presented using the original strings or equivalent rewritings and with two answer orders, yielding 64 judgments.

The model answered 58 correctly. Equivalent rewriting changed the semantic judgment in 4 of 32 matched representation comparisons; reversing answer order changed the judgment in 6 of 32 matched option-order comparisons. All six of the latter cases involved repeated selection of the letter A. One C\textsubscript{5}H\textsubscript{8} pair in Figure~\ref{fig:06} illustrates this behavior directly: an endocyclic and an exocyclic double bond define different connectivities, yet the model answered A under all four conditions, although A meant “same” in one order and “different” in the other.

Cloudra stores molecular connectivity separately from coordinates and uses atom correspondence to identify which atoms to measure, move, or compare. Product-endpoint alignment, path interpolation, and reaction-center measurements retain this correspondence as the structure's coordinates change.

\subsubsection{Geometric measurements reduce repeated numerical reasoning}

We selected four actual structures from the NB1 trajectories and asked the model which of two specified interatomic distances was shorter. Each structure was presented with both answer orders and with either a complete XYZ geometry or program-measured distances. These two information conditions were further crossed with direct-answer and thinking modes, producing 32 native chat calls (Figure~\ref{fig:07}).

In direct-answer mode, all eight responses to XYZ inputs selected A, with four correct answers. Providing measured distances produced eight correct answers and preserved the semantic answer when option order was reversed. With thinking enabled, the XYZ condition produced seven correct answers; the remaining response reached the output limit of 1,024 generated tokens. Increasing the budget for that input alone yielded a correct answer within 1,353 generated tokens. 

All eight answers in the measured-distance condition were correct. Across the eight paired comparisons with thinking enabled, the median generated token counts were 807.5 for XYZ inputs and 86 for measured distances.

Providing the distances removes the need for the model to calculate them from XYZ coordinates. Cloudra similarly computes bond lengths, bond angles, and dihedral angles in software and uses deterministic operators to rotate or translate the selected fragments. The model chooses the atoms and target geometry, then compares the measured distances or angles of the resulting candidates. Assigning these numerical operations to software follows the approach of program-aided language models.\citep{ref15}

\subsubsection{Feedback must match the specified atoms and geometric objective}

In an actual E1 trajectory, the model sought to bring the dihedral angle defined by the transferring hydrogen H7, C2, C3, and leaving atom Cl4 close to 180°. It rotated the fragment on the C3 side about the C2–C3 axis by +120° and \( - \)120°, producing candidates A and B, and set the C3–Cl4 distance to 2.35 Å. The representative dihedral in the default structure report involved H5, whereas the current objective involved H7. The model then explicitly measured H7–C2–C3–Cl4 in both candidates, obtaining approximately 59.4° and 179.4°, and used candidate B in its next actual submission.

Figure~\ref{fig:09} shows Newman projections along C2\(\to\)C3 drawn directly from the saved candidate coordinates, and Figure~\ref{fig:10} summarizes the corresponding target-angle residuals. Candidate B places the H7 target angle within 0.6° of 180°, compared with a residual of 120.6° for A. 

The corresponding residuals for the H5 comparison were 124.2° and 115.8° in A and B, respectively. Identifying the anti-periplanar H7--Cl4 arrangement in B required the H7-based dihedral.

When local probes provided only H5 measurements, the model requested an additional H7 measurement under all three cyclic option permutations. Adding H7 measurements led it to select B under all three permutations. Two further paired tests of information sufficiency also shifted from requesting additional measurements to judging the information sufficient when the target quantity was added.

\subsubsection{Activation replacement affects distance answers differently across option orders}

To test how a reported distance affects the answer, we performed activation replacement at a single token position in the same local quantized model.\citep{ref16} One distance in the input was changed from 2.029 Å to 1.029 Å, while the other remained at 1.248 Å, reversing the correct comparison answer. We then transferred the residual-stream state at the numeric token position from the original input into the modified input at layers 7, 15, 31, and 47, and read the full-vocabulary probability assigned to the answer supported by the original fact.

The effect varied with layer and option order (Figure~\ref{fig:13}). With reversed options, the probability of A, the original-fact answer, increased from 0.074 for the modified input to 0.956 and 0.718 after replacement at layers 7 and 15, respectively. Replacements at layers 31 and 47 yielded 0.072 and 0.070. With the original option order, the probability of answer B changed from 0.009 to 0.029 and 0.015 after replacement at the first two tested layers, while the last two remained close to the modified-input baseline. 

Self-replacement controls and control replacements at an unchanged prefix position produced zero logit differences across the full vocabulary.

With reversed options, replacing the early-layer state restored the original-fact answer as the preferred choice; with the original option order, the modified-fact answer remained preferred. Recovering the answer from the same numeric position therefore depended on option order, as did the identity judgments in Section~\ref{sec:identity-judgments}.

\subsection{Identifying the species and pathways established by calculations}

\subsubsection{Associating frequencies and IRC endpoints with the same saddle}

The frequency calculation and IRC must use the declared saddle, and each endpoint optimization must start from the corresponding IRC terminal frame, followed by frequency analysis of the resulting minimum. The DAG links these calculations to the species and step and records their execution, convergence and physical checks.

We extracted four source evidence chains from the archive and constructed versions that preserved the links or changed a single reference. Crossing these with flat and linked representations and two answer orders yielded 32 inputs. The altered references concerned the frequency geometry, IRC starting geometry, endpoint-optimization input, or target species; local termination, convergence, energy, and frequency values were held constant. Figure~\ref{fig:15} illustrates the E1 case: changing the frequency geometry reference from the saddle point to an existing endpoint changes how that frequency result supports the current step.

All 16 judgments with linked representations were correct, compared with 14 of 16 with flat representations. Both errors occurred when a reference had been changed and the options reversed, in E1 and the second NB1 step. In subsequent native follow-up calls on two selected source chains, enabling thinking yielded correct judgments for all 16 judgments of the evidence relationships, including the two previously incorrect cases. 

The model detected every altered reference in the linked format, even though the local numerical values were unchanged. This presentation effect is consistent with observations that graph encoding affects relational reasoning in language models.\citep{ref17} Cloudra links each species, step and calculation attempt to its calculation results, showing which saddle, IRC and endpoint calculations belong together.

\subsubsection{Diagnostic feedback on IRC connectivity and transition-state curvature}

Diagnostic feedback connects an evidence defect to the chemical object and calculation that require revision. Cloudra's structured diagnostics identify specific problems in quantum-chemical outputs, including steric-clash atom pairs, imaginary-mode misalignment, and disconnected trajectory endpoints. The agent can then change the clashing geometry, inspect the imaginary mode, or recalculate the affected endpoint.

Two archived trajectories illustrate this connection. In a Brief-errors HA1 trial, compact error codes replaced the detailed IRC-endpoint linkage diagnostics. The Computational Agent repeatedly substituted source-minimum references while the geometric disconnection remained unresolved. In a Guided NB1 trial, the verification found zero imaginary frequencies and returned a normal-mode diagnostic. The Coordinator requested a retry with revised transition-state guess parameters and optimization constraints, which the Computational Agent used in the rescheduled calculation.\citep{ref18}

\subsubsection{IA1: a completion request with a missing IRC calculation}

An archived IA1 trajectory shows how a completion request can conflict with the recorded scientific requirements. The reaction step had mandatory bidirectional short\_irc verification pending, and the waiver list was empty. The Supervisor had earlier commented positively on intermediate optimization convergence. The model subsequently requested completion while the IRC requirement remained unresolved.

The harness rejected that request and returned the outstanding short\_irc requirement to the model. In the next submission, the model changed its status to partial completion (Figure~\ref{fig:17}).

We also tested whether the model could identify the outstanding IA1 requirements from the recorded state. All 12 judgments were correct (100\%) across six pairs with reversed answer-option order, using both a minimal summary and an extended 12-message history.

\subsection{Workflow guidance, scientific judgment, and outlook}

Across the 50-task comparison, Guided had higher mean self-consistency and accuracy than Standard. On Hard tasks, their weighted contributions to the gain were +0.1353 and +0.1274. The HH1 and HI1 results show how IRC and endpoint calculations establish the connections behind reference-consistent barriers.

HA8 and NU6 show why energy references and saddle assignments must follow the computed structures and paths. In HA8, the stable reactant minimum defines the barrier reference, while the IRC connects the saddle to a local reactive conformer. In NU6, guesses for betaine-to-product closure repeatedly optimized toward the direct reactant-to-product saddle region. The DAG keeps the reactant conformer, saddle, IRC endpoints and assigned step together so that the agent can check which structure each energy belongs to after a new calculation. The calculations provide external feedback for the agent's planning and verification.\citep{valmeekam2023planning}

These cases call for different next steps: calculate a missing endpoint, revise a guess that returns to another route, or reassess a method that cannot describe the requested electronic state. The agent must update its reaction description when the computed structures or paths change.\citep{turpin2023unfaithful} Record retained the tools and obtained lower self-consistency than Guided, particularly on Normal and Hard tasks. The archived trajectories show how tool feedback directs the next geometric operation or calculation.\citep{ref18,huang2024selfcorrect}

Brief-errors scored higher on IA1 and IA9, whereas Guided scored higher on IG8. These crossed outcomes motivate a task-dependent hypothesis: detailed diagnostics may help repair a TS or IRC calculation, with limited benefit from further local repair when the method or target needs reconsideration.

In G2, repeated checks led the agent to delete numerical values already supported by native calculations. Supervisor feedback restored the fields before further rejection led to their final deletion. A numerical observation and its log should remain attached to the graph while its mechanistic assignment is being tested.

P2 contains one run per task and configuration; its task-cluster bootstrap intervals describe variation across tasks. P1b provides three repeats on each of four tasks under each configuration, totalling 48 runs. Some Hard-task Accuracy targets score qualitative channel identity, so the reported gain includes pathway identification and evidence completion as well as numerical barriers.

Standard sometimes ended research while calculations were still running, whereas Guided/HI1 remained waiting after the completion notification. Returning results therefore need to resume the unfinished study before final reporting. NU6's repeated convergence to one saddle region also motivates comparing new candidates with previous attempts before choosing another guess.

Default preorganization samples 32 rigid relative placements and selects locally relaxed MMFF94 candidates by energy and clustering. CREST provides additional sampling through an optional backend.

The benchmark scores electronic energy differences, including the prescribed references for NI5 in weakly acidic aqueous solution and NU5 in concentrated dichloromethane. Solution-phase $\Delta G^\ddagger$ profiles would include zero-point, thermal/entropic, standard-state and solvation contributions. Solvent participation can also change the chemical model used to describe a pathway; solvent sensitivity remains to be assessed.

The workflow comparisons use fixed atom-mapping and branch-crowding settings and the DeepSeek-v4.1-flash (1017) backbone, with API alias \texttt{deepseek-v4-flash}.

Cloudra is publicly available as open-source software under the GNU Affero General Public License (AGPL) at \url{https://github.com/yyywaa/Cloudra-mechanism-agent}, supporting inspection, reuse, and community development of the system. Reproducible tables and scores are in the companion data package.

\section*{Data availability}
The retained P2 benchmark records, the 48-run P1b study, Gold reference-calculation evidence, and supplementary reproduction materials are publicly available at \url{https://drive.google.com/drive/folders/1yZo3eJuCUuAgedQztgCtLyJcc8MZipf7}. Package contents, versions, and checksums are described in Supplementary Information Section~S7 and the accompanying download records. The source code is publicly available at \url{https://github.com/yyywaa/Cloudra-mechanism-agent} under the AGPL; revision 70daa5a identifies the described implementation. Reproducible tables and scores are in this companion data package.
\bibliographystyle{unsrtnat}
\bibliography{manuscript/references}
\end{document}


\begin{center}
{\LARGE\bfseries Supplementary Information\par}
\vspace{0.6em}
{\large\bfseries Cloudra: Geometric Reasoning and Scientific Validation for Autonomous Reaction Mechanism Exploration\par}
\vspace{0.6em}
{\normalsize Chengyi Yang\,\textsuperscript{1,*}, Qianxin Zhou\,\textsuperscript{2,\textdagger}, and Yang Li\,\textsuperscript{1,\textdagger}\par}
\vspace{0.4em}
{\small \textsuperscript{1}\,College of Chemistry and Pharmaceutical Engineering, Changsha University of Science and Technology, Changsha 410114, China\par
\textsuperscript{2}\,State Key Laboratory of Photocatalysis on Energy and Environment, College of Chemistry, Fuzhou University, Fuzhou 350002, China\par}
\vspace{0.3em}
{\footnotesize \textsuperscript{*}\,Corresponding author: Chengyi Yang. E-mail: \href{mailto:202509240114@csust.edu.cn}{\nolinkurl{202509240114@csust.edu.cn}}.\par}
{\footnotesize \textsuperscript{\textdagger}\,Qianxin Zhou and Yang Li contributed equally as second authors.\par}

\vspace{0.6em}
{\small 2 November 2026\par}
\end{center}
\noindent This supplement provides the experimental protocol, layered evaluation, complete task and score tables, expanded diagnostic analyses, literature source register, and machine-readable materials accompanying the manuscript. Standard denotes the same-model backbone comparison; Record and Brief are workflow ablations. Section, table, figure, equation, and page numbers carry the prefix S to distinguish them from the main text.
\tableofcontents
\section{Benchmark design and experimental protocol}
\label{sec:si-protocol}

\subsection{Task specification and study units}
CloudraBench tests whether agents can construct the requested species and reaction pathways, run the calculations, and draw conclusions supported by the results. It comprises 50 tasks: 20 Easy, 10 Normal, 10 Hard, and 10 Impossible. Easy tasks cover elementary transformations, conformational changes, and basic execution; Normal tasks require identifying intermediates and connecting successive elementary steps; Hard tasks compare competing pathways and stereochemical products; Impossible tasks require recognizing what the allowed electronic-structure methods can support. Each task specifies reactant and target identities, reaction conditions, total charge, spin multiplicity, requested results, and permitted methods. All configurations receive the same specifications for that task. The complete catalogue accompanies the per-task results.

The repeated-task study, P1b, measures repeated runs on four tasks; the full-benchmark study, P2, compares all 50 tasks (Table~\ref{tab:si-study-design}). P1b contains three independent runs for each task and configuration. E1 concerns elimination preorganization; NB1 concerns multistep decarboxylation; HA1 concerns stereochemical ring opening; and IA1 concerns an electronic-structure method limitation. P2 retains one run per task and configuration, so comparisons pair the four configurations on the same task. The 48 repeated-task runs and the 200 full-benchmark runs are analyzed separately.

\begin{table}[htbp]
\centering
\caption{Experimental units in the two studies.}
\label{tab:si-study-design}
\small
\begin{tabular}{p{0.21\textwidth}p{0.31\textwidth}p{0.34\textwidth}}
\toprule
Design & Repeated-task study (P1b) & Full-benchmark study (P2) \\
\midrule
Task set & E1, NB1, HA1, IA1 & All 50 CloudraBench tasks \\
Backbone & DeepSeek-v4.1-flash (1017) & DeepSeek-v4.1-flash (1017); official DeepSeek chat-completions API \\
Configurations & Guided, Standard, Record, Brief-errors & Guided, Standard, Record, Brief-errors \\
Repetitions & Three per task and configuration & One retained run per task and configuration \\
Task-level runs & $4\times4\times3=48$ & $50\times4=200$ \\
Elapsed-time unit & Four-task batch at concurrency two; three batches per configuration & Computational effort on the common 15-task subset \\
\bottomrule
\end{tabular}
\end{table}

\subsection{Workflow configurations and primary comparison}
The primary comparison is Guided against Standard with the same model backbone. Record and Brief-errors test the effects of removing required physical validation and shortening diagnostics, respectively (Table~\ref{tab:si-configurations}). Standard provides computational tools and neutral feedback while leaving physical checks to the agent; explicit delivery-DAG tools are hidden. Guided retains the graph, required scientific checks, and detailed diagnostics. Record retains the tools and records while making physical validation optional. Brief-errors retains guidance and physical validation with shorter diagnostics. Every configuration has the same task and method restrictions.

\begin{table}[htbp]
\centering
\caption{Operational definition of the four workflow configurations. Brief-errors is abbreviated as Brief in figures and tables; Standard is also labelled Backbone.}
\label{tab:si-configurations}
\small
\begin{tabular}{p{0.16\textwidth}p{0.25\textwidth}p{0.25\textwidth}p{0.22\textwidth}}
\toprule
Configuration & Delivery-graph interface & Required physical validation & Diagnostic presentation \\
\midrule
Guided & Explicit & Retained & Detailed \\
Standard & Explicit DAG tools hidden & Removed & Neutral computation feedback \\
Record & Available tools and records retained & Removed & Recording retained \\
Brief-errors & Retained & Retained & Shortened \\
\bottomrule
\end{tabular}
\end{table}

Both studies used DeepSeek-v4.1-flash (1017). All eight saved coordinator and follower configurations and their model-call records use the API alias \texttt{deepseek-v4-flash}.\citep{deepseekv42026} Each run retains its model name, campaign identifier, budget, stopping reason, and calculation files. Cloudra revision 70daa5a identifies the implementation described in the main-text methods.\citep{ref14} Execution, scoring, and offline organization have separate version records; the versions used in each step are taken from its archive.

\subsection{P2 model configuration and campaign versions}
\label{sec:si-p2-model-metadata}

P1b contains four tasks, four workflow configurations, and three runs per combination, giving 48 task runs. P2 contains 50 tasks under the four configurations, giving 200 retained task runs from the P2b and P2c campaigns. Three completed P2b runs were retained (Guided/D3, Record/D1, and Standard/D3); P2c supplied the remaining 197 runs.

The backbone was DeepSeek-v4.1-flash (1017). All eight saved P2 configurations, one per campaign and workflow combination, specify \texttt{deepseek-v4-flash} for both the coordinator and follower. The request-start and completion events in all eight runtime archives record that same identifier. Cloudra used the official DeepSeek endpoint with streaming OpenAI-compatible chat completions, the route selected by \texttt{native\_search=auto} in the campaign implementation. Table~\ref{tab:si-p2-model-metadata} lists the request settings and execution versions, including both the deployment name and the API alias stored in the records.

\begin{table}[htbp]
\centering
\small
\caption{Frozen P2 configuration and campaign metadata. Request settings describe the harness configuration; telemetry counts are provided in the accompanying \texttt{p2\_model\_metadata.json}.}
\label{tab:si-p2-model-metadata}
\begin{tabular}{p{0.27\linewidth}p{0.66\linewidth}}
\toprule
Item & Recorded value \\
\midrule
Deployed model & DeepSeek-v4.1-flash (1017) \\
Provider and API alias & DeepSeek; \texttt{deepseek-v4-flash} in all eight snapshots and their request/completion telemetry \\
Endpoint and transport & \url{https://api.deepseek.com}; streaming chat completions \\
Role slots & Coordinator and follower use the same provider, model identifier, and request settings \\
Reasoning setting & Requested \texttt{reasoning\_effort=minimal} \\
Temperature & 1.0 in the archived run-start settings \\
Context window & 400,000 tokens in the frozen harness configuration \\
Spawned-agent budget & \texttt{max\_tokens=32768} in the archived \texttt{agent.spawned} settings; this budget applies to each agent request. Total run token consumption is a separate quantity. \\
Tool invocation & JavaScript protocol \\
Retained task runs & Three from P2b and 197 from P2c \\
Task concurrency & Two per configuration, eight in total \\
Campaign starts & P2b: 25 October 2026, 18:01:17 UTC+8; P2c: 25 October 2026, 22:18:45 UTC+8 \\
Campaign base & \texttt{53309d0e} \\
Configuration commits & Guided: \texttt{6e89d20c}; Record: \texttt{746c33f0}; Standard: \texttt{c7fc94a1}; Brief-errors: \texttt{3b9831bb}. Both campaigns use these commits. \\
\bottomrule
\end{tabular}
\end{table}

The configuration commits identify the code used to run the experiments. The main-text implementation, later evaluator fixes, and publication-table reproduction have separate version records. The companion metadata file gives the archive path of each configuration and runtime record. Its request-start and completion counts include auxiliary LLM calls and cancelled P2b attempts; the benchmark includes only the 200 retained task runs.

\paragraph{Fixed geometric heuristics.}
Atom maps are ranked with fixed costs: 2 for each formed or broken bond, 1 for each bond-order change, 4 for a hydrogen-count mismatch, 8 for charge or radical-electron mismatches, and 2 for a stereochemical mismatch. Maps within $J_{\min}+2$ are retained when reporting ambiguity. The branch-crowding descriptor uses decay $\rho=0.5$, traversal through depth five, and a congestion flag when the four-branch total $B_{\mathrm{axis}}\geq9.0$. These heuristics rank maps and flag structures for inspection. Physical validation uses the quantum-chemical results. The imaginary-mode participation ratio is compared with 0.5 as an advisory diagnostic; validation requires the first-order saddle and IRC/endpoint calculations.

\paragraph{Conformational sampling.}
Default preorganization places molecular fragments using their topology, relaxes them with guided MMFF94 optimization, and selects candidates by energy and RMSD. For multicomponent systems, it samples 32 relative orientations and retains up to eight guided placements. A separate conformer-search tool uses RDKit by default and offers CREST\citep{crest2020} for additional conformer or noncovalent-complex sampling. The default CREST route uses GFN2-xTB with the quick setting. The resulting geometries are candidates for subsequent calculations and must pass the task's stationary-point and pathway checks.

\subsection{Allowed calculations and reference definitions}
Easy, Normal, and Hard tasks permit xTB, PM7, and B3LYP-D3(BJ)/def2-SVP.\citep{gfn2xtb2019,pm72013,becke1993,lyp1988,stephens1994,d32010,d3bj2011,def22005} Impossible tasks are restricted to xTB and PM7. Computational quantities and termination status are read from native ORCA, Gaussian, and MOPAC outputs.\citep{orca2025,gaussian16,mopac2026,ref11} Each result retains its program, calculation level, species, and task conditions.

The reference answer (Gold) specifies the mechanistic species and elementary steps, stereochemical identities, applicable numerical targets, and explicit negative targets. Numerical activation barriers use electronic energies at a consistent level of theory:
\begin{equation}
\Delta E_e^{\ddagger}=E_{\mathrm{TS},e}-E_{\mathrm{ref},e}.
\label{eq:si-barrier}
\end{equation}
These benchmark barriers use electronic energies alone. The energy reference is the verified stable minimum of the corresponding species. Declared conformational connections within the same species can share that stable reference; a conformational conversion itself uses its own starting conformer. Distinct stereochemical substrates and independent channel contact complexes retain separate references. The IRC endpoint and the energy reference are recorded separately, since they can be different conformers of the same species.

References follow the individual task definitions. NB1\_E2 and NQ5\_E2 are exempt from numerical references under the specified proton-transfer policy. IB9 combines a numerical first-step reference of 5.65~kcal~mol$^{-1}$ with a negative second-step target. Section~\ref{sec:si-scoring} specifies the qualitative selectivity targets and numerical tolerances. Each reference record separately states whether the task and Gold graph are available and whether the complete calculation chain has been physically validated.

\subsection{Physical evidence associated with a mechanism}
A mechanism graph represents species occurrences as nodes and elementary steps as edges. A species minimum requires a converged optimization and zero imaginary frequencies. An accessible transition-state step requires a first-order saddle with exactly one imaginary frequency associated with the intended reaction, plus IRC results connecting that saddle to the declared endpoint species.\citep{ref09,ref10} Normal termination, geometry convergence, frequencies, and pathway connectivity are checked separately.

Both IRC directions must start at the same transition state. The actual outer frame from each direction is then optimized to a minimum with zero imaginary frequencies and the declared chemical identity and stereochemistry. This outer-frame--optimization--frequency chain can establish the connection from a short IRC. The frames and links between calculations are archived. Native calculation results establish the endpoint connection; geometric interval measurements aid interpretation. For separated products, the required constituent minima are verified; an intact molecular complex is evaluated as a whole-system species.

These requirements apply to the species and steps requested by the task. A negative conclusion marks the target as physically inaccessible under the permitted methods and records the reason. A species-only negative question is evaluated directly on that species claim.

\subsection{Offline organization and publication records}
Formal evaluation represents the research completed by all four configurations in the same mechanism-graph format. Human organizers read each run's conversations, tool records, reports, geometries, and native calculation outputs, then encode the agent's stated species, steps, mechanistic intent, and conclusions. They recover graph entries, repair file references, and convert quantities already present in the run; each recovered field records its source. The evaluator searches that run's sandbox for native calculation results and applies the same physical criteria and scoring rules to every configuration. Organizers use each run's existing calculations and stated conclusions; Gold supplies the evaluation targets.

The full-benchmark table contains 200 runs, all 600 Graph, Self-consistency, and Accuracy values, and their evaluation sources. The weighted components reproduce the final totals to four decimal places; Section~S7 specifies the references and evaluator revision. The tables also record whether the run wrote a root mechanism graph and whether it was organized offline.

The repeated-task study retains its historical results for all four configurations. The later offline organization updates Standard separately; the other configurations in the 48-run study retain their historical scores. Analyses of this archive identify which version supplies each result.

\subsection{Paired statistical analysis and time measurements}
For a task $i$ and configuration $a$, let $S_{ia}$ denote its composite score. Configuration means give each task equal weight within the reported set. The primary paired difference is
\begin{equation}
\overline{\Delta}_{\mathrm{G-S}}=\frac{1}{n}\sum_{i=1}^{n}
\left(S_{i,\mathrm{Guided}}-S_{i,\mathrm{Standard}}\right).
\label{eq:si-paired-difference}
\end{equation}
The same calculation is used within each tier and for the combined Normal/Hard subset. Weighted component contributions are the mean paired component differences multiplied by 0.1, 0.3, and 0.6, respectively. They add to the composite difference up to the reporting precision of the final scores.

The 95\% percentile intervals use 100,000 task-cluster bootstrap samples with seed 20261031. Each draw selects a task with its four configuration records together, preserving the paired comparison.

Elapsed time in the repeated-task study runs from the recorded start to finish of each four-task batch at concurrency two: three batches per configuration, or 12 batches in total. It is measured in batch wall-clock hours. Full-benchmark effort is measured on the 15 tasks with records for all four configurations, comprising 60 task-level runs. Calculation counts and cumulative quantum-chemical job-hours are reported; job-hours sum the individual calculation durations within each run. For Standard/M1, native-log CPU time is divided by four to estimate job duration. The tables keep job-hours separate from core-hours and from the elapsed time of an entire P2 research run.

\section{Score components and scientific completion}
\label{sec:si-scoring}

\subsection{Composite score and reported outcomes}
The composite score combines graph structure $G$, physical self-consistency $\mathrm{SC}$, and accuracy $A$:
\begin{equation}
S=0.1G+0.3\mathrm{SC}+0.6A.
\label{eq:si-composite}
\end{equation}
The three components are reported separately, alongside execution status, original graph delivery, and completion of the requested steps. Complete scientific validation additionally requires every requested species and step to meet its physical criteria.

\subsection{Graph structure and chemical correspondence}
The graph component measures how well the delivered mechanism covers the reference species and steps. For comparison, transition-state nodes are folded into their elementary-step edges. Species are matched by chemical identity; steps are compared by their endpoints and atom-mapped bond changes. RDKit represents the species and NetworkX represents and matches the graphs.\citep{ref13,networkx2008} Missing reference species or steps, incorrect chemical identities, and mismatched accessibility declarations add edit costs. The graph component permits additional nodes and edges.

Graph comparison ignores stereochemistry for Easy and Normal tasks and retains it for Hard tasks. A species' linked native calculation results determine which evidence is assessed for it. Graph coverage receives its own score independently of the binary self-consistency result.

\subsection{Graded physical self-consistency}
Self-consistency is scored on the species and steps in the organized graph, independently of numerical agreement with Gold. Table~\ref{tab:si-self-consistency} lists the credit for each calculation stage of a transition-state step; a complete verified chain receives 1.0. The evaluator searches the same run's sandbox for the native files and links between calculations, recovering file associations where needed. It assigns credit from these results using the common physical criteria.

\begin{table}[htbp]
\centering
\caption{Graded self-consistency contribution for an accessible transition-state step. The endpoint contribution is 0.25 for one verified side and 0.50 for both sides.}
\label{tab:si-self-consistency}
\small
\begin{tabular}{p{0.65\textwidth}r}
\toprule
Evidence stage & Contribution \\
\midrule
Verified first-order saddle: exactly one imaginary frequency & 0.35 \\
IRC starting geometry consistent with the selected saddle & 0.15 \\
Logs linking both actual IRC terminal frames through Opt/Freq to the declared minima & 0.50 \\
\bottomrule
\end{tabular}
\end{table}

For graphs containing reaction edges, $\mathrm{SC}$ is the mean of the delivered edge scores. An edge represented by its endpoints alone requires coverage of both endpoint species. For a species-only graph, $\mathrm{SC}$ is the mean species-coverage score. Explicit targets judged physically inaccessible under the task's permitted methods are exempt from the corresponding stationary-point or pathway requirements and receive the corresponding self-consistency credit.

The archived binary self-consistency indicator requires all physical checks for the delivered species and steps, including minimum characterization and transition-state/IRC associations. Task-wide scientific completion requires these checks for every requested species and step. Counts of $\mathrm{SC}=1$ in the main text and score tables refer to full graded component scores for the delivered objects.

\subsection{Electronic barriers and numerical accuracy}
For a step with a numerical reference, the evaluator checks the declared barrier against the transition-state and starting-species energies in the native outputs. If the barrier field is empty, it can calculate the difference from those energies when both are reported and supported by their native records. The conversion factor from hartree to kcal~mol$^{-1}$ is 627.509474.

For a supported numerical value $\widehat{\Delta E}_e^{\ddagger}$ and reference $\Delta E_{e,\mathrm{Gold}}^{\ddagger}$, the underlying Gaussian score is
\begin{equation}
q_e=\exp\left[-\frac{\left(\widehat{\Delta E}_e^{\ddagger}-
\Delta E_{e,\mathrm{Gold}}^{\ddagger}\right)^2}{2\sigma_e^2}\right].
\label{eq:si-gaussian}
\end{equation}
The released task rubrics use base widths of 3.0~kcal~mol$^{-1}$ for Easy and Impossible tasks and 2.0~kcal~mol$^{-1}$ for Normal and Hard tasks. For a reference barrier at or below 10.0~kcal~mol$^{-1}$, the reference-comparison width is the larger of the base width and 5.0~kcal~mol$^{-1}$. This small-barrier width applies to comparison with Gold; the declaration-to-log consistency check retains the task's base width.

All reference edges remain in the accuracy denominator with their recorded weights. A numerical reference edge requires a matching delivered step, a declared barrier, and a log-derived value to receive credit. Missing any of these gives zero for that edge; other supported steps keep their credit. The combined penalty for numerical deviations on supported edges is capped at 0.3. Penalties for missing steps, missing evidence, and incorrect negative conclusions apply in full. For total reference-edge weight $W$, supported numerically graded edges $D$, and missing or incorrectly delivered targets $U$,
\begin{equation}
A=1-\min\!\left(\frac{\sum_{e\in D}w_e(1-q_e)}{W},\,0.3\right)
-\frac{\sum_{e\in U}w_e}{W}.
\label{eq:si-accuracy-aggregation}
\end{equation}
Correct negative conclusions and accessible edges exempt from numerical references receive their prescribed full credit through $W$. For the latter, credit concerns the specified nonnumerical result.

The evaluator checks the allowed combinations of transition-state and reference energies in the delivered results. If one combination agrees with Gold and the reported barrier uses a different supported conformer or energy reference, the answer-definition rule replaces $q_e$ with 0.9. An unsupported barrier declaration with zero reference-consistent alternatives triggers the integrity rule: the self-consistency and accuracy contributions become zero, leaving $0.1G$. An omitted step loses its own contribution.

\subsection{Qualitative selectivity and property targets}
For HA1, HA6, HB1, and HF1, Accuracy scores the qualitative choice of favored channel. The delivered mechanism must identify one favored edge matching the reference channel or its product. Product identity is used when the agent's channel label differs from the reference label. Numerical barrier spreads are retained as calculations; the scored target is the qualitative channel preference. HL1 scores its independently referenced channel barriers.

For species-property tasks, Accuracy compares the declared properties with their references. Numerical comparison widths are 0.02~\AA\ for bond lengths, 50~cm$^{-1}$ for frequencies, and 2.0~kcal~mol$^{-1}$ for energy differences. State-ordering text is compared after case and whitespace normalization. When several properties are requested, their scores are averaged.

\subsection{Impossible tasks and final scientific claims}
Impossible tasks are scored on the agent's final conclusion about the requested target. Correctly identifying that target as physically inaccessible under the allowed methods earns negative-target credit, including after exploratory calculations. Credit is lost for an incorrect claim that the target is accessible, a barrier or property presented as trustworthy for that negative target, or omission of the required negative conclusion. Endpoint species associated with a negative step receive the same treatment. A species-only negative target is evaluated directly on the species claim.

\subsection{Original delivery, organized evaluation, and score reproduction}
Original delivery completeness records whether the run generated its mechanism graph during execution. Evaluation uses the graph organized by the common offline procedure in Section~\ref{sec:si-protocol}. The publication table records each run's final component scores and field sources. All configurations use the same final references and scoring rules.

The reproduction procedure reads the evaluation records to rebuild the per-run table, summaries, and figure data. It checks for 200 unique task/configuration records, 600 nonmissing component values, and agreement between the weighted components and the 200 final totals. The maximum absolute difference from the published four-decimal composites is $4.9711\times10^{-5}$. Table reproduction uses the saved component scores; calculation reassessment uses the run files and evaluator revision.

\section{CloudraBench task catalogue}
\label{sec:task-catalogue}

CloudraBench comprises 20 Easy, 10 Normal, 10 Hard, and 10 Impossible tasks. Tables~\ref{tab:catalogue-easy}--\ref{tab:catalogue-impossible} list each task's chemistry, charge, multiplicity, conditions, and research question. The machine-readable catalogue contains task names and molecular inputs; the full task definitions also contain atom mappings, prompts, reference graphs, and scoring settings.

Method set A permits xTB, PM7, and B3LYP-D3(BJ)/def2-SVP for Easy, Normal, and Hard tasks. Method set B permits xTB and PM7 for Impossible tasks. Each answer is evaluated under its task's permitted methods. Mechanism tasks request the investigated species and elementary steps as a graph, electronic activation barriers for computed steps, and the corresponding stationary-point and pathway calculations. Stereochemical tasks also require assigning identities from the computed geometries. The tables list each task's specific question in addition to these shared requirements.

For Impossible tasks, a supported conclusion that the target is physically inaccessible under the allowed methods is a valid result.

\begingroup
\small
\setlength{\tabcolsep}{3.5pt}
\renewcommand{\arraystretch}{1.15}
\begin{longtable}{@{}p{0.065\linewidth}p{0.285\linewidth}p{0.60\linewidth}@{}}
\caption{Easy-tier task catalogue. Method set A applies to every task in this table.}\label{tab:catalogue-easy}\\
\toprule
ID & Chemical task & Stated conditions and task-specific research target \\
\midrule
\endfirsthead
\multicolumn{3}{l}{\tablename\ \thetable\ (continued)}\\
\toprule
ID & Chemical task & Stated conditions and task-specific research target \\
\midrule
\endhead
\midrule
\multicolumn{3}{r}{Continued on the next page}\\
\endfoot
\bottomrule
\endlastfoot
D1 & butadiene + ethylene Diels-Alder (parent DA)\newline {\footnotesize Charge 0; multiplicity 1.} & \textit{Conditions:} gas phase, 487--648 \({}^{\circ}\mathrm{C}\) kinetics.\newline \textit{Target:} Cycloaddition pathway, stationary points, and electronic activation barrier. \\
\addlinespace[3pt]
D3 & cyclopentadiene dimerization (self Diels-Alder)\newline {\footnotesize Charge 0; multiplicity 1.} & \textit{Conditions:} gas/liquid phase, slow at room temperature.\newline \textit{Target:} Dimerization pathway, stationary points, and electronic activation barrier. \\
\addlinespace[3pt]
D4 & cyclopentadiene + acrolein Diels-Alder\newline {\footnotesize Charge 0; multiplicity 1.} & \textit{Conditions:} gas phase or hydrocarbon solvent, heated.\newline \textit{Target:} Cycloaddition pathway, stationary points, and electronic activation barrier. \\
\addlinespace[3pt]
E1 & F- + CH3CH2Cl elimination -> ethylene + HF + Cl-\newline {\footnotesize Charge -1; multiplicity 1.} & \textit{Conditions:} gas phase.\newline \textit{Target:} Elimination pathway from the pre-reaction complex, stationary points, and electronic activation barrier. \\
\addlinespace[3pt]
F1 & BH4- + formaldehyde -> borane-methoxide adduct [CH3OBH3]-\newline {\footnotesize Charge -1; multiplicity 1.} & \textit{Conditions:} gas phase.\newline \textit{Target:} Hydride-transfer pathway to the borane--methoxide adduct, stationary points, and electronic barrier. \\
\addlinespace[3pt]
F4 & BH3 + formaldehyde -> methoxyborane via Lewis adduct\newline {\footnotesize Charge 0; multiplicity 1.} & \textit{Conditions:} gas phase.\newline \textit{Target:} Lewis-adduct-mediated conversion to methoxyborane, with the investigated species and elementary steps. \\
\addlinespace[3pt]
G2 & diimide cis reduction of ethylene\newline {\footnotesize Charge 0; multiplicity 1.} & \textit{Conditions:} gas phase.\newline \textit{Target:} Reduction pathway, stationary points, and electronic activation barrier. \\
\addlinespace[3pt]
G3 & propylene + ethylene ene reaction (Alder-ene)\newline {\footnotesize Charge 0; multiplicity 1.} & \textit{Conditions:} gas phase, high temperature.\newline \textit{Target:} Ene-reaction pathway, stationary points, and electronic activation barrier. \\
\addlinespace[3pt]
L1 & cyclobutene -> butadiene ring opening\newline {\footnotesize Charge 0; multiplicity 1.} & \textit{Conditions:} gas phase, 403-448 K.\newline \textit{Target:} Ring-opening pathway, stationary points, and electronic activation barrier. \\
\addlinespace[3pt]
L2 & hexatriene -> cyclohexadiene ring closure\newline {\footnotesize Charge 0; multiplicity 1.} & \textit{Conditions:} gas phase, 390-434 K.\newline \textit{Target:} Electrocyclic ring-closure pathway, stationary points, and electronic activation barrier. \\
\addlinespace[3pt]
M1 & 2,2'-dimethylbiphenyl aryl-aryl rotation\newline {\footnotesize Charge 0; multiplicity 1.} & \textit{Conditions:} gas phase.\newline \textit{Target:} Aryl--aryl rotational pathway, connected conformers, and rotational barrier. \\
\addlinespace[3pt]
M14 & acrolein s-trans <-> s-cis rotation\newline {\footnotesize Charge 0; multiplicity 1.} & \textit{Conditions:} gas phase.\newline \textit{Target:} s-trans/s-cis conformers, their connecting pathway, and rotational barrier. \\
\addlinespace[3pt]
M15 & HONO cis-trans isomerization\newline {\footnotesize Charge 0; multiplicity 1.} & \textit{Conditions:} gas phase.\newline \textit{Target:} cis/trans structures, their connecting pathway, and isomerization barrier. \\
\addlinespace[3pt]
M16 & formic acid dehydrogenation HCOOH -> CO2 + H2\newline {\footnotesize Charge 0; multiplicity 1.} & \textit{Conditions:} gas phase.\newline \textit{Target:} Dehydrogenation pathway and electronic activation barrier. \\
\addlinespace[3pt]
M17 & methyl isocyanide -> acetonitrile rearrangement\newline {\footnotesize Charge 0; multiplicity 1.} & \textit{Conditions:} gas phase.\newline \textit{Target:} Isomerization pathway to acetonitrile and electronic activation barrier. \\
\addlinespace[3pt]
M3 & NH3 umbrella inversion\newline {\footnotesize Charge 0; multiplicity 1.} & \textit{Conditions:} gas phase.\newline \textit{Target:} Inversion pathway, stationary points, and inversion barrier. \\
\addlinespace[3pt]
M8 & cyclohexane chair flip\newline {\footnotesize Charge 0; multiplicity 1.} & \textit{Conditions:} gas phase.\newline \textit{Target:} Chair-interconversion pathway, relevant conformers, and barrier. \\
\addlinespace[3pt]
P2 & cyclopentadiene degenerate H shift\newline {\footnotesize Charge 0; multiplicity 1.} & \textit{Conditions:} gas phase or solution, fast at room temperature.\newline \textit{Target:} Degenerate hydrogen-shift pathway, stationary points, and electronic activation barrier. \\
\addlinespace[3pt]
P3 & allyl vinyl ether Claisen rearrangement\newline {\footnotesize Charge 0; multiplicity 1.} & \textit{Conditions:} gas phase, \(\sim\)200 \({}^{\circ}\mathrm{C}\).\newline \textit{Target:} Claisen-rearrangement pathway, stationary points, and electronic activation barrier. \\
\addlinespace[3pt]
X1 & beta-propiolactone -> ethylene + CO2\newline {\footnotesize Charge 0; multiplicity 1.} & \textit{Conditions:} gas phase, \(\sim\)350 \({}^{\circ}\mathrm{C}\).\newline \textit{Target:} Decomposition pathway to ethylene and carbon dioxide, with stationary points and barrier. \\
\addlinespace[3pt]
\end{longtable}
\endgroup

\begingroup
\small
\setlength{\tabcolsep}{3.5pt}
\renewcommand{\arraystretch}{1.15}
\begin{longtable}{@{}p{0.065\linewidth}p{0.285\linewidth}p{0.60\linewidth}@{}}
\caption{Normal-tier task catalogue. Method set A applies to every task in this table.}\label{tab:catalogue-normal}\\
\toprule
ID & Chemical task & Stated conditions and task-specific research target \\
\midrule
\endfirsthead
\multicolumn{3}{l}{\tablename\ \thetable\ (continued)}\\
\toprule
ID & Chemical task & Stated conditions and task-specific research target \\
\midrule
\endhead
\midrule
\multicolumn{3}{r}{Continued on the next page}\\
\endfoot
\bottomrule
\endlastfoot
NB1 & malonic acid thermal decarboxylation: malonic acid -> acetic acid + CO2\newline {\footnotesize Charge 0; multiplicity 1.} & \textit{Conditions:} gas-phase thermolysis.\newline \textit{Target:} Identify intermediates and preserve elementary-step continuity through decarboxylation to the final products. \\
\addlinespace[3pt]
NI1 & imine formation: formaldehyde + methylamine -> N-methylmethanimine + water\newline {\footnotesize Charge 0; multiplicity 1.} & \textit{Conditions:} gas phase; optimal around pH 4-6 (bell-shaped rate profile).\newline \textit{Target:} Identify the addition/condensation intermediates and connecting elementary steps leading to the imine and water. \\
\addlinespace[3pt]
NI5 & imine formation (aqueous): formaldehyde + methylamine -> N-methylmethanimine + water\newline {\footnotesize Charge 0; multiplicity 1.} & \textit{Conditions:} aqueous solution, weakly acidic (pH 4-6, bell-shaped rate profile).\newline \textit{Target:} Resolve the aqueous imine-formation network and relevant proton-transfer steps under the stated conditions. \\
\addlinespace[3pt]
NQ1 & aromatic Claisen rearrangement: allyl phenyl ether -> 2-allylphenol\newline {\footnotesize Charge 0; multiplicity 1.} & \textit{Conditions:} gas phase; thermal, \(\sim\)200 \({}^{\circ}\mathrm{C}\).\newline \textit{Target:} Resolve rearrangement and subsequent conversion to 2-allylphenol, retaining the intermediate network. \\
\addlinespace[3pt]
NQ5 & oxy-Cope rearrangement: 1,5-hexadien-3-ol -> hex-5-enal ([3,3] then enol-keto tautomerization)\newline {\footnotesize Charge 0; multiplicity 1.} & \textit{Conditions:} gas phase, heated.\newline \textit{Target:} Resolve the [3,3] rearrangement and enol--keto tautomerization as the investigated network. \\
\addlinespace[3pt]
NR1 & formamide dehydration: formamide -> hydrogen cyanide + water\newline {\footnotesize Charge 0; multiplicity 1.} & \textit{Conditions:} Gas phase; isolated neutral, closed-shell formamide; no added water, acid, or catalyst.\newline \textit{Target:} Investigate formimidic-acid intermediates and conformational transitions in the neutral dehydration network. \\
\addlinespace[3pt]
NU3 & H3P=CH2 ylide + formaldehyde -> ethylene + H3P=O\newline {\footnotesize Charge 0; multiplicity 1.} & \textit{Conditions:} salt-free model conditions (gas-phase ylide, closest to model chemistry).\newline \textit{Target:} Resolve the salt-free model Wittig-reaction network to ethylene and phosphine oxide. \\
\addlinespace[3pt]
NU5 & Passerini reaction: formaldehyde + formic acid + methyl isocyanide -> N-methyl-alpha-formyloxyacetamide\newline {\footnotesize Charge 0; multiplicity 1.} & \textit{Conditions:} dichloromethane (DCM), room temperature, high concentration.\newline \textit{Target:} Resolve the three-component Passerini-reaction network to the specified alpha-acyloxy amide. \\
\addlinespace[3pt]
NU6 & Staudinger beta-lactam synthesis: ketene + methanimine -> 2-azetidinone\newline {\footnotesize Charge 0; multiplicity 1.} & \textit{Conditions:} ketene generated in situ; inert solvent.\newline \textit{Target:} Identify the ketene--imine addition and beta-lactam-forming network, retaining any identified intermediate. \\
\addlinespace[3pt]
NY3 & isocyanic acid hydrolysis: HNCO + one water -> ammonia + carbon dioxide\newline {\footnotesize Charge 0; multiplicity 1.} & \textit{Conditions:} Gas phase; neutral closed-shell system with exactly one water molecule; no added solvent, acid, counterion, or catalyst.\newline \textit{Target:} Investigate carbamic-acid intermediates and connecting conformers without presupposing a fixed number of steps. \\
\addlinespace[3pt]
\end{longtable}
\endgroup

\begingroup
\small
\setlength{\tabcolsep}{3.5pt}
\renewcommand{\arraystretch}{1.15}
\begin{longtable}{@{}p{0.065\linewidth}p{0.285\linewidth}p{0.60\linewidth}@{}}
\caption{Hard-tier task catalogue. Method set A applies to every task in this table.}\label{tab:catalogue-hard}\\
\toprule
ID & Chemical task & Stated conditions and task-specific research target \\
\midrule
\endfirsthead
\multicolumn{3}{l}{\tablename\ \thetable\ (continued)}\\
\toprule
ID & Chemical task & Stated conditions and task-specific research target \\
\midrule
\endhead
\midrule
\multicolumn{3}{r}{Continued on the next page}\\
\endfoot
\bottomrule
\endlastfoot
HA1 & 3-methylcyclobutene ring opening: determine the stereochemical outcome\newline {\footnotesize Charge 0; multiplicity 1.} & \textit{Conditions:} gas phase or inert solvent, \(\sim\)150--200 \({}^{\circ}\mathrm{C}\), thermal.\newline \textit{Target:} Determine stereochemical products and channel identities; report channel barriers and the requested relative preference. \\
\addlinespace[3pt]
HA6 & trans-3,4-dimethylcyclobutene ring opening: determine the stereochemical outcome\newline {\footnotesize Charge 0; multiplicity 1.} & \textit{Conditions:} gas phase or inert solvent, heating, thermal.\newline \textit{Target:} Determine stereochemical products and channel identities; report channel barriers and the requested relative preference. \\
\addlinespace[3pt]
HA8 & (2E,4Z,6E)-octa-2,4,6-triene thermal 6-pi electrocyclization: determine the stereochemical outcome\newline {\footnotesize Charge 0; multiplicity 1.} & \textit{Conditions:} Gas phase; first thermal ground-state electrocyclization only; no subsequent hydrogen shifts or photochemistry.\newline \textit{Target:} Resolve the product stereochemistry of the first thermal electrocyclization; report investigated channel identities and barriers. \\
\addlinespace[3pt]
HB1 & cyclopentadiene + maleic anhydride Diels-Alder: determine endo/exo outcome\newline {\footnotesize Charge 0; multiplicity 1.} & \textit{Conditions:} benzene, room temperature or mild heating.\newline \textit{Target:} Determine endo/exo products, channel barriers, and relative preference. \\
\addlinespace[3pt]
HB10 & cyclopentadiene + acrylonitrile Diels-Alder reaction: determine both stereochemical pathways\newline {\footnotesize Charge 0; multiplicity 1.} & \textit{Conditions:} Gas phase; concerted addition; common reactant-complex reference for the two channels.\newline \textit{Target:} Compute both endo/exo pathways using a common reactant-complex reference; report identities and barriers. \\
\addlinespace[3pt]
HD2 & cyclohexene oxide + hydroxide: determine the stereochemical outcome of the epoxide opening\newline {\footnotesize Charge -1; multiplicity 1.} & \textit{Conditions:} aqueous NaOH, heat.\newline \textit{Target:} Determine the stereochemistry of the epoxide-opening product from the computed structures. \\
\addlinespace[3pt]
HF1 & cyclohex-2-en-1-ol peracid epoxidation: determine the stereochemical outcome\newline {\footnotesize Charge 0; multiplicity 1.} & \textit{Conditions:} performic acid, CH2Cl2, 0--25 \({}^{\circ}\mathrm{C}\).\newline \textit{Target:} Determine the two facial pathways, product stereochemistry, barriers, and relative preference. \\
\addlinespace[3pt]
HH1 & 1,1'-binaphthyl thermal racemization\newline {\footnotesize Charge 0; multiplicity 1.} & \textit{Conditions:} thermal racemization, \(\sim\)50 \({}^{\circ}\mathrm{C}\), DMF.\newline \textit{Target:} Characterize the thermal racemization pathway, stereochemical endpoints, and barrier. \\
\addlinespace[3pt]
HI1 & Cope elimination of threo/erythro N,N-dimethyl-3-phenyl-2-butylamine oxide: determine the stereochemical mapping\newline {\footnotesize Charge 0; multiplicity 1.} & \textit{Conditions:} Gas phase; isolated neutral singlet amine oxide. Treat threo and erythro substrates as two separate systems and retain explicit stereochemical SMILES.\newline \textit{Target:} Compute threo and erythro substrates separately; establish substrate CIP assignments and alkene E/Z identity from 3D geometries. Use each substrate's own minimum for its barrier. \\
\addlinespace[3pt]
HL1 & formonitrile oxide + norbornene [3+2] cycloaddition: determine both facial pathways\newline {\footnotesize Charge 0; multiplicity 1.} & \textit{Conditions:} Gas phase; neutral singlet formonitrile oxide and unsubstituted norbornene. Compare approach from both faces relative to the methano bridge.\newline \textit{Target:} Compute both facial approaches relative to the methano bridge and assign products from 3D geometries. Use each channel's own stable contact minimum; report contact-conformer energies separately. \\
\addlinespace[3pt]
\end{longtable}
\endgroup

\begingroup
\small
\setlength{\tabcolsep}{3.5pt}
\renewcommand{\arraystretch}{1.15}
\begin{longtable}{@{}p{0.065\linewidth}p{0.285\linewidth}p{0.60\linewidth}@{}}
\caption{Impossible-tier task catalogue. Method set B applies to every task in this table.}\label{tab:catalogue-impossible}\\
\toprule
ID & Chemical task & Stated conditions and task-specific research target \\
\midrule
\endfirsthead
\multicolumn{3}{l}{\tablename\ \thetable\ (continued)}\\
\toprule
ID & Chemical task & Stated conditions and task-specific research target \\
\midrule
\endhead
\midrule
\multicolumn{3}{r}{Continued on the next page}\\
\endfoot
\bottomrule
\endlastfoot
IA1 & cyclobutadiene D4h automerization\newline {\footnotesize Charge 0; multiplicity 1.} & \textit{Conditions:} gas phase, low temperature (matrix isolation regime).\newline \textit{Target:} Thermal automerization of rectangular cyclobutadiene through the square transition structure: mechanism and activation barrier. \\
\addlinespace[3pt]
IA5 & cyclobutene disrotatory ring opening\newline {\footnotesize Charge 0; multiplicity 1.} & \textit{Conditions:} gas phase, thermal.\newline \textit{Target:} Thermal disrotatory ring opening of cyclobutene to butadiene: mechanism and activation barrier. \\
\addlinespace[3pt]
IA7 & Bergman cyclization of (Z)-hex-3-ene-1,5-diyne\newline {\footnotesize Charge 0; multiplicity 1.} & \textit{Conditions:} gas phase, thermal.\newline \textit{Target:} Bergman cyclization to p-benzyne: mechanism and activation barrier. \\
\addlinespace[3pt]
IA9 & 1,2-dioxetane thermolysis to two formaldehyde\newline {\footnotesize Charge 0; multiplicity 1.} & \textit{Conditions:} gas phase, thermal.\newline \textit{Target:} Thermolysis of 1,2-dioxetane to two formaldehyde molecules: mechanism and activation barrier. \\
\addlinespace[3pt]
IB9 & diazomethane and ethylene to cyclopropane via 1-pyrazoline\newline {\footnotesize Charge 0; multiplicity 1.} & \textit{Conditions:} gas-phase thermolysis.\newline \textit{Target:} Conversion of diazomethane and ethylene to cyclopropane and nitrogen through 1-pyrazoline. \\
\addlinespace[3pt]
IC3 & phenylnitrene open-shell singlet state\newline {\footnotesize Charge 0; multiplicity 3.} & \textit{Conditions:} gas phase.\newline \textit{Target:} Singlet--triplet energy gap and the electronic character of the lowest phenylnitrene singlet state. \\
\addlinespace[3pt]
ID2 & anomalous bonding in dioxygen difluoride (FOOF)\newline {\footnotesize Charge 0; multiplicity 1.} & \textit{Conditions:} gas phase, low temperature.\newline \textit{Target:} Equilibrium F--O and O--O distances of FOOF and its F--O bond-dissociation energy. \\
\addlinespace[3pt]
IE2 & nitrogen triple-bond dissociation curve\newline {\footnotesize Charge 0; multiplicity 1.} & \textit{Conditions:} gas phase.\newline \textit{Target:} Ground-state N2 dissociation curve and bond-dissociation energy. \\
\addlinespace[3pt]
IF3 & triangulene triplet ground state and singlet-triplet gap\newline {\footnotesize Charge 0; multiplicity 3.} & \textit{Conditions:} gas phase (surface-synthesis analogue).\newline \textit{Target:} Ground-state spin multiplicity and singlet--triplet gap of triangulene. \\
\addlinespace[3pt]
IG8 & chromium dimer ground-state dissociation curve\newline {\footnotesize Charge 0; multiplicity 1.} & \textit{Conditions:} gas phase.\newline \textit{Target:} Ground-state Cr2 potential-energy curve and bond-dissociation energy. \\
\addlinespace[3pt]
\end{longtable}
\endgroup

\section{Detailed benchmark results and reaction-path analysis}
\label{sec:si-detailed-results}
\subsection{NI1: two requested steps and their conformer references}
\label{sec:si-ni1-chemical-objects}

The Guided NI1 study delivered methylamine addition to formaldehyde and dehydration of the resulting carbinolamine. The retained run is \texttt{batch\_94084d85d0634ca0}, under \path{campaign_data/cloudra-p2c-20261025/guided/r1/sandboxes/NI1/} in the full P2 archive. Table~\ref{tab:si-ni1-energies} lists values extracted from that run's native Gaussian outputs. All calculations use B3LYP-D3(BJ)/def2-SVP in the gas phase, charge 0, multiplicity 1. Energies are electronic energies without thermal or zero-point corrections.

\begin{table}[!htbp]
\centering\small
\caption{Stationary points in Guided/NI1. Relative electronic energies use the reactant complex as zero; $n_{\mathrm{imag}}$ counts imaginary frequencies. $I'$ and $P'$ are the actual optimized endpoints of the second step's IRC.}
\label{tab:si-ni1-energies}
\begin{tabular}{lrrrl}
\toprule
Object & $E$ / Ha & $\Delta E$ / kcal mol$^{-1}$ & $n_{\mathrm{imag}}$ & Imaginary frequency / cm$^{-1}$ \\
\midrule
Reactant complex $R$ & $-210.218065844$ & 0.00000 & 0 & --- \\
Addition saddle TS1 & $-210.173256653$ & 28.11819 & 1 & $-1481.0579$ \\
Lowest carbinolamine $I$ & $-210.237884930$ & $-12.43666$ & 0 & --- \\
IRC-side carbinolamine $I'$ & $-210.236755311$ & $-11.72782$ & 0 & --- \\
Dehydration saddle TS2 & $-210.145578627$ & 45.48642 & 1 & $-1796.8431$ \\
Lowest imine--water $P$ & $-210.223036629$ & $-3.11921$ & 0 & --- \\
IRC-side imine--water $P'$ & $-210.223036529$ & $-3.11915$ & 0 & --- \\
\bottomrule
\end{tabular}
\end{table}

The addition barrier is 28.11819 kcal/mol. For dehydration, referencing the lowest carbinolamine gives 57.92308 kcal/mol, while referencing the actual optimized IRC-side conformer gives 57.21423 kcal/mol. $I'$ lies 0.70885 kcal/mol above $I$. The lowest product and the corresponding IRC endpoint differ by 0.00006 kcal/mol. The interconversion between $I$ and $I'$ was not calculated.

For addition, the TS1 frequency geometry is the starting geometry of \path{ts/ts_0001/jobs/declaration_0006/} IRC calculations. The two actual terminal frames are the inputs of \path{ts/repair/e1_fix/rc1_side_optfreq.log} and \path{ts/repair/e1_fix/int1_side_optfreq.log}. For dehydration, the TS2 frequency and IRC calculations share \path{ts/ts_0004/jobs/declaration_0003/}; its terminal frames are the inputs of \path{s2_fix/src_optfreq.log} and \path{s2_fix/tgt_optfreq.log}. Both IRC calculations followed 24 points per direction and stopped at the point limit. All four endpoint optimizations terminated normally at the declared species with zero imaginary frequencies.

The original study also explored an extra explicit-water-assisted dehydration branch, whose reactant and product compositions include an additional water molecule. Its barrier is 36.42945 kcal/mol from its lowest reactant complex and 32.54199 kcal/mol from its local IRC-side conformer. The main-text profile displays the two requested steps with the original molecular composition.

The directory \path{companion_data/revision_20261101/ni1_chemical_objects/} supplies the geometries, energy table, figure script, and archive-relative native-output locations and hashes.

\subsection{Uncapped electronic-barrier differences and archived physical records}
\label{sec:si-archived-barrier-errors}

The positive tiers contain 36 tasks with 53 numerical reference edges per workflow. HA1, HA6, HB1, and HF1 score qualitative channel preference; NB1\_E2 and NQ5\_E2 have no numerical reference. The Impossible tier is kept separate, including IB9's one numerical edge and its negative target. We recovered 160 explicit log-derived-versus-Gold comparison notes from the original benchmark databases, comprising 159 positive-target observations and one Guided/IB9 observation. Native TS and reference energies, their echoed methods, and file hashes are retained with the accompanying analysis.

For 156 positive-target observations, barriers recalculated from the selected, normally terminated TS and reference logs agreed with the original comparison notes to the recorded precision of 0.001~kcal~mol$^{-1}$. The TS and reference calculations used consistent methods and continuum settings. Three further observations belong to runs whose original score components differ from the final publication components; excluding these leaves 153 final-aligned observations: Guided 41, Standard 37, Record 40, and Brief 35, from 33, 29, 32, and 30 tasks, respectively. The 59 remaining opportunities out of 212 positive numerical run--edge pairs remain listed as unavailable or unresolved observations. Missing values receive neither zero error nor an imputed numerical penalty. The final F4 replay records are not included in this set, which was assembled from the original comparison notes.

The original comparison notes identify this observed cohort. Final target correspondence then selects the task-specified reference from the same run's existing calculations. This changes Standard/L2 from the local-reference edge \texttt{e1} to the stable-reference edge \texttt{e2}, with a native barrier of 28.27501~kcal~mol$^{-1}$. The statistics use this final selection within the original cohort; the original notes and earlier results are archived.

For an observed pair, we calculate the signed difference as $(E_{\mathrm{TS}}-E_{\mathrm{ref}})\,627.509474-\Delta E^{\ddagger}_{\mathrm{Gold}}$, in kcal~mol$^{-1}$, using native Hartree energies and the frozen task-file reference. Mean absolute error (MAE), root-mean-square error (RMSE), median absolute error, and maximum absolute error have no Gaussian transformation or penalty cap. Each observed reference edge has equal weight.

The intersection shared by all four workflows contains 26 reference edges from 23 positive tasks, giving 104 run--edge observations (Table~\ref{tab:si-uncapped-common}). Their native TS and reference pairs use B3LYP-D3(BJ)/def2-SVP with no explicit continuum keyword in the original input routes. Edge assignments follow the task's specified reaction and energy-reference definitions. Guided/G2 is excluded because its final structured delivery omits the numerical barrier and TS energy/log bindings, although its historical native calculations support an approximately 0.845-kcal~mol$^{-1}$ value.

\begin{table}[!htbp]
\centering
\small
\caption{Uncapped electronic-barrier differences on the 26 reference edges shared by all four workflows, covering 23 positive tasks. All entries are in kcal~mol$^{-1}$; every workflow uses the same edge set.}
\label{tab:si-uncapped-common}
\begin{tabular}{lrrrr}
\toprule
Workflow & MAE & RMSE & Median absolute error & Maximum absolute error \\
\midrule
Guided & 1.1676 & 2.8566 & 0.0573 & 10.7826 \\
Standard & 1.5423 & 4.6374 & 0.0574 & 21.4713 \\
Record & 1.5763 & 5.7082 & 0.0731 & 28.6373 \\
Brief & 1.1310 & 2.3080 & 0.0985 & 9.0486 \\
\bottomrule
\end{tabular}
\end{table}

For a common subset of 19 edges from 17 tasks, the saved Gold input files confirm that the method and continuum setting match those of the run calculations (Table~\ref{tab:si-uncapped-gold-input}). The input-file check covers method and continuum settings; complete reference-log and endpoint-chain coverage remains to be checked. Record has the lowest MAE in this subset. The changed ordering shows that the numerical ranking depends on which archived reference comparisons are available.

\begin{table}[!htbp]
\centering
\small
\caption{Uncapped differences for the 19 common reference edges with additional corroboration from retained selected Gold input files, covering 17 positive tasks. Units and statistics follow Table~\ref{tab:si-uncapped-common}.}
\label{tab:si-uncapped-gold-input}
\begin{tabular}{lrrrr}
\toprule
Workflow & MAE & RMSE & Median absolute error & Maximum absolute error \\
\midrule
Guided & 1.0105 & 2.7118 & 0.0388 & 10.7826 \\
Standard & 0.8015 & 2.2167 & 0.0424 & 9.0486 \\
Record & 0.3934 & 0.8609 & 0.0424 & 3.0918 \\
Brief & 1.2811 & 2.6213 & 0.0744 & 9.0486 \\
\bottomrule
\end{tabular}
\end{table}

We checked which physical-validation records were available for the 160 positive runs. Guided, Record, and Brief each retain delivery-DAG element states for all 40 tasks; Standard's configuration has no delivery-DAG objects. Frozen delivery-disposition reports exist for Guided 28, Record 39, Brief 5, and Standard 0 runs, totalling 72/160. Reports containing native scientific checks with a graph hash matching the archived final root graph exist for Guided 28 and Brief 5 runs, totalling 33/160. These records cover the species and steps delivered during each run. A uniform audit of all requested species and steps remains pending.

The companion directory \path{companion_data/revision_20261101/metrics/} provides every numerical opportunity, original comparison note, recovered native energy and method, source/version relation, exclusion, and cohort membership. Its reproduction script applies the final selections in \path{revision_20261102/target_matching/accuracy_replay.json} to the stored native energies and recomputes the statistics against the frozen Gold values.

\subsection{Complete workflow and tier comparisons}

The dataset contains 200 task runs: one per task under Guided, Standard, Record, and Brief. Each run has a composite and all three component scores. Table~\ref{tab:si-arm-tier} reports overall and tier means. Standard uses the same model backbone as Guided; Record removes required physical validation, and Brief shortens diagnostic feedback.

\begin{table}[!htbp]
\centering
\small
\caption{Composite and component means for each workflow and task tier.}
\label{tab:si-arm-tier}
\begin{tabular}{llrrrrr}
\toprule
Tier & Workflow & $n$ & Composite & Graph & SC & Accuracy \\
\midrule
All & Guided & 50 & 0.8050 & 0.8851 & 0.9233 & 0.7324 \\
 & Standard & 50 & 0.5836 & 0.7774 & 0.5303 & 0.5779 \\
 & Record & 50 & 0.7038 & 0.7813 & 0.7177 & 0.6839 \\
 & Brief & 50 & 0.7781 & 0.8493 & 0.8468 & 0.7320 \\
\addlinespace
Easy & Guided & 20 & 0.9125 & 0.9778 & 0.9250 & 0.8954 \\
 & Standard & 20 & 0.7044 & 0.8849 & 0.5858 & 0.7336 \\
 & Record & 20 & 0.9128 & 0.9419 & 0.9050 & 0.9118 \\
 & Brief & 20 & 0.9370 & 0.9681 & 0.9125 & 0.9440 \\
\addlinespace
Normal & Guided & 10 & 0.7370 & 0.8589 & 0.9021 & 0.6342 \\
 & Standard & 10 & 0.5731 & 0.7607 & 0.5217 & 0.5675 \\
 & Record & 10 & 0.5287 & 0.6786 & 0.5783 & 0.4789 \\
 & Brief & 10 & 0.5857 & 0.7472 & 0.7740 & 0.4646 \\
\addlinespace
Hard & Guided & 10 & 0.9577 & 0.9107 & 0.9542 & 0.9672 \\
 & Standard & 10 & 0.6888 & 0.8481 & 0.5033 & 0.7549 \\
 & Record & 10 & 0.8723 & 0.9243 & 0.6650 & 0.9672 \\
 & Brief & 10 & 0.8282 & 0.9286 & 0.7767 & 0.8372 \\
\addlinespace
Impossible & Guided & 10 & 0.5051 & 0.7006 & 0.9100 & 0.2700 \\
 & Standard & 10 & 0.2473 & 0.5083 & 0.4550 & 0.1000 \\
 & Record & 10 & 0.2925 & 0.4200 & 0.5350 & 0.1500 \\
 & Brief & 10 & 0.6029 & 0.6345 & 0.8583 & 0.4700 \\
\bottomrule
\end{tabular}
\par\smallskip\begin{minipage}{\linewidth}\footnotesize SC denotes the graded self-consistency component. The overall tier weights are 20/50 for Easy and 10/50 for each other tier.\end{minipage}
\end{table}

Table~\ref{tab:si-pairwise} gives all six pairwise comparisons on the same 50 tasks. Guided--Standard has a mean difference of 0.221366, a median difference of 0.114200, and 35 task wins, eight losses, and seven ties. The 95\% task-bootstrap interval is [0.134870, 0.312368]. Within the Normal tier, the means are 0.737020 and 0.573050, with a difference of 0.163970 [0.036520, 0.292230] and seven wins in ten tasks. Within Hard, the means are 0.957650 and 0.688750; the difference is 0.268900 [0.121300, 0.445930], with Guided higher in all ten task pairs. Easy has means of 0.912520 and 0.704405, with 11 wins, five losses, and four ties; its paired median difference is 0.020850. Impossible has means of 0.505060 and 0.247330 and is examined at the level of the requested target below.

\begin{table}[!htbp]
\centering
\small
\caption{All pairwise composite-score comparisons across the 50 tasks.}
\label{tab:si-pairwise}
\begin{tabular}{lrrr}
\toprule
Comparison & Mean difference & 95\% interval & W/L/T \\
\midrule
Guided--Record & +0.1012 & [+0.0396, +0.1695] & 28/7/15 \\
Guided--Standard & +0.2214 & [+0.1349, +0.3124] & 35/8/7 \\
Guided--Brief & +0.0268 & [-0.0486, +0.0996] & 26/11/13 \\
Record--Standard & +0.1202 & [+0.0224, +0.2197] & 28/13/9 \\
Record--Brief & -0.0743 & [-0.1455, -0.0066] & 14/24/12 \\
Standard--Brief & -0.1946 & [-0.2990, -0.0914] & 11/34/5 \\
\bottomrule
\end{tabular}
\par\smallskip\begin{minipage}{\linewidth}\footnotesize The difference is the first workflow minus the second. W/L/T gives the first workflow's task wins, losses, and ties. Percentile task-bootstrap intervals use 100,000 resamples and seed 20261031; the four configurations of each sampled task remain together.\end{minipage}
\end{table}

Table~\ref{tab:si-weighted-gain} decomposes the Guided--Standard difference into weighted score components. Across all tasks, self-consistency contributes +0.117875 and accuracy +0.092717. In Normal and Hard together, the composite difference is +0.216435, including +0.124688 from self-consistency and +0.083712 from accuracy. In Hard, accuracy contributes +0.127391 and self-consistency +0.135250: both the requested answers and the calculations supporting them contribute to the gain.

\begin{table}[!htbp]
\centering
\small
\caption{Weighted decomposition of the Guided--Standard composite difference.}
\label{tab:si-weighted-gain}
\begin{tabular}{lrrrrr}
\toprule
Scope & $n$ & Composite & $0.1\Delta G$ & $0.3\Delta\mathrm{SC}$ & $0.6\Delta A$ \\
\midrule
All & 50 & +0.221366 & +0.010776 & +0.117875 & +0.092717 \\
Easy & 20 & +0.208115 & +0.009287 & +0.101750 & +0.097082 \\
Normal & 10 & +0.163970 & +0.009823 & +0.114125 & +0.040033 \\
Hard & 10 & +0.268900 & +0.006259 & +0.135250 & +0.127391 \\
Impossible & 10 & +0.257730 & +0.019226 & +0.136500 & +0.102000 \\
Normal + Hard & 20 & +0.216435 & +0.008041 & +0.124688 & +0.083712 \\
\bottomrule
\end{tabular}
\par\smallskip\begin{minipage}{\linewidth}\footnotesize Component contributions use unrounded values; composite differences use four-decimal scores.\end{minipage}
\end{table}

Guided--Record averages +0.101156 [0.039606, 0.169478] across all tasks. In Normal and Hard together, its +0.146845 composite difference decomposes into +0.008338 from graph, +0.091938 from self-consistency, and +0.046574 from accuracy. Guided--Brief averages +0.026816 [$-0.048638$, 0.099584], with mean differences of +0.151350 on Normal and +0.129470 on Hard. Individual multistep tasks differ: NB1 has composite scores of 0.9628, 0.9128, and 0.9900 for Guided, Standard, and Brief; Guided and Brief both have self-consistency 1. NI1 has Guided and Standard composites of 0.9958 and 0.6196, with self-consistency 1 and 0.35. NQ5 favors Record over Guided, while NU3 favors Brief. The leading configuration varies among multistep tasks.

\subsection{Graph delivery and evaluation of the run results}

During execution, Guided, Record, and Brief each wrote 50 root mechanism graphs; Standard wrote 43. Standard's remaining seven studies---F1, HA1, HF1, IA9, M17, NR1, and P3---were organized from their own conversations, reports, tool results, and calculation files. The common evaluation procedure thus produced 50 graphs per workflow and 200 complete score profiles, with the source of each recovered field recorded.

The seven Standard studies have a mean organized composite of 0.153743, compared with 0.828529 for Guided on the same tasks. They account for 42.7\% of the total Guided--Standard difference. Table~\ref{tab:si-sc-distribution} separately reports self-consistency for the 40 non-Impossible tasks. Guided has 24 full-score records (60.0\%) and 16 partial-score records; Standard has seven full-score records (17.5\%), 29 partial-score records, and four zero-score records.

\begin{table}[!htbp]
\centering
\small
\caption{Self-consistency distribution over all 40 non-Impossible tasks.}
\label{tab:si-sc-distribution}
\begin{tabular}{lrrrr}
\toprule
Workflow & SC $=1$ & $0<$ SC $<1$ & SC $=0$ & Mean SC \\
\midrule
Guided & 24 & 16 & 0 & 0.9266 \\
Standard & 7 & 29 & 4 & 0.5492 \\
Record & 16 & 24 & 0 & 0.7633 \\
Brief & 20 & 20 & 0 & 0.8439 \\
\bottomrule
\end{tabular}
\end{table}

\subsection{Elapsed research time and accumulated calculation effort}

Table~\ref{tab:si-batch-times} preserves every batch observation from the 48-run repeated-task experiment. A batch contains the same four tasks at a concurrency of two, and its elapsed time spans the batch's actual start and finish. Guided's mean is 4.4514 h, Standard's 3.9577 h, Record's 3.8942 h, and Brief's 4.8273 h. Relative to Standard, Guided's observed mean is 12.5\% longer; relative to Brief, it is 7.8\% shorter. These observations include waiting, calculation execution, and continuation of the research after results return.

\begin{table}[!htbp]
\centering
\small
\caption{All 12 elapsed-time observations in the repeated-task study, in hours.}
\label{tab:si-batch-times}
\begin{tabular}{lrrrr}
\toprule
Workflow & Batch 1 & Batch 2 & Batch 3 & Mean \\
\midrule
Guided & 4.2383 & 3.8669 & 5.2489 & 4.4514 \\
Standard & 5.3381 & 2.5044 & 4.0306 & 3.9577 \\
Record & 5.4189 & 2.6203 & 3.6433 & 3.8942 \\
Brief & 7.2689 & 3.6475 & 3.5656 & 4.8273 \\
\bottomrule
\end{tabular}
\par\smallskip\begin{minipage}{\linewidth}\footnotesize Each observation is a four-task batch (E1, NB1, HA1, IA1) at concurrency two. Three batches per workflow contain 48 task runs in total.\end{minipage}
\end{table}

The P2 effort comparison uses the common 15-task set D3, M1, M8, NB1, NQ5, NR1, NU3, NU5, NU6, HA1, HA6, HB1, HB10, HD2, and HF1. This yields 60 run-level effort records, with all four configurations represented for each task. Table~\ref{tab:si-common-effort} gives the total and mean job-hours, job counts, terminal outcomes, and corresponding final scores. Job-hours sum the durations recorded for individual quantum-chemical jobs. For Standard/M1, the archived effort reconstruction uses native-log CPU time divided by four, as specified in the source record. The measurement unit throughout this comparison is accumulated job-hours.

\begin{table}[!htbp]
\centering
\small
\caption{Recorded quantum-chemical effort on the 15 tasks shared by all configurations.}
\label{tab:si-common-effort}
\begin{tabular}{lrrrrrr}
\toprule
Workflow & Job-h total & Job-h/task & Jobs & Failed & Cancelled & Mean $S$ \\
\midrule
Guided & 166.19 & 11.079 & 789 & 114 & 94 & 0.8124 \\
Standard & 113.32 & 7.555 & 419 & 45 & 50 & 0.6495 \\
Record & 118.16 & 7.877 & 431 & 62 & 38 & 0.7360 \\
Brief & 233.85 & 15.590 & 883 & 167 & 120 & 0.7273 \\
\bottomrule
\end{tabular}
\par\smallskip\begin{minipage}{\linewidth}\footnotesize Each workflow contributes 15 task runs. Job-hours sum recorded process durations. Failed and cancelled jobs are separate terminal categories.\end{minipage}
\end{table}

Guided used a mean of 11.0793 job-hours and obtained a mean composite of 0.812413. Relative to Standard, it used 46.7\% more job time and scored 0.162933 higher; relative to Brief, it used 28.9\% less job time and scored 0.085080 higher. Guided had both a higher score and lower job time than Brief on seven of the 15 paired tasks. Guided recorded 789 jobs, including 114 failures and 94 cancellations; Brief recorded 883 jobs, including 167 failures and 120 cancellations. In the wider 63-run archive, NU3, NU5, NU6, and NR1 account for 163.5 of 673.82 job-hours across their 16 runs. These trajectories record repeated searches for the required intermediates and connecting pathways.

\subsection{Conformer references and reaction-step assignments}

Table~\ref{tab:si-case-profiles} gives all score components for the following cases. We examine which reactant conformer defines each barrier and which reaction step each optimized saddle connects.

\begin{table}[!htbp]
\centering
\small
\caption{Complete score profiles for the four scientific-attribution cases.}
\label{tab:si-case-profiles}
\begin{tabular}{llrrrr}
\toprule
Task & Workflow & Composite & Graph & SC & Accuracy \\
\midrule
HA8 & Guided & 0.9625 & 1.0000 & 0.8750 & 1.0000 \\
 & Standard & 0.6250 & 1.0000 & 0.3500 & 0.7000 \\
HH1 & Guided & 0.9838 & 1.0000 & 1.0000 & 0.9731 \\
 & Standard & 0.8338 & 1.0000 & 0.5000 & 0.9731 \\
HI1 & Guided & 0.9999 & 0.9992 & 1.0000 & 1.0000 \\
 & Standard & 0.7032 & 0.3325 & 0.2333 & 1.0000 \\
NU6 & Guided & 0.6719 & 1.0000 & 0.9375 & 0.4844 \\
 & Standard & 0.6892 & 0.8167 & 0.6250 & 0.7000 \\
\bottomrule
\end{tabular}
\end{table}

\paragraph{HA8: stable reactant reference and reactive conformer.}
Guided reports 29.46 kcal/mol and Standard 20.43 kcal/mol, against a reference of 29.48 kcal/mol. Their 9.03 kcal/mol difference comes from the reactant energy: Guided uses the lowest all-s-trans minimum, whereas Standard uses a reactive cisoid conformer. The IRC identifies the local conformer connected to the saddle; the task's barrier is measured from the specified stable reactant minimum.

\paragraph{HH1: the same barrier with different IRC support.}
Both workflows report a binaphthyl racemization barrier of 23.48 kcal/mol; the reference is 23.01 kcal/mol. Their accuracy components are 0.973067 and 0.973064 and both graph components are 1. Guided has self-consistency 1, whereas Standard has 0.5. Their composites are consequently 0.9838 and 0.8338. Standard's archive contains candidate channels and IRC inputs; Guided completes the required chain from the saddle through the IRC to the endpoint calculations. This difference in path support contributes 0.1500 to the composite despite the matching barriers.

\paragraph{HI1: numerical agreement with different species and path coverage.}
Guided and Standard both have accuracy approximately 1. Their graph components are 0.999176 and 0.332509, and self-consistency components are 1 and 0.233333. The corresponding composites are 0.9999 and 0.7032.

\paragraph{NU6: the optimized saddle and the intended second step.}
The Guided record gives $E_{\mathrm{TS}}=-247.030042392$ Ha for its concerted saddle region and $E_{\mathrm{betaine}}=-247.060060699$ Ha. Using 627.5094740631 kcal mol$^{-1}$ Ha$^{-1}$ gives
\begin{equation}
\left[E_{\mathrm{TS}}-E_{\mathrm{betaine}}\right]\,627.5094740631
=18.836772\ \mathrm{kcal\,mol^{-1}}.
\end{equation}
Standard reports 18.837 kcal/mol as the second-step ring-closure barrier. Guided leaves the betaine-to-product ring-closure saddle unresolved and describes the energy difference relative to the concerted saddle region. Its searches from interpolated betaine geometries and constrained-scan peaks repeatedly return toward the direct reactant-to-product saddle region.

\subsection{Impossible-tier results and the requested negative conclusions}

Table~\ref{tab:si-imp-pairs} resolves Brief's mean composite of 0.602940 and Guided's 0.505060 into all ten task differences. Brief is higher on four tasks, Guided on five, and one tie. The Brief-minus-Guided mean is +0.097880, with the task-bootstrap interval [$-0.160$, $+0.361$]. Its weighted contributions are $-0.006607$ from graph, $-0.015500$ from self-consistency, and +0.120000 from accuracy. IA1, IA9, and IC3 provide the large positive differences, while IG8 has the opposite accuracy ordering. Across the six tasks with equal accuracy components in these two workflows, Brief-minus-Guided averages $-0.058333$.

\begin{table}[!htbp]
\centering
\small
\caption{Task-resolved comparison of Brief and Guided in the Impossible tier.}
\label{tab:si-imp-pairs}
\begin{tabular}{lrrrrr}
\toprule
Task & Guided C & Brief C & Brief $-$ Guided & Guided A & Brief A \\
\midrule
IA1 & 0.1717 & 1.0000 & +0.8283 & 0.0 & 1.0 \\
IA5 & 1.0000 & 0.9667 & -0.0333 & 1.0 & 1.0 \\
IA7 & 0.2583 & 0.2583 & +0.0000 & 0.0 & 0.0 \\
IA9 & 0.3667 & 0.9333 & +0.5666 & 0.0 & 1.0 \\
IB9 & 0.8200 & 0.7700 & -0.0500 & 0.7 & 0.7 \\
IC3 & 0.4000 & 1.0000 & +0.6000 & 0.0 & 1.0 \\
ID2 & 0.3339 & 0.1839 & -0.1500 & 0.0 & 0.0 \\
IE2 & 0.3333 & 0.1833 & -0.1500 & 0.0 & 0.0 \\
IF3 & 0.3667 & 0.4000 & +0.0333 & 0.0 & 0.0 \\
IG8 & 1.0000 & 0.3339 & -0.6661 & 1.0 & 0.0 \\
\bottomrule
\end{tabular}
\par\smallskip\begin{minipage}{\linewidth}\footnotesize C denotes the composite and A the accuracy component for the task-defined numerical or negative target. All values are retained from the publication score table.\end{minipage}
\end{table}

Brief reaches accuracy 1 for IA1 (square-geometry cyclobutadiene automerization), IA5 (thermal disrotatory ring opening), IA9 (dioxetane thermolysis), and IC3 (open-shell singlet phenylnitrene). Guided reaches accuracy 1 for IA5 and IG8 (chromium dimer); Record reaches accuracy 1 for IA5. Both Guided and Brief have accuracy 0.7 in IB9, which contains positive and negative target components. Across all 40 Impossible runs, eight have accuracy 1 and ten have self-consistency 1 with accuracy 0. Offline target correspondence assigns Accuracy before publication scoring; Graph and Self-consistency remain those of the organized records. The following cases illustrate remaining local-support versus requested-target disagreements.

Guided/IC3 retains Graph $=1$, Self-consistency $=1$, and Accuracy $=0$: the run distinguishes the open-shell target from a closed-shell branch but does not receive negative-target Accuracy credit under the final correspondence rules applied to that graph. Brief/IC3 receives Accuracy $=1$ for the same open-shell target, with composite 1.0.

Record/IA5, Guided/IA5, and Brief/IA5 distinguish the requested disrotatory ring opening from an allowed conrotatory path and receive full negative-target Accuracy credit after correspondence. IF3 remains Accuracy 0 with Self-consistency 1 in Guided and Brief.

\subsection{Easy-tier energy references and delivery trajectories}
\label{sec:si-easy-case-audit}

These five Easy cases examine the reactant energy used for a barrier, the chemical identity assigned to a graph node, and whether calculated values remain in the final graph. We read each run's final graph, native calculation logs, evaluation notes, and runtime events. Table~\ref{tab:si-easy-case-audit} reports the final scores. $\mathrm{SC}_{\mathrm{all}}$ grades every delivered graph element, including additions to Gold; $\mathrm{SC}_{\mathrm{Gold}}$ grades only the reference-matched subgraph. The composite uses $\mathrm{SC}_{\mathrm{all}}$.

\begin{table}[!htbp]
\centering
\small
\setlength{\tabcolsep}{4pt}
\caption{Final component scores for the five inspected Easy tasks, rounded to four decimal places. Brief denotes Brief-errors. Accuracy is the final component after reference matching and the specified numerical penalty cap.}
\label{tab:si-easy-case-audit}
\begin{tabular}{llrrrrr}
\toprule
Task & Workflow & Graph & $\mathrm{SC}_{\mathrm{all}}$ & $\mathrm{SC}_{\mathrm{Gold}}$ & Accuracy & Composite \\
\midrule
F4 & Guided & 0.9167 & 1.0000 & 1.0000 & 1.0000 & 0.9917 \\
 & Standard & 0.5176 & 0.7500 & 0.0000 & 0.0000 & 0.2768 \\
 & Record & 0.6009 & 0.7500 & 0.0000 & 0.0000 & 0.2851 \\
 & Brief & 0.9167 & 1.0000 & 1.0000 & 1.0000 & 0.9917 \\
\addlinespace
G2 & Guided & 0.8889 & 0.5000 & 0.5000 & 0.0000 & 0.2389 \\
 & Standard & 1.0000 & 0.5000 & 0.5000 & 0.8782 & 0.7769 \\
 & Record & 1.0000 & 0.5000 & 0.5000 & 0.8782 & 0.7769 \\
 & Brief & 1.0000 & 0.5000 & 0.5000 & 1.0000 & 0.8500 \\
\addlinespace
D1 & Guided & 1.0000 & 1.0000 & 1.0000 & 0.7000 & 0.8200 \\
 & Standard & 0.7917 & 0.5000 & 0.5000 & 1.0000 & 0.8291 \\
 & Record & 0.7917 & 1.0000 & 1.0000 & 0.7000 & 0.7992 \\
 & Brief & 1.0000 & 0.7500 & 0.7500 & 0.7000 & 0.7450 \\
\addlinespace
D3 & Guided & 1.0000 & 0.7500 & 0.7500 & 0.7000 & 0.7450 \\
 & Standard & 1.0000 & 0.7500 & 0.7500 & 0.9995 & 0.9247 \\
 & Record & 0.9444 & 0.3500 & 0.3500 & 0.9909 & 0.7940 \\
 & Brief & 0.8333 & 0.7500 & 0.7500 & 0.7000 & 0.7283 \\
\addlinespace
L2 & Guided & 1.0000 & 0.7500 & 0.7500 & 0.7000 & 0.7450 \\
 & Standard & 1.0000 & 0.5000 & 0.5000 & 0.8832 & 0.7799 \\
 & Record & 1.0000 & 0.7500 & 0.7500 & 0.8832 & 0.8549 \\
 & Brief & 1.0000 & 0.8333 & 0.7500 & 0.8832 & 0.8799 \\
\bottomrule
\end{tabular}
\end{table}

\paragraph{F4: adduct identity in the final mechanism graph.}
Guided reported hydride transfer from the Lewis adduct of borane and formaldehyde to methoxyborane. Its linked reactant and saddle calculations gave energies of $-141.041124115$ and $-141.015601446$~hartree, a barrier of 16.01572~kcal~mol$^{-1}$ supporting the declared 16.02. Independent saddle frequency analysis found one imaginary frequency, $-398.7218$~cm$^{-1}$. Optimizations starting from both actual IRC outer frames converged to the corresponding endpoint minima with zero imaginary frequencies. All four workflows obtained essentially the same saddle and adduct energies.

The F4 reference specifies hydride transfer from the bonded Lewis adduct, \texttt{[BH3-][O+]=C}, with a 16.02-kcal~mol$^{-1}$ barrier. Guided and Brief named this adduct in their final graphs and obtained Accuracy close to 1. Standard and Record kept the separated-fragment representation \texttt{[BH3].C=O} for the reactant node. This failed to match the reference adduct and received Accuracy~0, although the native calculations supported essentially the same pathway.

\paragraph{G2: computed evidence and final structured delivery.}
Guided's native logs contain a concerted double-H-transfer saddle at $-189.093172517$~hartree with one imaginary frequency, $-444.9504$~cm$^{-1}$. The optimized cis reactant complex and the IRC-derived reactant minimum have energies of $-189.094518484$ and $-189.094518735$~hartree. They give barriers of 0.84461 and 0.84476~kcal~mol$^{-1}$, respectively, close to the 0.87 reference.

The program rejected a numerical barrier on a companion complex-product edge four times because its scientific checks remained open. Meanwhile, reports for the selected separated-products target returned \texttt{complete=true} with all targets marked resolved. After an initial deletion, the Supervisor asked the agent to restore all numerical barriers and the saddle's energy/log fields. After the fourth rejection, the agent deleted the companion-edge values. A subsequent artifact check rejected the saddle-evidence claim; the next edit removed the selected edge's remaining barrier and the saddle's energy/log fields. The final graph kept the values in descriptive text, but the numerical target received Accuracy~0. Its graded SC contribution remained 0.5.

Brief retained a numerical barrier of 0.844 and obtained Accuracy~0.99999. Standard and Record declared approximately 3.4 using a lower-energy reactant reference; Record's native report identifies this as a trans-diimide stack and separately records the approximately 0.84 cis-reference result.

\paragraph{D1 and D3: numerical penalties with different physical origins.}
The evaluation notes explicitly attribute Accuracy~0.7 to the maximum numerical-deviation penalty of 0.3. For D1, 15.738 versus 18.830~kcal~mol$^{-1}$ gives an underlying Gaussian score of 0.5880 before this cap. Guided and Standard obtained nearly identical saddle energies, but their reactant energies differ by 3.11859~kcal~mol$^{-1}$. Their butadiene dihedrals are approximately $26.33^\circ$ and $-179.93^\circ$, corresponding to reactive s-cis and lower s-trans references. Record also delivered the s-cis barrier and retained SC~1 with Accuracy~0.7.

In D3, Guided and Standard reported barriers of 18.14687 and 14.55472~kcal~mol$^{-1}$. Their reactant energies differ by only 0.09899~kcal~mol$^{-1}$, whereas their saddle energies differ by 3.69113~kcal~mol$^{-1}$. Both saddle logs contain exactly one imaginary frequency. Both evaluations assign SC~0.75 and warn about the product-endpoint association; the numerical barrier difference determines the capped Accuracy score.

\paragraph{L2: two reference barriers and the selected comparison edge.}
Guided used the reactive cZc minimum and obtained 18.98740~kcal~mol$^{-1}$; its graph states that the stable-to-reactive conformer contribution was uncomputed. Record and Brief used the lower tZt reference, giving approximately 28.275. The conformer energy difference is 9.28761~kcal~mol$^{-1}$.

Standard delivered both references and both barriers: 18.99 on edge \texttt{e1} and 28.28 on \texttt{e2}. The earlier evaluation selected \texttt{e1}: the candidates have identical bond-change lists, the unspecified-stereochemistry reference gives equal endpoint substitution costs, and the edge-identifier tie-break selects the local-reference value, with capped Accuracy~0.7. The final target-correspondence evaluation instead selects \texttt{e2}, whose verified stable-minimum reference matches the task definition. Its native barrier is 28.27501~kcal~mol$^{-1}$, giving Accuracy~0.883233 and composite~0.7799, as reported in Table~\ref{tab:si-easy-case-audit} and the complete score table.

\subsection{Official target-correspondence Accuracy overlay}
\label{sec:matcher-sensitivity}

The published evaluation applies chemical target correspondence before Accuracy scoring. Explicit reaction channels and electronic states distinguish otherwise identical connectivity, and the task-specified energy reference selects among references of the same native saddle. Graph and Self-consistency remain those recorded for the organized runs. Twelve runs whose task definitions involve these distinctions were replayed against the updated rules; five Accuracy values change. The overlay updates those Accuracy components, recomputes the corresponding composites, and leaves the other 195 runs unchanged. The companion package \path{revision_20261102/target_matching/} retains the before/after Accuracy reports, matching audit, and correction table applied by the publication preparation script.

The five corrected runs are Standard/L2 (lowest verified stable-minimum barrier reference), Guided/IA5, Record/IA5, Brief/IA5 (thermal disrotatory cyclobutene ring opening), and Brief/IC3 (open-shell singlet phenylnitrene). After correspondence, Guided, Record, and Brief receive full negative-target Accuracy credit on IA5, and Brief receives full credit on IC3. Table~\ref{tab:matcher-sensitivity} reports the resulting Impossible-tier means.

\begin{table}[htbp]
\centering
\small
\caption{Impossible-tier mean composite scores after the official target-correspondence Accuracy overlay.}
\label{tab:matcher-sensitivity}
\begin{tabular}{@{}lrrrr@{}}
\toprule
Track & Guided & Standard & Record & Brief \\
\midrule
Official evaluation values & 0.50506 & 0.24733 & 0.29250 & 0.60294 \\
\bottomrule
\end{tabular}
\end{table}

Relative to the pre-overlay Impossible means (Guided 0.44506, Standard 0.24733, Record 0.23250, Brief 0.48294), the overlay raises Guided by 0.06000, Record by 0.06000, and Brief by 0.12000, while Standard is unchanged in this tier. Primary Guided--Standard comparisons elsewhere in the manuscript use the same overlay, including the Easy-tier Standard/L2 correction outside this Impossible summary.

\subsection{Discussion of the supplementary comparisons}

Cloudra's Delivery DAG constrains the agents' submitted mechanism representation through species-occurrence nodes and elementary-step edges that carry molecular geometries, bonding patterns, and reaction connectivity. Cheminformatics algorithms establish and compare identities and atom correspondences; frequency and IRC/endpoint calculations determine the supported stationary points and connections. The Normal and Hard cases show why these assignments matter: HA8 requires correcting the reactant energy reference, and NU6 requires recognizing that a search has returned to a different saddle region. Agents use the represented species and steps to choose new geometries, continue a path calculation, or revise the mechanism, complementing external planning and tool feedback.\citep{valmeekam2023planning,ref18,huang2024selfcorrect}

P2 retains one run per task and configuration. Its task-cluster bootstrap intervals measure variation across the sampled tasks. P1b supplies three repeats on each of four tasks under each configuration; these 48 runs provide repeat measurements on those four tasks within the 50-task P2 study. The P2 table includes all 200 runs and their three components. Some Hard tasks score qualitative channel identity, so the gain includes pathway identification and completion of the supporting calculations as well as numerical barriers. Repeating the full task set would measure run-to-run variation and how often the agents select each route.

The benchmark reports electronic energy differences. NI5 specifies weakly acidic aqueous solution, and NU5 specifies dichloromethane at high concentration; both use the prescribed electronic-energy references. A solution-phase free-energy treatment would add zero-point, thermal/entropic, standard-state, and solvation contributions.

The Easy cases show two different problems. In F4, the final graphs assign different reactant identities despite calculations supporting essentially the same pathway. In G2, validation feedback prompts deletion of structured numerical fields while the values remain in prose. Recording validation status separately would let the agent state a quantity's scientific qualification while retaining the quantity and its source.

\subsection{Complete per-task and per-configuration scores}
\label{sec:full-scores}

Tables~\ref{tab:full-scores-easy}--\ref{tab:full-scores-impossible} report all 200 runs and all four reported quantities, comprising 200 composite scores and 600 component values. Standard is the primary comparison using the same model backbone; Record and Brief are workflow ablations. Every task has one result for each configuration.

Composite scores are displayed to four decimal places and component scores to six decimal places. The accompanying \texttt{cloudrabench\_200\_run\_scores.csv} retains the original full-precision numeric strings. Weighted composites use the full-precision components; displayed values are rounded separately.

\begingroup
\small
\setlength{\tabcolsep}{6pt}
\renewcommand{\arraystretch}{1.05}
\begin{longtable}{@{}llrrrr@{}}
\caption{Complete Easy-tier results for all four configurations. $S$: composite score; $G$: graph structure; $\mathrm{SC}$: scientific self-consistency score; $A$: accuracy.}\label{tab:full-scores-easy}\\
\toprule
Task & Configuration & $S$ & $G$ & $\mathrm{SC}$ & $A$ \\
\midrule
\endfirsthead
\multicolumn{6}{l}{\tablename\ \thetable\ (continued)}\\
\toprule
Task & Configuration & $S$ & $G$ & $\mathrm{SC}$ & $A$ \\
\midrule
\endhead
\midrule
\multicolumn{6}{r}{Continued on the next page}\\
\endfoot
\bottomrule
\endlastfoot
D1 & Guided & 0.8200 & 1.000000 & 1.000000 & 0.700000 \\
D1 & Standard & 0.8291 & 0.791667 & 0.500000 & 0.999960 \\
D1 & Record & 0.7992 & 0.791667 & 1.000000 & 0.700000 \\
D1 & Brief & 0.7450 & 1.000000 & 0.750000 & 0.700000 \\
\addlinespace[3pt]
D3 & Guided & 0.7450 & 1.000000 & 0.750000 & 0.700000 \\
D3 & Standard & 0.9247 & 1.000000 & 0.750000 & 0.999496 \\
D3 & Record & 0.7940 & 0.944444 & 0.350000 & 0.990939 \\
D3 & Brief & 0.7283 & 0.833333 & 0.750000 & 0.700000 \\
\addlinespace[3pt]
D4 & Guided & 0.9250 & 1.000000 & 0.750000 & 0.999982 \\
D4 & Standard & 0.8500 & 1.000000 & 0.500000 & 0.999982 \\
D4 & Record & 0.9104 & 1.000000 & 0.750000 & 0.975699 \\
D4 & Brief & 0.9049 & 0.944444 & 0.750000 & 0.975698 \\
\addlinespace[3pt]
E1 & Guided & 1.0000 & 1.000000 & 1.000000 & 0.999972 \\
E1 & Standard & 0.7828 & 0.777778 & 0.350000 & 0.999973 \\
E1 & Record & 1.0000 & 1.000000 & 1.000000 & 0.999973 \\
E1 & Brief & 1.0000 & 1.000000 & 1.000000 & 0.999973 \\
\addlinespace[3pt]
F1 & Guided & 0.9417 & 0.916667 & 0.833333 & 1.000000 \\
F1 & Standard & 0.0333 & 0.333333 & 0.000000 & 0.000000 \\
F1 & Record & 0.9917 & 0.916667 & 1.000000 & 1.000000 \\
F1 & Brief & 0.9117 & 0.916667 & 0.750000 & 0.991654 \\
\addlinespace[3pt]
F4 & Guided & 0.9917 & 0.916667 & 1.000000 & 0.999999 \\
F4 & Standard & 0.2768 & 0.517569 & 0.750000 & 0.000000 \\
F4 & Record & 0.2851 & 0.600902 & 0.750000 & 0.000000 \\
F4 & Brief & 0.9917 & 0.916667 & 1.000000 & 0.999999 \\
\addlinespace[3pt]
G2 & Guided & 0.2389 & 0.888889 & 0.500000 & 0.000000 \\
G2 & Standard & 0.7769 & 1.000000 & 0.500000 & 0.878159 \\
G2 & Record & 0.7769 & 1.000000 & 0.500000 & 0.878161 \\
G2 & Brief & 0.8500 & 1.000000 & 0.500000 & 0.999986 \\
\addlinespace[3pt]
G3 & Guided & 0.9832 & 0.833333 & 1.000000 & 0.999830 \\
G3 & Standard & 0.1417 & 0.666667 & 0.250000 & 0.000000 \\
G3 & Record & 0.9832 & 0.833333 & 1.000000 & 0.999830 \\
G3 & Brief & 0.9749 & 0.750000 & 1.000000 & 0.999833 \\
\addlinespace[3pt]
L1 & Guided & 0.9999 & 1.000000 & 1.000000 & 0.999899 \\
L1 & Standard & 0.9749 & 1.000000 & 0.916667 & 0.999900 \\
L1 & Record & 0.9999 & 1.000000 & 1.000000 & 0.999900 \\
L1 & Brief & 0.9999 & 1.000000 & 1.000000 & 0.999899 \\
\addlinespace[3pt]
L2 & Guided & 0.7450 & 1.000000 & 0.750000 & 0.700000 \\
L2 & Standard & 0.7799 & 1.000000 & 0.500000 & 0.883233 \\
L2 & Record & 0.8549 & 1.000000 & 0.750000 & 0.883231 \\
L2 & Brief & 0.8799 & 1.000000 & 0.833333 & 0.883239 \\
\addlinespace[3pt]
M1 & Guided & 0.9478 & 1.000000 & 1.000000 & 0.913062 \\
M1 & Standard & 0.7526 & 1.000000 & 0.350000 & 0.912639 \\
M1 & Record & 0.9476 & 1.000000 & 1.000000 & 0.912639 \\
M1 & Brief & 0.9476 & 1.000000 & 1.000000 & 0.912638 \\
\addlinespace[3pt]
M14 & Guided & 1.0000 & 1.000000 & 1.000000 & 1.000000 \\
M14 & Standard & 0.8050 & 1.000000 & 0.350000 & 1.000000 \\
M14 & Record & 1.0000 & 1.000000 & 1.000000 & 1.000000 \\
M14 & Brief & 1.0000 & 1.000000 & 1.000000 & 1.000000 \\
\addlinespace[3pt]
M15 & Guided & 1.0000 & 1.000000 & 1.000000 & 0.999999 \\
M15 & Standard & 1.0000 & 1.000000 & 1.000000 & 0.999999 \\
M15 & Record & 1.0000 & 1.000000 & 1.000000 & 0.999983 \\
M15 & Brief & 1.0000 & 1.000000 & 1.000000 & 0.999999 \\
\addlinespace[3pt]
M16 & Guided & 0.9748 & 1.000000 & 0.916667 & 0.999691 \\
M16 & Standard & 0.9998 & 1.000000 & 1.000000 & 0.999692 \\
M16 & Record & 0.9915 & 0.916667 & 1.000000 & 0.999694 \\
M16 & Brief & 0.9998 & 1.000000 & 1.000000 & 0.999692 \\
\addlinespace[3pt]
M17 & Guided & 0.9999 & 1.000000 & 1.000000 & 0.999914 \\
M17 & Standard & 0.0778 & 0.777778 & 0.000000 & 0.000000 \\
M17 & Record & 0.9999 & 1.000000 & 1.000000 & 0.999914 \\
M17 & Brief & 0.9999 & 1.000000 & 1.000000 & 0.999914 \\
\addlinespace[3pt]
M3 & Guided & 1.0000 & 1.000000 & 1.000000 & 0.999996 \\
M3 & Standard & 1.0000 & 1.000000 & 1.000000 & 0.999996 \\
M3 & Record & 1.0000 & 1.000000 & 1.000000 & 0.999996 \\
M3 & Brief & 1.0000 & 1.000000 & 1.000000 & 0.999996 \\
\addlinespace[3pt]
M8 & Guided & 0.9996 & 1.000000 & 1.000000 & 0.999335 \\
M8 & Standard & 0.9996 & 1.000000 & 1.000000 & 0.999339 \\
M8 & Record & 0.9996 & 1.000000 & 1.000000 & 0.999294 \\
M8 & Brief & 0.9996 & 1.000000 & 1.000000 & 0.999294 \\
\addlinespace[3pt]
P2 & Guided & 1.0000 & 1.000000 & 1.000000 & 0.999948 \\
P2 & Standard & 0.9833 & 0.833333 & 1.000000 & 0.999947 \\
P2 & Record & 0.9833 & 0.833333 & 1.000000 & 0.999947 \\
P2 & Brief & 0.9750 & 1.000000 & 0.916667 & 0.999947 \\
\addlinespace[3pt]
P3 & Guided & 0.9380 & 1.000000 & 1.000000 & 0.896739 \\
P3 & Standard & 0.1000 & 1.000000 & 0.000000 & 0.000000 \\
P3 & Record & 0.9380 & 1.000000 & 1.000000 & 0.896735 \\
P3 & Brief & 0.8309 & 1.000000 & 1.000000 & 0.718109 \\
\addlinespace[3pt]
X1 & Guided & 0.9999 & 1.000000 & 1.000000 & 0.999916 \\
X1 & Standard & 0.9999 & 1.000000 & 1.000000 & 0.999914 \\
X1 & Record & 0.9999 & 1.000000 & 1.000000 & 0.999916 \\
X1 & Brief & 0.9999 & 1.000000 & 1.000000 & 0.999914 \\
\addlinespace[3pt]
\end{longtable}
\endgroup

\begingroup
\small
\setlength{\tabcolsep}{6pt}
\renewcommand{\arraystretch}{1.05}
\begin{longtable}{@{}llrrrr@{}}
\caption{Complete Normal-tier results for all four configurations. $S$: composite score; $G$: graph structure; $\mathrm{SC}$: scientific self-consistency score; $A$: accuracy.}\label{tab:full-scores-normal}\\
\toprule
Task & Configuration & $S$ & $G$ & $\mathrm{SC}$ & $A$ \\
\midrule
\endfirsthead
\multicolumn{6}{l}{\tablename\ \thetable\ (continued)}\\
\toprule
Task & Configuration & $S$ & $G$ & $\mathrm{SC}$ & $A$ \\
\midrule
\endhead
\midrule
\multicolumn{6}{r}{Continued on the next page}\\
\endfoot
\bottomrule
\endlastfoot
NB1 & Guided & 0.9628 & 0.628571 & 1.000000 & 0.999972 \\
NB1 & Standard & 0.9128 & 0.628571 & 0.833333 & 0.999970 \\
NB1 & Record & 0.8625 & 0.600000 & 0.675000 & 0.999972 \\
NB1 & Brief & 0.9900 & 0.900000 & 1.000000 & 0.999972 \\
\addlinespace[3pt]
NI1 & Guided & 0.9958 & 1.000000 & 1.000000 & 0.993046 \\
NI1 & Standard & 0.6196 & 0.946119 & 0.350000 & 0.700000 \\
NI1 & Record & 0.7767 & 0.816667 & 0.916667 & 0.700000 \\
NI1 & Brief & 0.9889 & 1.000000 & 1.000000 & 0.981480 \\
\addlinespace[3pt]
NI5 & Guided & 0.7715 & 0.928571 & 1.000000 & 0.631094 \\
NI5 & Standard & 0.3136 & 0.571429 & 0.675000 & 0.089878 \\
NI5 & Record & 0.1542 & 0.442375 & 0.366667 & 0.000000 \\
NI5 & Brief & 0.3571 & 0.571429 & 1.000000 & 0.000000 \\
\addlinespace[3pt]
NQ1 & Guided & 0.7395 & 0.945455 & 0.750000 & 0.700000 \\
NQ1 & Standard & 0.9040 & 0.797980 & 0.750000 & 0.998691 \\
NQ1 & Record & 0.7131 & 0.680808 & 0.750000 & 0.700000 \\
NQ1 & Brief & 0.1731 & 0.680808 & 0.350000 & 0.000000 \\
\addlinespace[3pt]
NQ5 & Guided & 0.7075 & 1.000000 & 0.625000 & 0.700000 \\
NQ5 & Standard & 0.7510 & 0.920000 & 0.550000 & 0.823369 \\
NQ5 & Record & 0.8725 & 1.000000 & 0.625000 & 0.974999 \\
NQ5 & Brief & 0.7810 & 1.000000 & 0.625000 & 0.822502 \\
\addlinespace[3pt]
NR1 & Guided & 0.6167 & 0.666667 & 0.833333 & 0.500000 \\
NR1 & Standard & 0.1643 & 0.592593 & 0.350000 & 0.000002 \\
NR1 & Record & 0.5378 & 0.777778 & 0.533333 & 0.500000 \\
NR1 & Brief & 0.6142 & 0.666667 & 0.875000 & 0.475000 \\
\addlinespace[3pt]
NU3 & Guided & 0.5527 & 0.776832 & 0.916667 & 0.333330 \\
NU3 & Standard & 0.3621 & 0.571429 & 0.350000 & 0.333330 \\
NU3 & Record & 0.4071 & 0.571429 & 0.500000 & 0.333330 \\
NU3 & Brief & 0.6829 & 0.828571 & 0.666667 & 0.666657 \\
\addlinespace[3pt]
NU5 & Guided & 0.3518 & 0.642857 & 0.958333 & 0.000000 \\
NU5 & Standard & 0.2654 & 0.904762 & 0.116667 & 0.233136 \\
NU5 & Record & 0.2966 & 0.779762 & 0.233333 & 0.247741 \\
NU5 & Brief & 0.1983 & 0.662546 & 0.440000 & 0.000000 \\
\addlinespace[3pt]
NU6 & Guided & 0.6719 & 1.000000 & 0.937500 & 0.484393 \\
NU6 & Standard & 0.6892 & 0.816667 & 0.625000 & 0.700000 \\
NU6 & Record & 0.1450 & 0.400000 & 0.350000 & 0.000000 \\
NU6 & Brief & 0.3400 & 0.400000 & 1.000000 & 0.000000 \\
\addlinespace[3pt]
NY3 & Guided & 1.0000 & 1.000000 & 1.000000 & 0.999998 \\
NY3 & Standard & 0.7485 & 0.857143 & 0.616667 & 0.796239 \\
NY3 & Record & 0.5217 & 0.716855 & 0.833333 & 0.333333 \\
NY3 & Brief & 0.7312 & 0.761905 & 0.783333 & 0.700000 \\
\addlinespace[3pt]
\end{longtable}
\endgroup

\begingroup
\small
\setlength{\tabcolsep}{6pt}
\renewcommand{\arraystretch}{1.05}
\begin{longtable}{@{}llrrrr@{}}
\caption{Complete Hard-tier results for all four configurations. $S$: composite score; $G$: graph structure; $\mathrm{SC}$: scientific self-consistency score; $A$: accuracy.}\label{tab:full-scores-hard}\\
\toprule
Task & Configuration & $S$ & $G$ & $\mathrm{SC}$ & $A$ \\
\midrule
\endfirsthead
\multicolumn{6}{l}{\tablename\ \thetable\ (continued)}\\
\toprule
Task & Configuration & $S$ & $G$ & $\mathrm{SC}$ & $A$ \\
\midrule
\endhead
\midrule
\multicolumn{6}{r}{Continued on the next page}\\
\endfoot
\bottomrule
\endlastfoot
HA1 & Guided & 0.9625 & 1.000000 & 0.875000 & 1.000000 \\
HA1 & Standard & 0.1000 & 1.000000 & 0.000000 & 0.000000 \\
HA1 & Record & 0.9250 & 1.000000 & 0.750000 & 1.000000 \\
HA1 & Brief & 0.9250 & 1.000000 & 0.750000 & 1.000000 \\
\addlinespace[3pt]
HA6 & Guided & 0.9750 & 1.000000 & 0.916667 & 1.000000 \\
HA6 & Standard & 0.9625 & 1.000000 & 0.875000 & 1.000000 \\
HA6 & Record & 0.9750 & 1.000000 & 0.916667 & 1.000000 \\
HA6 & Brief & 0.9750 & 1.000000 & 0.916667 & 1.000000 \\
\addlinespace[3pt]
HA8 & Guided & 0.9625 & 1.000000 & 0.875000 & 0.999961 \\
HA8 & Standard & 0.6250 & 1.000000 & 0.350000 & 0.700000 \\
HA8 & Record & 0.9250 & 1.000000 & 0.750000 & 0.999961 \\
HA8 & Brief & 0.7450 & 1.000000 & 0.750000 & 0.700000 \\
\addlinespace[3pt]
HB1 & Guided & 0.9325 & 0.700000 & 0.875000 & 1.000000 \\
HB1 & Standard & 0.8250 & 0.750000 & 0.500000 & 1.000000 \\
HB1 & Record & 0.9700 & 0.700000 & 1.000000 & 1.000000 \\
HB1 & Brief & 0.9800 & 0.800000 & 1.000000 & 1.000000 \\
\addlinespace[3pt]
HB10 & Guided & 0.9995 & 1.000000 & 1.000000 & 0.999225 \\
HB10 & Standard & 0.9928 & 0.932344 & 1.000000 & 0.999217 \\
HB10 & Record & 0.8044 & 0.999011 & 0.350000 & 0.999226 \\
HB10 & Brief & 0.7845 & 0.800000 & 0.350000 & 0.999239 \\
\addlinespace[3pt]
HD2 & Guided & 0.7867 & 0.666667 & 1.000000 & 0.700000 \\
HD2 & Standard & 0.6977 & 0.666667 & 0.350000 & 0.876795 \\
HD2 & Record & 0.5950 & 1.000000 & 0.250000 & 0.700000 \\
HD2 & Brief & 0.6699 & 0.999176 & 0.500000 & 0.700000 \\
\addlinespace[3pt]
HF1 & Guided & 0.9742 & 0.741868 & 1.000000 & 1.000000 \\
HF1 & Standard & 0.3425 & 0.799505 & 0.875000 & 0.000000 \\
HF1 & Record & 0.9072 & 0.821728 & 0.750000 & 1.000000 \\
HF1 & Brief & 0.2937 & 0.687405 & 0.750000 & 0.000000 \\
\addlinespace[3pt]
HH1 & Guided & 0.9838 & 1.000000 & 1.000000 & 0.973067 \\
HH1 & Standard & 0.8338 & 1.000000 & 0.500000 & 0.973064 \\
HH1 & Record & 0.7888 & 1.000000 & 0.350000 & 0.973066 \\
HH1 & Brief & 0.9838 & 1.000000 & 1.000000 & 0.973066 \\
\addlinespace[3pt]
HI1 & Guided & 0.9999 & 0.999176 & 1.000000 & 0.999998 \\
HI1 & Standard & 0.7032 & 0.332509 & 0.233333 & 0.999998 \\
HI1 & Record & 0.9350 & 0.999588 & 0.783333 & 0.999998 \\
HI1 & Brief & 0.9999 & 0.999176 & 1.000000 & 0.999998 \\
\addlinespace[3pt]
HL1 & Guided & 0.9999 & 0.999176 & 1.000000 & 0.999999 \\
HL1 & Standard & 0.8050 & 1.000000 & 0.350000 & 0.999999 \\
HL1 & Record & 0.8972 & 0.722222 & 0.750000 & 0.999995 \\
HL1 & Brief & 0.9250 & 1.000000 & 0.750000 & 0.999999 \\
\addlinespace[3pt]
\end{longtable}
\endgroup

\begingroup
\small
\setlength{\tabcolsep}{6pt}
\renewcommand{\arraystretch}{1.05}
\begin{longtable}{@{}llrrrr@{}}
\caption{Complete Impossible-tier results for all four configurations. $S$: composite score; $G$: graph structure; $\mathrm{SC}$: scientific self-consistency score; $A$: accuracy.}\label{tab:full-scores-impossible}\\
\toprule
Task & Configuration & $S$ & $G$ & $\mathrm{SC}$ & $A$ \\
\midrule
\endfirsthead
\multicolumn{6}{l}{\tablename\ \thetable\ (continued)}\\
\toprule
Task & Configuration & $S$ & $G$ & $\mathrm{SC}$ & $A$ \\
\midrule
\endhead
\midrule
\multicolumn{6}{r}{Continued on the next page}\\
\endfoot
\bottomrule
\endlastfoot
IA1 & Guided & 0.1717 & 0.666667 & 0.350000 & 0.000000 \\
IA1 & Standard & 0.1717 & 0.666667 & 0.350000 & 0.000000 \\
IA1 & Record & 0.1717 & 0.666667 & 0.350000 & 0.000000 \\
IA1 & Brief & 1.0000 & 1.000000 & 1.000000 & 1.000000 \\
\addlinespace[3pt]
IA5 & Guided & 1.0000 & 1.000000 & 1.000000 & 1.000000 \\
IA5 & Standard & 0.1417 & 0.666667 & 0.250000 & 0.000000 \\
IA5 & Record & 1.0000 & 1.000000 & 1.000000 & 1.000000 \\
IA5 & Brief & 0.9667 & 0.666667 & 1.000000 & 1.000000 \\
\addlinespace[3pt]
IA7 & Guided & 0.2583 & 0.333333 & 0.750000 & 0.000000 \\
IA7 & Standard & 0.1383 & 0.333333 & 0.350000 & 0.000000 \\
IA7 & Record & 0.2583 & 0.333333 & 0.750000 & 0.000000 \\
IA7 & Brief & 0.2583 & 0.333333 & 0.750000 & 0.000000 \\
\addlinespace[3pt]
IA9 & Guided & 0.3667 & 0.666667 & 1.000000 & 0.000000 \\
IA9 & Standard & 0.2583 & 0.333333 & 0.750000 & 0.000000 \\
IA9 & Record & 0.3667 & 0.666667 & 1.000000 & 0.000000 \\
IA9 & Brief & 0.9333 & 0.333333 & 1.000000 & 1.000000 \\
\addlinespace[3pt]
IB9 & Guided & 0.8200 & 1.000000 & 1.000000 & 0.700000 \\
IB9 & Standard & 0.1800 & 0.750000 & 0.350000 & 0.000000 \\
IB9 & Record & 0.5450 & 0.200000 & 0.750000 & 0.500000 \\
IB9 & Brief & 0.7700 & 1.000000 & 0.833333 & 0.700000 \\
\addlinespace[3pt]
IC3 & Guided & 0.4000 & 1.000000 & 1.000000 & 0.000000 \\
IC3 & Standard & 0.1833 & 0.333333 & 0.500000 & 0.000000 \\
IC3 & Record & 0.2167 & 0.666667 & 0.500000 & 0.000000 \\
IC3 & Brief & 1.0000 & 1.000000 & 1.000000 & 1.000000 \\
\addlinespace[3pt]
ID2 & Guided & 0.3339 & 0.339244 & 1.000000 & 0.000000 \\
ID2 & Standard & 0.0000 & 0.000000 & 0.000000 & 0.000000 \\
ID2 & Record & 0.0000 & 0.000000 & 0.000000 & 0.000000 \\
ID2 & Brief & 0.1839 & 0.339244 & 0.500000 & 0.000000 \\
\addlinespace[3pt]
IE2 & Guided & 0.3333 & 0.333333 & 1.000000 & 0.000000 \\
IE2 & Standard & 0.0333 & 0.333333 & 0.000000 & 0.000000 \\
IE2 & Record & 0.0333 & 0.333333 & 0.000000 & 0.000000 \\
IE2 & Brief & 0.1833 & 0.333333 & 0.500000 & 0.000000 \\
\addlinespace[3pt]
IF3 & Guided & 0.3667 & 0.666667 & 1.000000 & 0.000000 \\
IF3 & Standard & 0.3667 & 0.666667 & 1.000000 & 0.000000 \\
IF3 & Record & 0.3333 & 0.333333 & 1.000000 & 0.000000 \\
IF3 & Brief & 0.4000 & 1.000000 & 1.000000 & 0.000000 \\
\addlinespace[3pt]
IG8 & Guided & 1.0000 & 1.000000 & 1.000000 & 1.000000 \\
IG8 & Standard & 1.0000 & 1.000000 & 1.000000 & 1.000000 \\
IG8 & Record & 0.0000 & 0.000000 & 0.000000 & 0.000000 \\
IG8 & Brief & 0.3339 & 0.339329 & 1.000000 & 0.000000 \\
\addlinespace[3pt]
\end{longtable}
\endgroup

\subsection{Expanded task-score matrices}
The two vector panels below expand the complete per-task score view for reading alongside the exact values in Tables~\ref{tab:full-scores-easy}--\ref{tab:full-scores-impossible}. Both use the same color scale and configuration order as the manuscript.
\begin{figure}[!t]
\centering
\includegraphics[width=\textwidth]{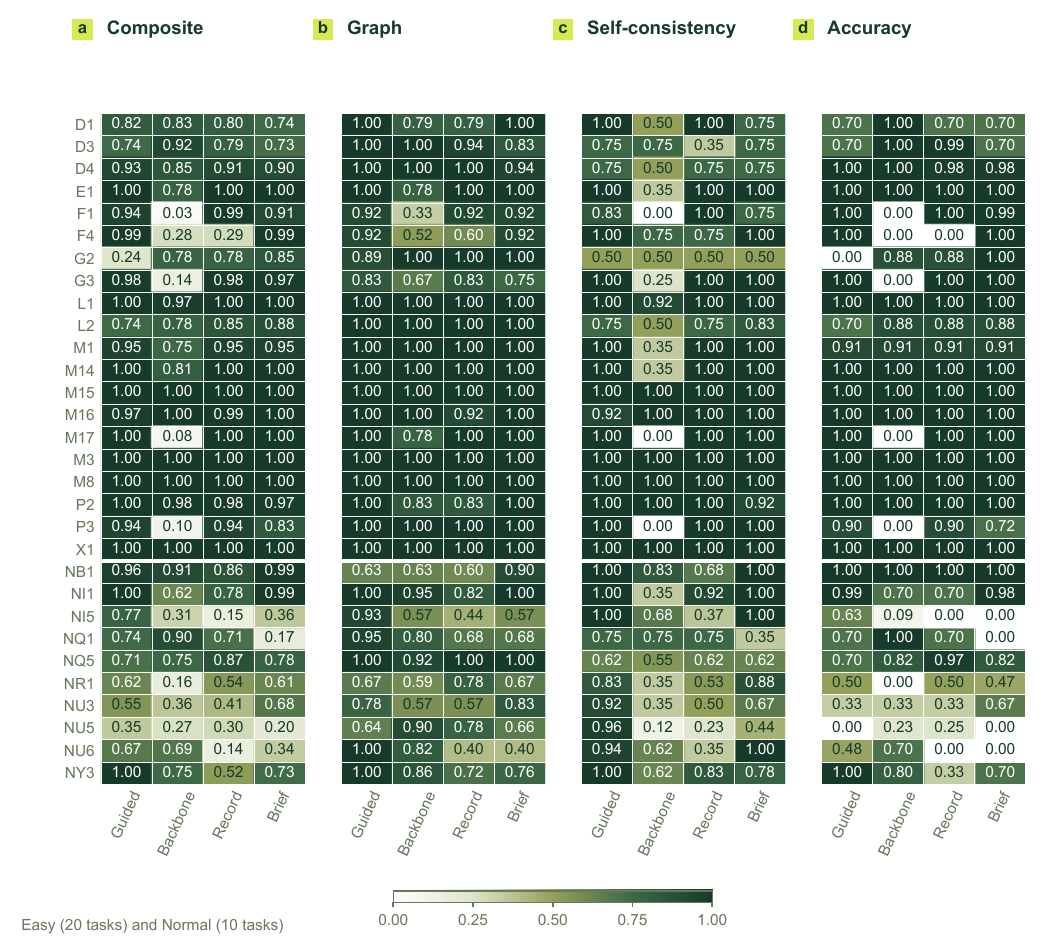}
\caption{All Easy and Normal task scores. Each task has four configuration rows and four score layers: composite, graph, scientific self-consistency, and accuracy. Cells are rounded to two decimals; the preceding tables and companion CSV retain greater precision. Backbone denotes Standard.}
\label{fig:si-task-matrix-en}
\end{figure}
\begin{figure}[!t]
\centering
\includegraphics[width=\textwidth]{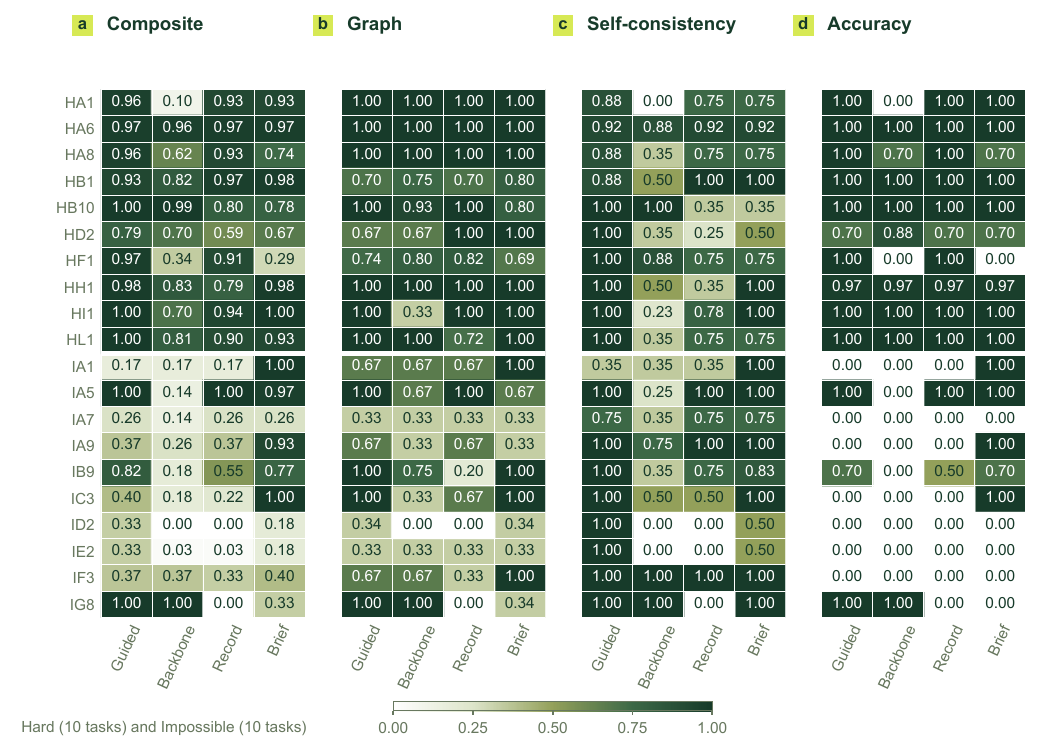}
\caption{Scores for all 80 Hard and Impossible runs, with the same component and configuration ordering as Figure~\ref{fig:si-task-matrix-en}.}
\label{fig:si-task-matrix-hi}
\end{figure}

\section{Behavioral diagnostics and archived research trajectories}
\label{sec:si-behavioral}

\subsection{Study materials and local inference settings}

We examined Cloudra's recorded research trajectories and constructed local questions about the molecules, geometries, and calculations in them. The trajectories retain the original model statements, tool inputs and outputs, and subsequent actions. The local tests vary which information the model receives and the order of its answer options. The supplementary data contain every question, expected answer, actual answer, request configuration, and source identifier.

The local deployment was archived as \texttt{qwen3.8:27b}, using a GGUF Q4\_K\_M quantization with a reported parameter size of 27.3B. The deployment digest is
\begin{quote}\small\ttfamily
22130167c4c20e20c7b71454612966ca8e8171e9b3cc8ab6ce8aa6cbfec79643.
\end{quote}
The local API reports a maximum context of 262,144 tokens. The short diagnostic requests use an 8,192-token context. Temperature is 0, seed 42, top-$k$ 0, top-$p$ 1, minimum-$p$ 0, repeat penalty 1, presence penalty 0, and draft prediction count 0. The initial identity and evidence-binding tests use the saved explicit ChatML rendering, an empty thinking block, thinking disabled, and a maximum of eight generated tokens; top-20 token log probabilities are retained. Native-chat controls use separate system and user messages, the model's native rendering, and a maximum of 1,024 generated tokens. The thinking setting is recorded for each request. A final choice is valid only if it exactly matches a permitted uppercase letter. The letter and its meaning are recorded separately.

\begin{table}[!htbp]
\centering\small
\caption{Experimental units and crossed conditions of the local diagnostic studies.}
\label{tab:si-diagnostic-design}
\begin{tabular}{p{0.24\linewidth}p{0.57\linewidth}r}
\toprule
Study & Frozen inputs and experimental factors & Calls\\
\midrule
Chemical identity & 16 pairs $\times$ 2 representations $\times$ 2 answer orders & 64\\
Distance judgment & 4 geometries $\times$ 2 information conditions $\times$ 2 answer orders $\times$ 2 thinking settings & 32\\
Target-angle feedback & 2 information conditions; 3 cyclic next-action permutations and 2 sufficiency permutations per condition & 10\\
Evidence associations & 4 chains $\times$ 2 association conditions $\times$ 2 presentations $\times$ 2 answer orders & 32\\
Native diagnostic follow-up & 20 identity inputs and 16 association inputs, each under 2 thinking settings & 72\\
Outstanding requirements & 3 state questions $\times$ 2 context conditions $\times$ 2 answer orders & 12\\
\bottomrule
\end{tabular}
\par\smallskip\begin{minipage}{\linewidth}\footnotesize
Representation, answer-order, and information variants use the same source molecule, geometry, or calculation chain. The single expanded-budget distance follow-up is additional to the 32-call study.
\end{minipage}
\end{table}

\subsection{Chemical identity under equivalent representations and answer reversal}

Sixteen molecular pairs were derived from four task systems. They include pairs representing the same molecule, different connectivities with the same formula, changes in connectivity and formal-charge location, different double-bond stereochemistry, and protonation changes. Each pair was presented in its original notation and an equivalent rewriting, with both answer orders. In the forward order, A means the same chemical identity and B means different identities; in the swapped order the meanings exchange. The questions instruct the model to compare connectivity, bond orders, formal charges, and specified stereochemistry, allowing differences in traversal order, ring digits, and disconnected-component order.

Across the resulting 64 judgments, 58 answers were correct. For the same pair and answer order, equivalent rewriting changed the chemical judgment in four of 32 comparisons. For the same pair and representation, reversing the options changed the judgment in six of 32 comparisons. All six selected A in both option orders. In the same-formula pair \texttt{CC1CC=C1} and \texttt{C=C1CCC1}, the double bond lies within and outside the ring, respectively. The model answered A under all four representation/order conditions, even though A changed from ``same'' to ``different'' when the options were reversed.

Identical strings test whether the model recognizes a representation as matching itself. Equivalent rewritings test recognition of the same molecule in another notation. Different-identity controls test whether it detects changed connectivity or stereochemistry despite a shared molecular formula. All strings and expected judgments are saved with the diagnostic data.

\subsection{XYZ coordinates and explicitly measured distances}

Four NB1 geometries were selected from the archived research trajectories. The local question asks which of two specified atom pairs has the shorter Euclidean distance. The original-XYZ condition supplies the complete geometry in its saved atom order. The measured-distance condition supplies program-calculated values for the same atom pairs. Each condition appears in both answer orders and in direct-answer and thinking modes, yielding 32 native chat calls. Table~\ref{tab:si-distance-stimuli} gives the measured values used to construct the paired conditions; all atom indices are one-based.

\begin{table}[!htbp]
\centering\small
\caption{Distance comparisons derived from the four archived NB1 geometries.}
\label{tab:si-distance-stimuli}
\begin{tabular}{llrlr}
\toprule
Source workflow & Pair 1 & $d_1$ (\AA) & Pair 2 & $d_2$ (\AA)\\
\midrule
Guided & C2--C4 & 2.029 & O3--H8 & 1.248\\
Brief & C4--H11 & 2.197 & O7--H11 & 0.981\\
Brief & C4--H11 & 1.388 & O7--H11 & 1.362\\
Brief & C4--H11 & 2.386 & O7--H11 & 0.968\\
\bottomrule
\end{tabular}
\end{table}

\begin{table}[!htbp]
\centering\small
\caption{Complete outcomes of the native distance-judgment controls.}
\label{tab:si-distance-outcomes}
\begin{tabular}{llrrrr}
\toprule
Thinking & Information & Calls & Correct & Valid wrong & Budget-truncated\\
\midrule
Off & XYZ coordinates & 8 & 4 & 4 & 0\\
Off & Measured distances & 8 & 8 & 0 & 0\\
On & XYZ coordinates & 8 & 7 & 0 & 1\\
On & Measured distances & 8 & 8 & 0 & 0\\
\bottomrule
\end{tabular}
\par\smallskip\begin{minipage}{\linewidth}\footnotesize
A budget-truncated generation reached the 1,024-token ceiling while still awaiting a final choice. It remains in the eight-input denominator. A separate 2,048-token follow-up of that one input produced the correct choice after 1,353 generated tokens.
\end{minipage}
\end{table}

All eight direct-answer XYZ responses selected A, giving four correct answers; reversing the options changed the judgment for every source geometry. Measured distances gave eight correct answers in each thinking setting, with the same judgment under reversal. With thinking enabled, median generated-token counts over the eight paired inputs were 807.5 for XYZ and 86 for measured distances, including thinking tokens. Program-calculated distances let the model compare the requested quantities directly, as in program-aided language models.\citep{ref15}

\subsection{Target-specific geometric feedback in an E1 operation chain}

In the E1 trajectory, the agent seeks an anti-periplanar H7--C2--C3--Cl4 dihedral near $180^\circ$ for the transferring hydrogen H7. It rotates the C3-side fragment about C2--C3 by $+120^\circ$ and $-120^\circ$ to produce candidates A and B, and sets the C3--Cl4 distance to 2.35 \AA. The default report gives a dihedral involving H5. The agent requests the H7 dihedral needed to choose between the candidates.

\begin{table}[!htbp]
\centering\small
\caption{Dihedral measurements and circular residuals for the saved E1 candidates.}
\label{tab:si-target-dihedrals}
\begin{tabular}{llrr}
\toprule
Candidate & Atom quadruplet & Signed angle ($^\circ$) & Distance to $180^\circ$\\
\midrule
A & H5--C2--C3--Cl4 & $-55.7831$ & 124.2169\\
A & H7--C2--C3--Cl4 & 59.3752 & 120.6248\\
B & H5--C2--C3--Cl4 & 64.2169 & 115.7831\\
B & H7--C2--C3--Cl4 & 179.3752 & 0.6248\\
\bottomrule
\end{tabular}
\end{table}

The manuscript's Newman projections use the saved coordinates viewed along C2$\to$C3, with projected bond directions normalized for display. Table~\ref{tab:si-target-dihedrals} gives the signed angles and circular distances from the target. The agent subsequently submits candidate B as the transition-state guess, citing the H7 anti-periplanar arrangement and stretched C3--Cl4 bond.

Ten local tests ask what to do next and whether the available measurements suffice. The H5-only inputs withhold H7 values and XYZ coordinates; the H5-plus-H7 inputs add the measured H7 values. In all three cyclic option permutations, adding H7 changes the next-action answer from requesting that measurement to selecting candidate B. In two paired sufficiency tests, it changes the answer from requesting more measurements to judging the information sufficient. The paired inputs use the same geometries and differ in the supplied measurements.

\subsection{Single-token transfer of a reported numerical fact}

The activation experiment uses the archived local quantized model for a reported-distance comparison.\citep{ref16} We changed the first reported distance from 2.029 to 1.029~\AA. The second distance remains 1.248~\AA. This reverses the correct comparison. The clean and edited prompts each contain 124 tokens; their token IDs differ only at global position 53, corresponding to the initial digit of the changed measurement. The original-fact answer is B in the forward option order and A in the reversed order.

For each option order, the residual-stream state at position 53 is captured from the original prompt and transferred into the edited prompt at zero-based layers 7, 15, 31, and 47. The four layers are fixed in the experiment manifest. The readout is the full-vocabulary softmax probability mass assigned to the original-fact answer, summing its bare-letter and leading-space spellings. Table~\ref{tab:si-fact-patch} preserves the clean, edited, and transferred-state values.

\begin{table}[!htbp]
\centering\small
\caption{Probability of the original-fact answer under single-position activation replacement.}
\label{tab:si-fact-patch}
\begin{tabular}{lrr}
\toprule
Input / intervention & Forward: original answer B & Reversed: original answer A\\
\midrule
Original input & 0.954032 & 0.996932\\
Edited input & 0.008694 & 0.073626\\
Transfer at layer 7 & 0.029099 & 0.955908\\
Transfer at layer 15 & 0.015033 & 0.717660\\
Transfer at layer 31 & 0.007961 & 0.072117\\
Transfer at layer 47 & 0.008763 & 0.069810\\
\bottomrule
\end{tabular}
\par\smallskip\begin{minipage}{\linewidth}\footnotesize
The readout uses the full vocabulary, with bare and leading-space answer tokens combined.
\end{minipage}
\end{table}

Two controls are executed in both option orders. Self-replacement transfers the edited prompt's own state at the changed token; the prefix control transfers at unchanged token position 0. Both preserve every vocabulary logit, with a maximum absolute difference of 0 from the edited baseline. With reversed options, transfers at layers 7 and 15 substantially increase the probability of the original-fact answer. With forward options, the changes are smaller and the edited answer remains preferred. Changing the state at the distance token therefore affects the answer in this test, with an effect that depends on option order.

\subsection{Evidence associations under fixed local calculation facts}

We used four calculation chains from three tasks to construct records with correct links and records with one changed link. The two NB1 chains come from the same run. Each record is presented as a flat list or with explicit links between calculations, in both answer orders, yielding $4\times2\times2\times2=32$ inputs. The changed link concerns the saddle-frequency geometry, IRC starting geometry, terminal-frame optimization input, or target species. Termination flags, convergence flags, energies, and imaginary-frequency values remain fixed.

The model is asked whether the calculations support the indicated reaction step. Linked inputs specify which geometry each frequency or IRC calculation uses, which terminal frame each endpoint optimization starts from, and which species the final minimum represents. The model answers all 16 linked inputs and 14 of 16 flat inputs correctly. Both flat-input errors occur with changed links and reversed options, in E1 and the second NB1 step. Explicit links state the calculation dependencies that the model must otherwise reconstruct from the list.\citep{ref17}

The native-chat follow-up uses the two chains on which these errors occurred, retaining every correct/changed-link, flat/linked, and option-order variant: 16 association questions per thinking mode. It also retains all four variants of five identity pairs with first-stage errors: 20 identity questions per mode. Table~\ref{tab:si-native-followup} reports all outcomes of these 72 calls. With thinking disabled, native chat gives 14/20 correct identity answers and 14/16 correct association answers, as in the initial tests. With thinking enabled, association answers reach 16/16; identity questions produce eight correct final choices and 12 generations truncated at the common 1,024-token limit.

\begin{table}[!htbp]
\centering\small
\caption{Complete outcomes for the selected native diagnostic follow-up.}
\label{tab:si-native-followup}
\begin{tabular}{llrrrr}
\toprule
Rendering / mode & Family & Calls & Correct & Valid wrong & Budget-truncated\\
\midrule
First-stage raw & Identity & 20 & 14 & 6 & 0\\
First-stage raw & Association & 16 & 14 & 2 & 0\\
Native, thinking off & Identity & 20 & 14 & 6 & 0\\
Native, thinking off & Association & 16 & 14 & 2 & 0\\
Native, thinking on & Identity & 20 & 8 & 0 & 12\\
Native, thinking on & Association & 16 & 16 & 0 & 0\\
\bottomrule
\end{tabular}
\par\smallskip\begin{minipage}{\linewidth}\footnotesize
The first-stage rows are the selected records from the existing 96-input identity/association study. The four native rows contain 72 additional calls. All saved input variants are included for each selected case; responses truncated by the output limit remain in the denominators.
\end{minipage}
\end{table}

\subsection{From a detected gap to the next research action}

The HA1 and NB1 trajectories show what the agents did after receiving diagnostics. In a Brief HA1 run, the agent repeatedly changed the source-minimum reference after compact feedback, while the IRC endpoint link remained unresolved. In a Guided NB1 run, the frequency calculation found zero imaginary modes and returned a normal-mode diagnostic. The Coordinator requested changes to the transition-state guess and optimization constraints; the next calculation used those changes. The saved sequences connect each diagnostic to the next instruction and calculation request.\citep{ref18}

In IA1, the delivery report has \texttt{complete=false}, a pending bidirectional short-IRC verification, and an empty formal-waiver list. After earlier positive comments about intermediate optimization, the Coordinator requests to stop with the study marked complete. The completion check returns the pending IRC requirement to the model. The next stop request marks the study partial.

The local IA1 test asks three questions: whether the latest report declares completion, whether a formal short-IRC waiver exists, and whether the missing calculation remains required. Each question is presented with the report alone or a quoted 12-message history, in both answer orders. All 12 answers match the recorded report fields and waiver list; all six option-reversed pairs preserve the same judgment.

\subsection{Discussion of the diagnostic scope}

The native follow-up includes all input variants of cases selected after first-stage errors.

Activation replacement measures the effect of a changed state on the quantized Qwen model's distance-comparison answer. In the E1 trajectory, Supervisor/workspace input intervenes between the measurements and the agent's use of candidate B. The separate local tests vary the supplied target measurements to examine their effect on the choice.

The completed IA1 tests use the report-only and 12-message inputs. A separate request with the full 306-message history contained 242,803 counted input tokens in a 262,144-token context. It reached the 3,600-s HTTP waiting limit while awaiting an answer, leaving the saved response empty. Of six planned full-history inputs, the first was attempted and the other five remain unattempted. This incomplete request is recorded separately from the 12 completed short-context judgments.

\section{Literature sample and source-linked capability register}
\label{sec:si-literature}
\subsection{Selection and coding procedure}
The literature sample contains 85 chemical AI agent systems, architectures, or workflows in the sources collected by 27 October 2026. Each included an LLM that planned calculations, used tools, modified inputs or code, or acted on feedback during an executed molecular, atomistic, or molecular coarse-grained calculation. Each separately described system or workflow counts once; journal and preprint versions and aliases are merged. The register lists the original sources, supporting passages, and coding decisions.

Systems were classified by what their examples studied: materials, crystals, and surfaces (39); finite molecules and complexes (23); liquids, soft matter, and biomolecules (12); or multiple classes (11). A periodic box containing a protein, liquid, or polymer was classified by its contents. Cloudra is shown separately in the manuscript figures.

Task codes mark activities demonstrated in the sources: T, structure construction/editing; E, separately demonstrated static energy, electronic structure, or material-property calculations; O, geometry optimization/relaxation; F, vibrations/frequencies and associated thermochemistry; R, explicit reaction paths or transition-state searches, including solid-state NEB; D, molecular or quantum dynamics. A record can have multiple codes. Their counts are 61, 61, 53, 22, 11, and 37, respectively. A blank leaves the activity unencoded and the system's capability in that category open. The register identifies coarse-grained normal modes and MD-derived free-energy profiles as such.

\subsection{Source register and counting variants}
Table~\ref{tab:si-literature} links each system to its primary article or official project record and any additional sources. The files \path{companion_data/literature/systems.json} and \path{companion_data/literature/source_index.csv} give the source location and supporting description for each code. They also list the demonstrated targets, boundary conditions, aliases, system families, and access limitations. Screening and search records include a crosswalk to the 49-system review used to find sources.

The main manuscript uses separately described systems/workflows. Grouping the specified El Agente, MDAgent, MatAgent/ChatMat, and DynaMate/DynaMate2 families gives 75 records (37 materials/surfaces, 17 molecules/complexes, 11 soft/biomolecular, and 10 multiple-class records). Excluding the three official-demonstration-only systems gives 82 records (38, 21, 12, 11); additionally excluding MLP Copilot gives 81 (37, 21, 12, 11). These counting variants are retained with their definitions in the source data.

\begingroup
\small
\setlength{\tabcolsep}{3pt}
\renewcommand{\arraystretch}{1.12}
\begin{longtable}{@{}p{0.05\linewidth}p{0.35\linewidth}p{0.17\linewidth}p{0.16\linewidth}p{0.17\linewidth}@{}}
\caption{Complete source-linked register of the 85 chemical AI agent systems/workflows used in the literature sample. Task codes are defined in the text. L01--L85 link to the primary records; lettered links give additional sources.}\label{tab:si-literature}\\
\toprule
No. & System/workflow & Demonstrated object class & Task codes & Original sources \\
\midrule
\endfirsthead
\multicolumn{5}{l}{\tablename\ \thetable\ (continued)}\\
\toprule
No. & System/workflow & Demonstrated object class & Task codes & Original sources \\
\midrule
\endhead
\midrule
\multicolumn{5}{r}{Continued on the next page}\\
\endfoot
\bottomrule
\endlastfoot
1 & Buehler molecular-DFT multi-agent demonstration & Molecules/complexes & T E & \href{https://pubs.acs.org/doi/abs/10.1021/acsengineeringau.3c00058}{L01}, \href{https://arxiv.org/abs/2310.19998}{a} \\
\addlinespace[2pt]
2 & ProtAgents & Soft/biomolecular & T F & \href{https://arxiv.org/html/2402.04268v1}{L02}, \href{https://arxiv.org/abs/2402.04268}{a} \\
\addlinespace[2pt]
3 & LLaMP & Materials/surfaces & T D & \href{https://arxiv.org/html/2401.17244v3}{L03} \\
\addlinespace[2pt]
4 & ChemReasoner & Materials/surfaces & E O T & \href{https://arxiv.org/html/2402.10980v3}{L04} \\
\addlinespace[2pt]
5 & AtomAgents & Materials/surfaces & T E O R & \href{https://arxiv.org/html/2407.10022v1}{L05} \\
\addlinespace[2pt]
6 & Adsorb-Agent & Materials/surfaces & T E O & \href{https://arxiv.org/html/2410.16658v4}{L06} \\
\addlinespace[2pt]
7 & LangSim & Materials/surfaces & T E O & \href{https://arxiv.org/html/2411.15221v1}{L07}, \href{https://www.mpie.de/5063016/LangSim}{a} \\
\addlinespace[2pt]
8 & MDCrow & Soft/biomolecular & D & \href{https://arxiv.org/html/2502.09565v1}{L08} \\
\addlinespace[2pt]
9 & MDAgent & Materials/surfaces & D & \href{https://www.nature.com/articles/s41598-025-92337-6}{L09} \\
\addlinespace[2pt]
10 & El Agente Q & Molecules/complexes & T E O F & \href{https://arxiv.org/abs/2505.02484}{L10} \\
\addlinespace[2pt]
11 & Aitomia & Molecules/complexes & T E O F & \href{https://arxiv.org/pdf/2505.08195}{L11} \\
\addlinespace[2pt]
12 & DynaMate (Mendible-Barreto et al.) & Multiple classes & T O D & \href{https://pubs.rsc.org/en/content/articlehtml/2025/me/d5me00062a}{L12}, \href{https://www.researchgate.net/publication/391996468_DynaMate_leveraging_AI-agents_for_customized_research_workflows}{a}, \href{https://github.com/omendibleba/DynaMate}{b} \\
\addlinespace[2pt]
13 & MatAgent (Lv et al.) & Molecules/complexes & T E & \href{https://onlinelibrary.wiley.com/doi/full/10.1002/mgea.70013}{L13} \\
\addlinespace[2pt]
14 & ChemGraph & Molecules/complexes & T E O F & \href{https://www.nature.com/articles/s42004-025-01776-9}{L14} \\
\addlinespace[2pt]
15 & TopoMAS & Materials/surfaces & T E & \href{https://arxiv.org/html/2507.04053v1}{L15} \\
\addlinespace[2pt]
16 & NAMD-Agent & Soft/biomolecular & T D & \href{https://arxiv.org/html/2507.07887v1}{L16}, \href{https://arxiv.org/html/2507.07887v2}{a} \\
\addlinespace[2pt]
17 & DREAMS & Materials/surfaces & T E O & \href{https://arxiv.org/html/2507.14267v2}{L17} \\
\addlinespace[2pt]
18 & SparksMatter & Materials/surfaces & T E O & \href{https://arxiv.org/html/2508.02956v1}{L18} \\
\addlinespace[2pt]
19 & AutoSolvate-Agent & Molecules/complexes & T E O F & \href{https://openreview.net/pdf?id=ATEtqdGn6W}{L19}, \href{https://openreview.net/pdf/395d4c9eda83111a078ba05d86adcf84ec77e4aa.pdf}{a}, \href{https://www.linkedin.com/in/fren5}{b} \\
\addlinespace[2pt]
20 & VASPilot & Materials/surfaces & E O & \href{https://arxiv.org/abs/2508.07035}{L20} \\
\addlinespace[2pt]
21 & ChemAgents molecular simulation workflow & Molecules/complexes & F D & \href{https://pubs.rsc.org/en/content/articlehtml/2026/sc/d5sc08794e}{L21}, \href{https://doi.org/10.26434/chemrxiv-2025-4kh1r}{a} \\
\addlinespace[2pt]
22 & AMLP & Materials/surfaces & E O D & \href{https://arxiv.org/html/2509.21647v1}{L22} \\
\addlinespace[2pt]
23 & ToPolyAgent & Soft/biomolecular & T D & \href{https://arxiv.org/html/2510.12091v1}{L23}, \href{https://arxiv.org/abs/2510.12091}{a} \\
\addlinespace[2pt]
24 & GENIUS & Materials/surfaces & E O & \href{https://arxiv.org/html/2512.06404v1}{L24}, \href{https://www.nature.com/articles/s43246-026-01167-0}{a} \\
\addlinespace[2pt]
25 & LAMMPS-Agents & Materials/surfaces & T E O F D & \href{https://github.com/ANL-NST/LAMMPS-Agents}{L25}, \href{https://doi.org/10.1039/D5DD00435G}{a}, \href{https://www.alcf.anl.gov/publications/multi-agentic-ai-framework-end-end-atomistic-simulations}{b} \\
\addlinespace[2pt]
26 & DynaMate (protein--ligand) & Soft/biomolecular & T O D & \href{https://arxiv.org/html/2512.10034v1}{L26}, \href{https://openreview.net/pdf?id=NkLAjIzwei}{a} \\
\addlinespace[2pt]
27 & AGAPI-Agents & Materials/surfaces & T E O & \href{https://arxiv.org/html/2512.11935v1}{L27} \\
\addlinespace[2pt]
28 & MASTER & Materials/surfaces & T E O & \href{https://arxiv.org/html/2512.13930v1}{L28} \\
\addlinespace[2pt]
29 & El Agente Cuántico & Molecules/complexes & E & \href{https://arxiv.org/html/2512.18847v3}{L29}, \href{https://arxiv.org/abs/2512.18847}{a} \\
\addlinespace[2pt]
30 & VASP Agent & Materials/surfaces & E O R & \href{https://arxiv.org/html/2512.19458v1}{L30} \\
\addlinespace[2pt]
31 & Masgent & Materials/surfaces & T E O & \href{https://arxiv.org/pdf/2512.23010}{L31} \\
\addlinespace[2pt]
32 & CASCADE & Materials/surfaces & E O & \href{https://arxiv.org/html/2512.23880v1}{L32} \\
\addlinespace[2pt]
33 & MDAgent2-RUNTIME & Materials/surfaces & D & \href{https://arxiv.org/html/2601.02075v1}{L33}, \href{https://github.com/FredericVAN/PKU_MDAgent2}{a} \\
\addlinespace[2pt]
34 & CatMaster & Materials/surfaces & T E O F R & \href{https://arxiv.org/html/2601.13508v3}{L34} \\
\addlinespace[2pt]
35 & QUASAR & Multiple classes & T E O D & \href{https://arxiv.org/html/2602.00185v1}{L35} \\
\addlinespace[2pt]
36 & ChatMat & Molecules/complexes & T E & \href{https://www.researchgate.net/publication/404118299_ChatMat_A_Multi-Agent_Chemist_for_Autonomous_Material_Prediction_and_Exploration}{L36}, \href{https://doi.org/10.1039/D5DD00582E}{a}, \href{https://zenodo.org/records/18459314}{b}, \href{https://github.com/Xiaobu-USTC/ChatMat}{c} \\
\addlinespace[2pt]
37 & El Agente Estructural & Molecules/complexes & T O & \href{https://arxiv.org/html/2602.04849v1}{L37} \\
\addlinespace[2pt]
38 & El Agente Quntur & Molecules/complexes & T E O F R & \href{https://arxiv.org/html/2602.04850v3}{L38} \\
\addlinespace[2pt]
39 & El Agente Gráfico & Multiple classes & T E O F & \href{https://arxiv.org/html/2602.17902v1}{L39}, \href{https://arxiv.org/html/2602.17902v2}{a} \\
\addlinespace[2pt]
40 & El Agente Sólido & Materials/surfaces & T E O F & \href{https://arxiv.org/html/2602.17886v1}{L40} \\
\addlinespace[2pt]
41 & Catalyst-Agent & Materials/surfaces & T E O & \href{https://arxiv.org/html/2603.01311v1}{L41} \\
\addlinespace[2pt]
42 & TritonDFT & Materials/surfaces & E O F & \href{https://arxiv.org/html/2603.03372v1}{L42} \\
\addlinespace[2pt]
43 & Chelatron & Molecules/complexes & T E & \href{https://science.osti.gov/-/media/About/pdf/scac/meetings/2026/lab-contributions-to-GM-03-24-2026v2.pdf}{L43} \\
\addlinespace[2pt]
44 & Electrolyte Agent / Reasoning-to-Simulation & Soft/biomolecular & T D & \href{https://openreview.net/pdf/837f62f17b0f4210dbc1260e813b7fd1ebabd467.pdf}{L44}, \href{https://zenodo.org/records/19198509}{a} \\
\addlinespace[2pt]
45 & MatSciAgent & Materials/surfaces & T D & \href{https://www.nature.com/articles/s43246-025-00994-x}{L45} \\
\addlinespace[2pt]
46 & OpenClaw computational-chemistry skills workflow & Molecules/complexes & T O D & \href{https://arxiv.org/html/2603.25522v1}{L46} \\
\addlinespace[2pt]
47 & Protomia & Molecules/complexes & T E F & \href{https://aitomistic.com/protomia/docs/research-workflow}{L47}, \href{https://arxiv.org/html/2608.18508v1}{a} \\
\addlinespace[2pt]
48 & SimMOF & Materials/surfaces & T E O D & \href{https://arxiv.org/pdf/2603.29152}{L48} \\
\addlinespace[2pt]
49 & PolyJarvis & Soft/biomolecular & T D & \href{https://arxiv.org/html/2604.02537v3}{L49} \\
\addlinespace[2pt]
50 & FermiLink & Multiple classes & E D & \href{https://arxiv.org/html/2604.03460v1}{L50} \\
\addlinespace[2pt]
51 & MatClaw & Materials/surfaces & E D & \href{https://arxiv.org/html/2604.02688v1}{L51} \\
\addlinespace[2pt]
52 & El Agente Forjador & Molecules/complexes & T E O F D & \href{https://arxiv.org/html/2604.14609v1}{L52} \\
\addlinespace[2pt]
53 & MDAgent (biomolecular, 2026) & Soft/biomolecular & T O D & \href{https://arxiv.org/pdf/2604.18622}{L53}, \href{https://arxiv.org/abs/2604.18622}{a} \\
\addlinespace[2pt]
54 & VirtualLab\_CC & Molecules/complexes & T E O F & \href{https://www.sciencedirect.com/science/article/pii/S2949747726000230}{L54}, \href{https://github.com/CCBG-Lab/VirtualLab_CC}{a}, \href{https://doi.org/10.26434/chemrxiv.15002407/v1}{b} \\
\addlinespace[2pt]
55 & XDFT & Materials/surfaces & T E & \href{https://arxiv.org/pdf/2604.26703v1}{L55} \\
\addlinespace[2pt]
56 & CatGo & Multiple classes & T E O F R & \href{https://www.researchgate.net/publication/410982627_CatGo_Bridging_CLI_coding_agents_with_interactive_structure_and_workflow_management_for_computational_chemistry}{L56}, \href{https://doi.org/10.1039/D6DD00273K}{a}, \href{https://zenodo.org/records/21552845}{b}, \href{https://doi.org/10.26434/chemrxiv.15002984/v1}{c} \\
\addlinespace[2pt]
57 & TSAgent & Materials/surfaces & E O F R & \href{https://arxiv.org/html/2605.14154v1}{L57} \\
\addlinespace[2pt]
58 & Lang2MLIP & Materials/surfaces & T E D & \href{https://arxiv.org/html/2605.14527v1}{L58} \\
\addlinespace[2pt]
59 & AtomisticSkills & Multiple classes & T E O D & \href{https://arxiv.org/html/2605.24002v1}{L59} \\
\addlinespace[2pt]
60 & DynaMate2 & Soft/biomolecular & T D & \href{https://arxiv.org/pdf/2605.20819v2}{L60}, \href{https://arxiv.org/abs/2605.20819}{a} \\
\addlinespace[2pt]
61 & ChatMOSP & Materials/surfaces & T & \href{https://arxiv.org/pdf/2605.24265}{L61} \\
\addlinespace[2pt]
62 & AutoDFT & Multiple classes & E O F D & \href{https://arxiv.org/html/2605.26179v1}{L62} \\
\addlinespace[2pt]
63 & MLIPilot & Multiple classes & E O D & \href{https://arxiv.org/html/2605.30889v1}{L63} \\
\addlinespace[2pt]
64 & CatDT & Materials/surfaces & T E O F R & \href{https://arxiv.org/html/2606.05050v2}{L64} \\
\addlinespace[2pt]
65 & Paimon & Multiple classes & T O D & \href{https://arxiv.org/html/2606.09422v1}{L65} \\
\addlinespace[2pt]
66 & MDForge & Soft/biomolecular & D & \href{https://arxiv.org/html/2606.12916v2}{L66} \\
\addlinespace[2pt]
67 & URSA LAMMPS agent & Materials/surfaces & T E D & \href{https://arxiv.org/html/2607.22596v1}{L67}, \href{https://arxiv.org/abs/2607.22596}{a} \\
\addlinespace[2pt]
68 & AdsMind & Materials/surfaces & T E O & \href{https://arxiv.org/html/2606.19152v1}{L68} \\
\addlinespace[2pt]
69 & LADeQ & Molecules/complexes & E & \href{https://arxiv.org/html/2606.20729v1}{L69} \\
\addlinespace[2pt]
70 & MutexaGPT & Soft/biomolecular & T D & \href{https://pmc.ncbi.nlm.nih.gov/articles/PMC13308407/}{L70}, \href{https://doi.org/10.1038/s43588-026-01049-y}{a} \\
\addlinespace[2pt]
71 & Agent-MD & Materials/surfaces & D & \href{https://arxiv.org/html/2608.07637v1}{L71} \\
\addlinespace[2pt]
72 & CGMas & Soft/biomolecular & T O D & \href{https://arxiv.org/html/2608.06694v1}{L72} \\
\addlinespace[2pt]
73 & matmcp & Materials/surfaces & T E O & \href{https://github.com/Crystalhihihi/matmcp}{L73}, \href{https://raw.githubusercontent.com/Crystalhihihi/matmcp/main/docs/experiments.md}{a}, \href{https://github.com/Crystalhihihi/matmcp/tree/main/evidence/20261009-225541_mos2_d84a}{b} \\
\addlinespace[2pt]
74 & Evolving TS workflows & Molecules/complexes & O F R & \href{https://www.nature.com/articles/s41524-026-02301-9}{L74} \\
\addlinespace[2pt]
75 & ARCHE & Molecules/complexes & T E O F R & \href{https://arxiv.org/abs/2609.11147}{L75} \\
\addlinespace[2pt]
76 & Multireference QC agent & Molecules/complexes & E & \href{https://arxiv.org/html/2609.13357v1}{L76} \\
\addlinespace[2pt]
77 & El Agente Potente & Multiple classes & T E O F R D & \href{https://arxiv.org/html/2609.14840v1}{L77} \\
\addlinespace[2pt]
78 & PVmatAgent & Materials/surfaces & T E O & \href{https://onlinelibrary.wiley.com/doi/10.1002/mgea.70102}{L78}, \href{https://github.com/Albertdeng23/PVMatAgent}{a}, \href{https://doi.org/10.26434/chemrxiv.15003915/v1}{b} \\
\addlinespace[2pt]
79 & Agent-E2MD & Materials/surfaces & E O D & \href{https://arxiv.org/html/2609.26657v1}{L79} \\
\addlinespace[2pt]
80 & SQD-Agent & Molecules/complexes & E & \href{https://arxiv.org/html/2609.39302v1}{L80} \\
\addlinespace[2pt]
81 & QAOS / Quantum Agentic Operating System & Molecules/complexes & T E O & \href{https://arxiv.org/html/2610.05304v1}{L81}, \href{https://arxiv.org/abs/2610.05304}{a} \\
\addlinespace[2pt]
82 & MLP Copilot & Materials/surfaces & E D & \href{https://pmc.ncbi.nlm.nih.gov/articles/PMC13684298/}{L82}, \href{https://github.com/flarecentury/mlpcopilot}{a} \\
\addlinespace[2pt]
83 & Magnolia & Multiple classes & T E O D & \href{https://arxiv.org/html/2610.27230v1}{L83} \\
\addlinespace[2pt]
84 & GBClaw & Materials/surfaces & T E O R & \href{https://arxiv.org/pdf/2610.28629}{L84}, \href{https://zhugroup.gatech.edu/gbclaw-demo}{a} \\
\addlinespace[2pt]
85 & PULSES & Molecules/complexes & T E O D & \href{https://arxiv.org/pdf/2610.32392}{L85}, \href{https://arxiv.org/abs/2610.32392}{a} \\
\addlinespace[2pt]
\end{longtable}
\endgroup

\section{Data, software, and reproduction materials}
\label{sec:si-reproduction}
\subsection{Public data availability}
The study data are publicly available with read-only access at \url{https://drive.google.com/drive/folders/1yZo3eJuCUuAgedQztgCtLyJcc8MZipf7}. This directory provides the full P2 campaign archive \path{Cloudra-P2-final-20261031-FULL.tar.gz}, the 48-run P1b repeated-task archive and its offline-organization supplements, the Gold reference-calculation evidence snapshot, and the compact SI companion-data archive \path{Cloudra_SI_Data.zip}. The package versions, inventories, and checksums are provided with the downloads.

The P2 archive contains the 200 full-benchmark runs; the P1b archive contains the separate 48 repeated-task runs. The Gold archive contains reference calculation inputs, native outputs, geometries, Hessians, and curation records. The SI companion archive contains task definitions, full-precision scores, behavioral records, and the materials used to reproduce the publication tables and figures.

The separate \href{https://drive.google.com/drive/folders/1Zkn6_2aDlloAe-bmfd7w27zxmQKj5L9K}{complete behavioral-source archive (20261030)} contains frozen inputs and responses, activation states and logits, readable event traces, source databases and calculation records, analysis scripts, and vector figures. Its topic index and manuscript-context files identify the section and figure numbers used in that manuscript version.

\subsection{Machine-readable task and score records}
The supplementary source package contains \path{companion_data/cloudrabench_50_task_catalogue.csv}, the 50 task JSON files in \path{companion_data/task_definitions/}, and \path{companion_data/cloudrabench_200_run_scores.csv}. The task files contain molecular inputs, task text, permitted methods, Gold graphs, and scoring settings. The score table stores all 200 composites and 600 components at their original precision. A manifest records file hashes and the precision displayed in the printed tables.

The P2 publication-source package contains the final evaluations, paired summaries, Brief/Impossible analyses, extracted timing data, and field sources used in the figures. The manifest \path{companion_data/evaluation_manifest.json} records the base evaluator revision \texttt{0961d9ab}, final reference task-file hashes, and the 2~November target-correspondence postprocess. All 200 organized graphs underwent candidate, channel, and electronic-state matching checks; the 12 runs for three affected tasks underwent native-evidence Accuracy replay, changing five Accuracy values and their composites. Graph and Self-consistency remain those of the organized records. The reproduction script reads these records and applies the correction CSV to assemble the final 200-run table; the plotting scripts use the resulting tables.

\subsection{Software responsibilities and calculation records}
Cloudra revision 70daa5a identifies the implementation described in the main-text methods.\citep{ref14} The source is publicly available at \url{https://github.com/yyywaa/Cloudra-mechanism-agent} under the GNU AGPL. Reproducible tables and scores are in the companion data package. RDKit and ETKDG supply molecular representations and coordinate embedding; MMFF94 supplies force-field relaxation.\citep{ref12,etkdg2020,ref13,mmff941996} NumPy supplies numerical array/linear-algebra operations, NetworkX supplies graph representation and matching, and cclib works with engine-specific readers to extract quantum-chemical quantities.\citep{numpy2020,networkx2008,cclib2008} ASE supplies the NEB/MDMin implementation used in standard IDPP, and Adam is used in the lightweight interpolation path.\citep{ref08,neb2000,ase2017,adam2015} Task-name semantic projections use BGE-M3 embeddings and scikit-learn PCA.\citep{bgem32024,sklearn2011} Publication tables are assembled with pandas, and the quantitative figures are drawn with Matplotlib.\citep{pandas2010,matplotlib2007}

Each run records the program, model/API name, charge, spin, method, basis, and keywords used for its calculations and model requests. These records identify the experimental versions; dependency lower bounds specify installation requirements. Native outputs contain stationary-point energies, frequencies, normal modes, IRC geometries, and the links to endpoint optimization and frequency calculations. Numerical barriers use the electronic-energy difference in Equation~\ref{eq:si-barrier}.

\subsection{Reproduction entry points}
The Cloudra source is publicly available at \url{https://github.com/yyywaa/Cloudra-mechanism-agent}. Reconstruct the retained P2 publication tables and figures from the companion data package by running the following commands in \path{companion_data/publication_reproduction/}:
\begin{quote}\small\ttfamily
python scripts/benchmark\_p2/prepare\_publication\_data.py\\
python scripts/benchmark\_p2/plot\_benchmark\_results.py\\
python scripts/benchmark\_p2/plot\_speed\_delivery.py\\
python scripts/benchmark\_p2/plot\_imp.py
\end{quote}
That directory in the companion-data archive has the same layout as the repository and includes these scripts. The archive also includes the timing extractions and frozen behavioral requests and responses. The activation archive contains the prompts, intervention commands, captured states, and full-vocabulary logits used to calculate the reported probabilities and control logit differences. Replay and analysis scripts are in \path{companion_data/behavioral_scripts/}; model weights remain outside the archive. The literature directory contains the complete register, screening records, and search history. Editable TeX files, vector figures, and the bibliography accompany this SI.

\subsection{Data-package inventory}
The \href{https://drive.google.com/file/d/1lXNR6zhWTP2kJ1B5_-A9elXmblQX5ALf/view}{SI companion-data archive} is distributed separately from the compact Overleaf typesetting project. Unpack \path{Cloudra_SI_Data.zip} beside the SI source to obtain the referenced \path{companion_data/} directory. Its subdirectories contain the following materials:
\begin{itemize}[nosep,leftmargin=1.5em]
\item \path{task_definitions/}: all 50 final reference task definitions, including molecular inputs and rubrics.
\item \path{publication_reproduction/}: publication input records, field provenance, bootstrap summaries, timing records, and table/figure scripts.
\item \path{behavioral/}: frozen input sets, actual request/response records, deployment snapshots, geometric/evidence source cases, and activation readouts.
\item \path{behavioral_scripts/}: the corresponding intervention and diagnostic code.
\item \path{revision_20261031/}: deployment/version metadata, archive-relative configuration and request summaries, inspected Easy-task trajectories, and historical Impossible assignment audits retained for provenance.
\item \path{revision_20261101/}: the original-run NI1 geometries and native-output provenance, uncapped barrier-error analysis, and inventories of the archived completion records.
\item \path{revision_20261102/target_matching/}: the official target-correspondence Accuracy overlay applied to the publication table, including before/after Accuracy reports, the matching audit, and \path{target_matching_accuracy_corrections.csv}.
\item \path{literature/}: all 85 source-linked systems and the coding/search records.
\end{itemize}
A file-level SHA-256 inventory accompanies the archive.

\bibliographystyle{unsrtnat}
\bibliography{supplement/references}